\documentclass[journal=aesccq]{achemso}
\pdfoutput=1
\usepackage{lineno,hyperref}
\usepackage{mathtools}
\usepackage{enumitem}
\usepackage{fancyhdr}
\usepackage{textcomp}
\usepackage{multirow}
\usepackage{arydshln} 
\usepackage[version=4]{mhchem}
\usepackage{url}
\usepackage{soul,color}
\usepackage{colortbl}
\usepackage{multirow}
\usepackage{float}
\usepackage{caption}
\usepackage{graphicx}
\usepackage{array}
\usepackage{enumitem}
\usepackage{booktabs}
\usepackage{makecell} 
\usepackage{microtype}
\usepackage{lmodern}
\usepackage{mathpazo}
\usepackage{xkeyval}
\usepackage{setspace}
\usepackage{natbib}
\usepackage{mciteplus}
\usepackage{listings}
\usepackage{wasysym}

\newcommand{\equationame}{Eq.}

\usepackage{amssymb}

\makeatletter
\newcommand*{\smallrel}[2][.8]{%
  \mathrel{\mathpalette{\smallrel@{#1}}{#2}}%
}
\newcommand*{\smallrel@}[3]{%
  \sbox0{$#2\vcenter{}$}%
  \dimen@=\ht0 %
  \raise\dimen@\hbox{%
    \scalebox{#1}{%
      \raise-\dimen@\hbox{$#2#3\m@th$}%
    }%
  }%
}
\makeatother

\title{\Large Thermal desorption of interstellar ices \\\large A review on the controlling parameters and their implications from snowlines to chemical complexity}

\author{Marco Minissale}\email{marco.minissale@univ-amu.fr}\affiliation{\scriptsize Aix Marseille Univ, CNRS, PIIM, Marseille, France}
\author{Yuri Aikawa}\affiliation{Department of Astronomy, Graduate School of Science, The University of Tokyo, 7-3-1 Hongo, Bunkyo-ku, Tokyo 113-0033, Japan}
\author{Edwin Bergin}\affiliation{Department of Astronomy, University of Michigan, Ann Arbor, MI, USA} \author{M. Bertin}\affiliation{Sorbonne Universit\'e, Observatoire de Paris, PSL University, CNRS, LERMA, Paris, France}
\author{Wendy A. Brown}\affiliation{Division of Chemistry, University of Sussex, Falmer, Brighton, BN1 9QJ, UK}
\author{Stephanie Cazaux}\affiliation{Faculty of Aerospace Engineering, Delft University of Technology, Delft, Netherlands}
\author{Steven B. Charnley}\affiliation{NASA Goddard Space Flight Center, USA }
\author{Audrey Coutens}\affiliation{Institut de Recherche en Astrophysique et Plan\'etologie, Universit\'e de Toulouse, UPS-OMP, CNRS, CNES, 9 Av. du Colonel Roche, 31028 Toulouse Cedex 4, France}
\author{Herma M. Cuppen}\affiliation{Institute for Molecules and Materials, Radboud University, Heyendaalseweg 135, 6525 AJ Nijmegen, The Netherlands}
\author{Victoria Guzman}\affiliation{Joint ALMA Observatory (JAO), Alonso de C\'ordova 3107, Vitacura, Santiago de Chile, Chile}
\author{Harold Linnartz}\affiliation{Laboratory for Astrophysics, Leiden Observatory, Leiden University, PO Box 9513, 2300 RA, Leiden, The Netherlands}
\author{Martin R. S. McCoustra}\affiliation{Institute of Chemical Sciences, Heriot-Watt University, Edinburgh, EH14 4AS, UK}
\author{Albert Rimola}\affiliation{Departament de Qu\'imica, Universitat Aut\`onoma de Barcelona, 08193 Bellaterra, Catalonia, Spain}
\author{Johanna G.M. Schrauwen}\affiliation{Institute for Molecules and Materials, Radboud University, Heyendaalseweg 135, 6525 AJ Nijmegen, The Netherlands}
\author{Celine Toubin}\affiliation{Univ. Lille, CNRS, UMR 8523 - PhLAM - Physique des Lasers Atomes et Mol\'ecules, F-59000 Lille, France }
\author{Piero Ugliengo}\affiliation{Dipartimento di Chimica and NIS Centre, Universit\'a degli Studi di Torino, 10125 Torino, Italy}
\author{Naoki Watanabe}\affiliation{Institute of Low Temperature Science, Hokkaido University, Sapporo, Hokkaido, 060-0819, Japan} 
\author{Valentine Wakelam} \affiliation[8]{Laboratoire d'astrophysique de Bordeaux, Univ. Bordeaux, CNRS, B18N, all\'ee Geoffroy Saint-Hilaire, 33615 Pessac , France} 
\author{Francois Dulieu}\affiliation{CY Cergy Paris Universit\'e, Sorbonne Universit\'e, Observatoire de Paris, PSL University, CNRS, LERMA, F-95000 Cergy, France}

\begin{document}

\begin{abstract}

The evolution of star-forming regions and their thermal balance are strongly influenced by their chemical composition, that, in turn, is determined by the physico-chemical processes that govern the transition between the gas phase and the solid state, specifically icy dust grains (e.g., particles adsorption and desorption). Gas-grain and grain-gas transitions as well as formation and sublimation of interstellar ices are thus essential elements of understanding astrophysical observations of cold environments (e.g., pre-stellar cores) where unexpected amounts of a large variety of chemical species have been observed in the gas phase. Adsorbed atoms and molecules also undergo chemical reactions which are not efficient in the gas phase. Therefore the parameterization of the physical properties of atoms and molecules interacting with dust grain particles is clearly a key aspect to interpret astronomical observations and to build realistic and predictive astrochemical models. 
In this consensus evaluation, we focus on parameters controlling the thermal desorption of ices and how these determine pathways towards molecular complexity and define the location of snowlines, which ultimately influence the planet formation process.\\
We review different crucial aspects of desorption parameters both from a theoretical and experimental point of view. We critically assess the desorption parameters (the  binding energies $E_b$ and the pre-exponential factor $\nu$) commonly used in the astrochemical community for astrophysically relevant species and provide tables with recommended values. The aim of these tables is to provide a coherent set of critically assessed  desorption parameters for common use in future work. In addition, we show that a non-trivial determination of the pre-exponential factor $\nu$ using the Transition State Theory can affect the binding energy value. \\
The primary focus is on pure ices, but we also discuss the desorption behavior of mixed, i.e. astronomically more realistic ices. This allows discussion of segregation effects.  Finally, we conclude this work by discussing the limitations of theoretical and experimental approaches currently used to determine the desorption properties with suggestions for future improvements. 
\end{abstract}

\section{Introduction}

\begin{figure}[ht]
\centering
\includegraphics[width=0.5\textwidth]{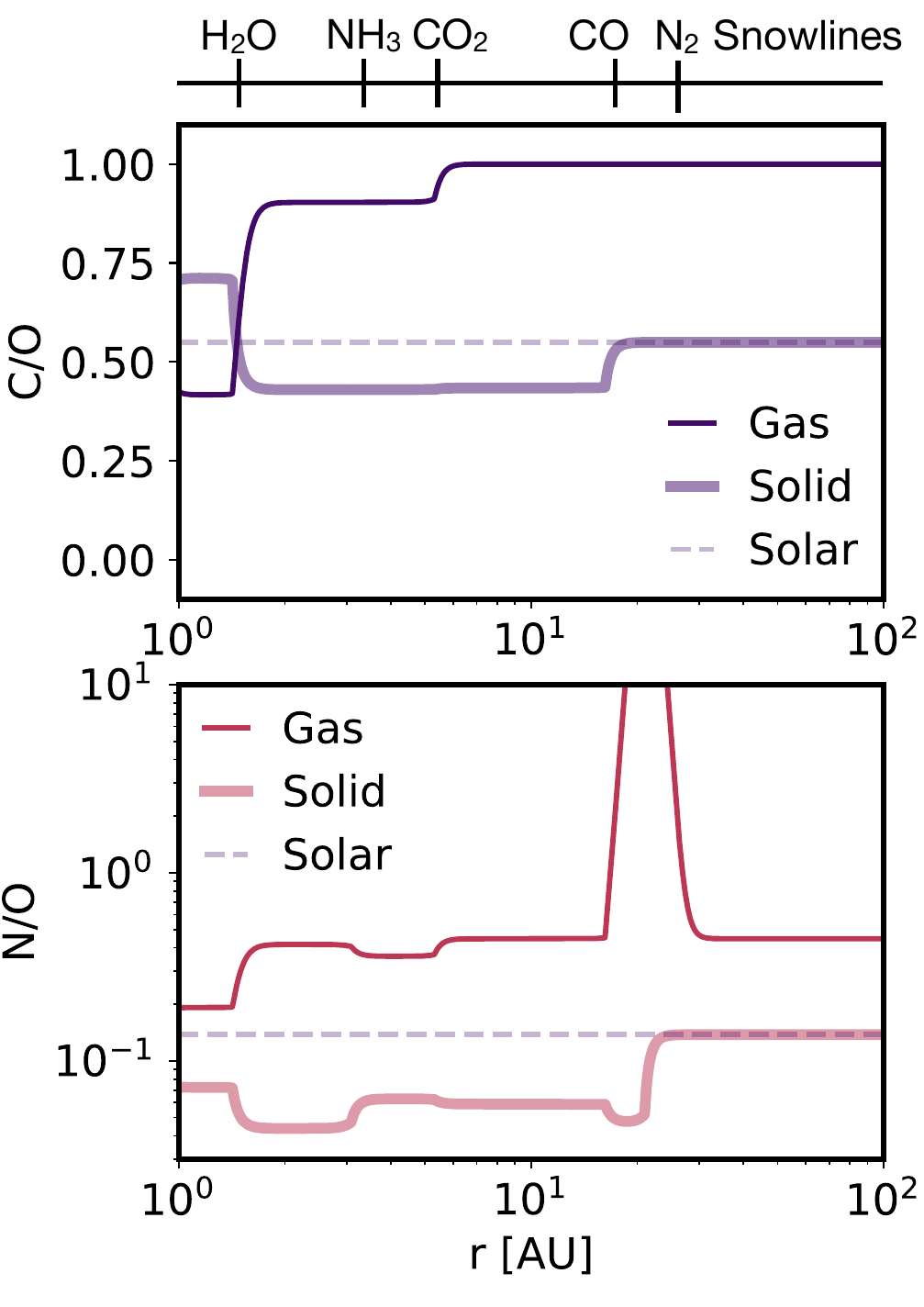}
\caption{Model of the bulk elemental composition ratios within a typical protoplanetary disk.  The temperature of gas and solids in the disk decays with radius leading to sequential freeze-out of volatile carriers of O (H$_2$O, CO, CO$_2$, organics, silicates), C (CO, CO$_2$, organics, carbonaceous grains), and N (NH$_3$, N$_2$, organics).  This sequential freeze-out produces sharp transitions in the bulk elemental composition ratios as these carriers transition from vapor to the solid state across ice lines.  Figure reproduced from \"{O}berg \& Bergin 2021\citep{Oberg21} with model description provided there.} \label{fig:01_01}
\end{figure}

Interactions between the gas phase and the solid state are key processes that play a critical role during the full cosmochemical evolution, from translucent and dark clouds to prestellar and protoplanetary disks up to the formation of new planets. Ingredients of terrestrial worlds, the solid cores of gas giants, asteroids, moons, and comets include icy solids that act as the chemical memory of this evolution and may inherit the pristine interstellar material comprising of a mixture of refractory and ices\citep{Encrenaz08, Oberg21}. 
Once a young stellar object is born, different gas phase / solid state abundances can be found for a range of molecules, varying as a function of distance to the new star. This is largely determined by the temperatures for which ice constituents are still frozen out or already have thermally desorbed. In the infalling envelope of protostars, for example, bright emission of various organic molecules such as CH$_3$OH are found in the central hot region ($\ge$ 100 K, $\sim$100 au), while unsaturated hydrocarbons are enhanced at the radius of CH$_4$ sublimation ($\sim$25 K, a few 1000 au) \cite{sakai2013}. In protoplanetary disks, the border regions of ice sublimation are known as snowlines. 
The compositional mixture for the initial stages of rocky planets is set by the snowlines of key volatile species \citep{Oberg11a} and this has been linked to the question of the origin of Earth's own water \citep{Hayashi81, DAngelo19}.
 A reverse picture is seen for gas-giants whose atmospheres reflect the gas composition within their birth zone \citep{parker2020}. Figure~\ref{fig:01_01} depicts the drastic change of the elemental C/O and N/O abundance ratios in the gas and solids due to the ice-to-gas transition of the major carrier of these elements as the mass accretes towards the central star. The bulk elemental composition of solids (refractories and ices) and the gas changes within a planet-forming disk as a function of distance from the star as the temperature decays produces vapor to ice transitions that depend on the sublimation temperature of key carriers.
 Further, it has also long been theorized that the water snowline at least is a special location to initiate planet formation itself, due the expectation of pressure changes across the boundary \citep{lunine1985}.
Thus, the physical and chemical response of icy dust grains and ices is of fundamental relevance in the context of planet formation.

A central aspect of this composition is the volatility of a given material - that is set by the binding energy of the involved molecules - which is highly dependent upon composition and the local conditions.   Because the pressures ($\sim 10^{-15}$ bar\footnote{1 bar is equivalent to 10$^6$ Dynes/cm$^2$.}) and temperatures (10-20~K) of molecular clouds are both low compared to Earth laboratory standards (1 bar, 300~K), efficient chemical pathways completely differ from terrestrial ones. Under interstellar conditions within regions of star and planetary birth, most molecules are adsorbed onto grain surfaces via weak van der Waals interactions, governed by electrostatic and dispersion interactions and referred to as physisorption, with an interaction strength of the order of $0.1$~eV \citep{Williams1968, watson1972}.\footnote{For reference adsorption is defined as the adhesion of an atom or molecule from a gas to a surface which is in contrast to desorption, or sublimation, which is the release of this trapped molecule from the ice/surface to the gas.}
Molecular ice abundances can be sufficiently large to enhance chemical reactivity. Some species, such as silicate minerals that form the cores of grains, are bound together via tighter covalent interactions (with interaction strengths of around 2~eV) as these species formed under higher temperature conditions.
Thus, refractory material (silicates and carbonaceous grains) in this context has the strongest interactions requiring the highest temperatures to activate sublimation.\footnote{At pressures in the interstellar medium and in the solar nebula, the liquid phase does not exist and the transition is direct from solid state to the gas.}  Ices, growing on top of such dust grains, on the other hand, have greater volatility with water and organics exhibiting dipole and hydrogen-bonding interactions being more strongly bound than those expressing only polarization and dispersion interactions, such as CO, N$_2$, and H$_2$.   
At low temperatures, typically below 20 K, lighter adsorbed particles, such as H, can diffuse across the surface and find reactive partners (e.g. at the outermost radius in Fig.~\ref{fig:01_01}a). If the grain temperature exceeds the sublimation temperature then
molecules are released from the surface, a process that is largely determined by the involved binding energies.
This puts the micro-physics of the gas-dust interaction at the center of our understanding of the evolution of dense molecule-dominated regions such as planet-forming disks, but also for dense (n $>$ 10$^4$~cm$^{-3}$) molecular cores that are the sites of star formation.  At the center of this is the binding energy for a given species, to species in the surrounding ice matrix or (in the case of submonolayer thick ices) with the bare grain surface. These values are only known for a handful of species, and in fact, also vary for different ice constituents.

This article will summarize the present state-of-the-art, linking physical chemical insights, derived both from experimental and theoretical work, to astronomical applications. It will summarize the values currently available for a large number of interstellar ice analogues. In this introductory section, we will first introduce the concept of binding energies in the solid state and discuss the role of thermal desorption parameters for different evolutionary stages. We stress that in the present consensus work we focus purely on thermal desorption of solids (zero order desorption) and monolayers (first order desorption) as these dominate in the literature and in the hope that further works would follow relating to other desorption orders (i.e. second order) and mechanisms (reactive desorption,
photodesorption, ...).

The primary focus in this work is on the chemical physics of the gas-grain interaction for typical astronomical conditions, more than linking these to specific stages in the star and planet formation process. Here we refer the reader to the following reviews\citep{Caselli12, Henning13, vanDishoeck13, Jorgensen20, Oberg21}.

The paper is organized as follows: in section 1 we give a simple definition of the controlling parameters of thermal desorption and we describe the astrophysical implications of the desorption parameters. In section 2, we present the methods used in experiments and computations to determine the desorption parameters and how such parameters are used in astrochemical models. Some relevant case studies are discussed in section 3 to illustrate both the issues that arise in experimentally and computationally exploring representative systems and the impact of such studies on astrophysical models.
In section 4, we critically assess the desorption parameters (the binding energies E$_b$ and the pre-exponential factor $\nu$) commonly used in the astrochemical community for astrophysically relevant species and provide tables with recommended values. 
The desorption behavior of astronomically more realistic ices is discussed in section 5. 
In section 6, we present the limitations and the frontiers of methods (both experimental and computational) and astrochemical models dealing with desorption parameters. In the last section, we discuss the main conclusions of this review.


\section{1. Thermal desorption: a simple definition of the controlling parameters}
\begin{figure}[ht]
\centering
 \includegraphics[width=0.6\textwidth]{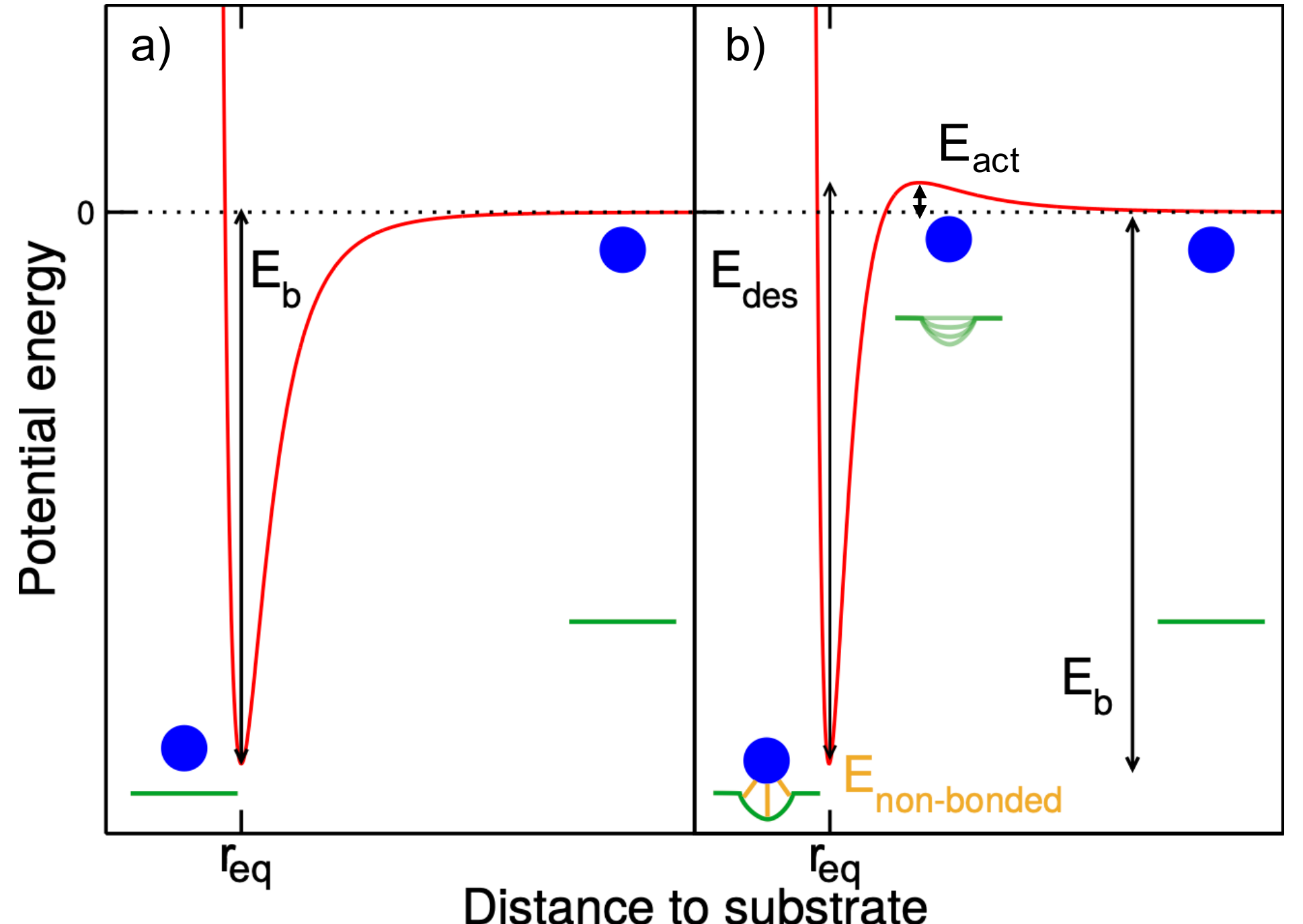}
 \caption{a) Schematic representation of a potential energy surface as a function of the distance to the substrate. The substrate is represented by the green line and the adsorbate by the blue sphere. There is a physisorption well at $r_\text{eq}$ with a depth $E_\text{b}$. The activation energy for desorption, $E_\text{des}$, is equal to the depth of the well. b) A similar schematic for the case of surface accommodation of the adsorbate. This can cause a small (additional) activation barrier for adsorption and desorption. In some cases, the binding energy is approximated by the sum of the non-bonded interaction between the adsorbate and the surface depicted in orange.}
 \label{fig2.1-pes}
\end{figure}

Gas-grain processes, as schematically illustrated in Fig.~\ref{fig2.1-pes}a, involve the interaction of an atom or molecule with a solid surface and are governed by the potential energy surface describing the surface-adsorbate system \citep{lennard-jones1932}. If the interaction involves simple electrostatics and dispersion, \textit{i.e.} van der Waals interactions, and hydrogen bonding rather than electron exchange, a shallow \textit{physisorption well} is observed at a ``van der Waals distance'', $r_\text{eq}$, typically at 2--4~\AA $\,$ as in Fig.~\ref{fig2.1-pes}a. In this situation
\begin{equation}
{E}_\text{b}={-E}_{\text{des}}
\label{eq:Eb=-Edes}
\end{equation}
\textit{i.e.} the \textit{binding energy}, $E_\text{b}$, and \textit{activation energy for desorption}, $E_\text{des}$, are equal in magnitude but opposite in sign as the system reference energy (i.e., $E=0$) corresponds to the adsorbate and substrate at infinite separation.

In the presence of a dynamical barrier associated with accommodation of the adsorbate and surface relaxation, $E_\text{act}$, which might be considered \textit{precursor-mediated adsorption}, as in Fig.~\ref{fig2.1-pes}b, then Eq.~\ref{eq:Eb=-Edes} no longer holds and 
\begin{equation}
{E}_{b}={{E}_{\text{act}}-E}_{\text{des}}.
\end{equation}

The thermal desorption rate per molecule/atom $i$, $k_{td, i}$
can be defined by: 
\begin{equation}
k_{\text{td},i} \simeq \nu _i e^{-\frac{E_{b,i}}{k_B T_{s}}}.
\end{equation}
where $\nu$ is the pre-exponential factor and $T_{s}$ is the surface temperature, 

It does not matter whether we investigate gas-grain interactions experimentally or theoretically, in the end E$_b$ and $\nu$ are the quantities that have to be determined. 
For practical purposes, the binding energy is sometimes approximated by the non-bonding interactions between the adsorbate and the substrate, schematically shown in orange in Fig.~\ref{fig2.1-pes}b. The non-bonded interaction is the sum of all physisorption contributions. The pre-exponential factor $\nu$ is often defined as the vibrational frequency of a given species $i$ in the surface potential well.

\section{Astrophysical implications of the desorption parameters}
\subsection{Dense interstellar medium}

\begin{figure*}[ht]
\centering
\includegraphics[width=0.7\textwidth]{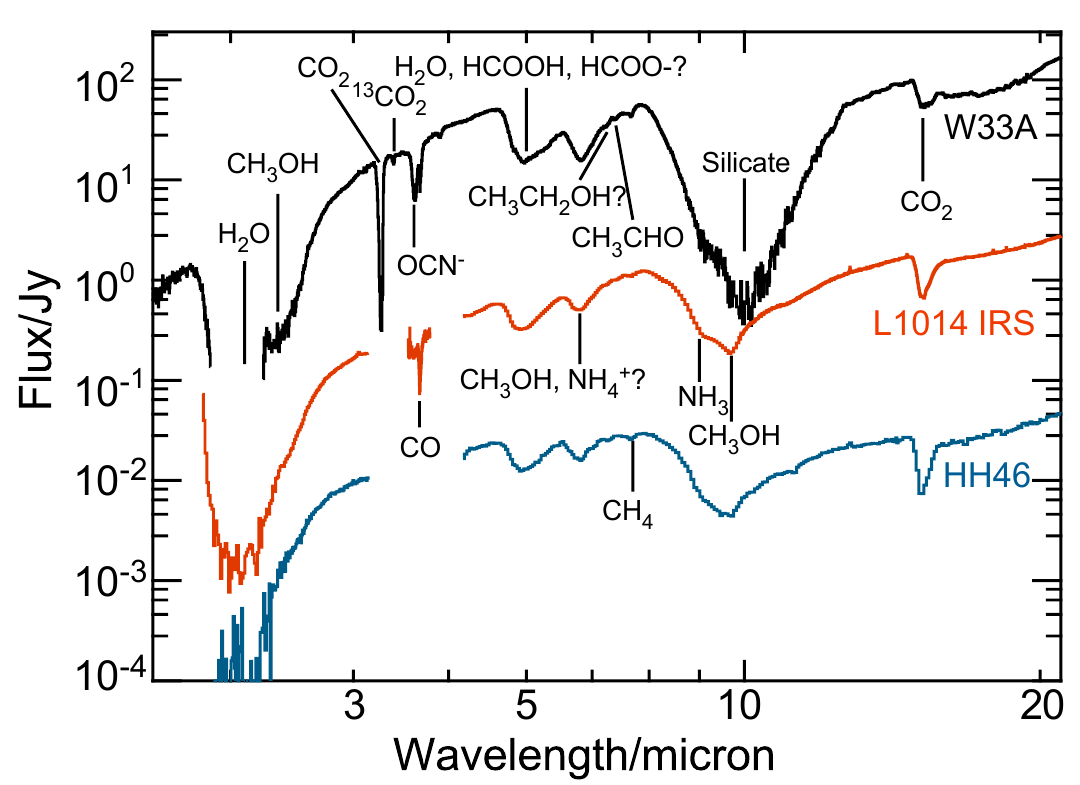}
\caption{Spectra of interstellar ice absorptions seen towards 3 protostellar objects. Spectrum taken from Oberg et al.\cite{Oberg11b}.}\label{fig:oberg}
\end{figure*}

Ices are readily observed in the interstellar medium. Ices present in a molecular cloud can be seen in absorption as such clouds are located in the lines-of-sight of stars. Such absorption bands are shown in Fig.~\ref{fig:oberg} towards the stars W33, L1014 IRS, and HH46. These spectra show the strong and deep absorption by water ice at 2.8 $\mu$m as the most abundant ice component. Narrower features are seen for other ices that are present at a few to tens of percent relative to water in abundance, such as CO, CO$_2$, CH$_4$, NH$_3$ \cite{Oberg11b, Boogert15}. Beyond these major ice components, there is evidence for developing chemical complexity as seen with the clear detection of CH$_3$OH and possibly of HCOOH or CH$_3$CHO\cite{Boogert15, linnartz2015, terwisscha2018}. Ices form in a variety of manners from direct deposition from the gas to assembly from
constituent atoms and/or molecules in the ice mantle.  For instance, CO forms primarily in the gas \cite{caselli1999, bergin02} concurrently with the initial development of hydrogen-rich polar ices H$_2$O, NH$_3$, CH$_4$ \citep{Boogert15}.  The subsequent adsorption of CO onto grains provides the fuel for CO$_2$ and CH$_3$OH formation.  These molecules, and their energetic processing, provide radicals that recombine within warming mantles to form larger complex organic molecules (COMs).\citep{Garrod2013}.   The mere presence of these ices confirms that the dust temperature in these objects is below the sublimation temperature set by the balance of gas phase adsorption and the sublimation governed by the intermolecular binding energy for a given molecule.    

Hollenbach et al.\cite{Hollenbach09} provide a useful description of this formalism where the thermal desorption rate per molecule/atom $i$, $k_{td, i}$($T_{s}, E_{b,i}$) is balanced by the flux of molecules deposited onto grain surfaces from the gas, $F_{ad}$ = 0.25$n_i v_i S$ (where $n_i$ is the volume density of the given species, $v_i$ the thermal velocity, and S is the sticking coefficient).
The temperature for which these two quantities ($k_{td, i}$ and $F_{ad}$) are equal provides the local sublimation temperature, $T_{sub}$, of that species, 
\begin{equation}
T_{sub,i} \simeq \frac{E_{b,i}}{k_B}\left[ 57 + \ln \left[ \left(\frac{N_{s,i}}{10^{15}\rm cm^{-2}} \right)\left({\frac{\nu _i}{ 10^{13}\rm s^{-1}}}\right)\left(\frac{1\rm cm^{-
3}}{n_i}\right)
\left(\frac{10^4\rm cm\ s^{-1}}{v_i}\right)\right]\right]^{-1},
\end{equation}
where N$_{s,i}$/10$^{15}$ is the surface coverage. 
\noindent Thus, the sublimation temperature depend strongly on the desorption parameters (e.g., binding energy and pre-exponential factor) and weakly on the local gas density.   Further the binding energy itself depends on inter-molecular forces set by the dominant species on grain surfaces.   
For instance, we can take the case of CO; in pure CO ice (E$_b \simeq 900 K$ ) it has a sublimation temperature near 16~K, while any CO bound to a surrounding water ice matrix has a higher binding energy (E$_b \simeq 1200 K$ ) and a higher sublimation temperature around $23$~K. 
Thus, the detection of CO ice towards L1014 IRS in Fig.~\ref{fig:oberg} confirms the presence of very cold gas and grains along this line-of-sight, i.e. well below 25 K. At the same time absorption signals in the 4-7 $\mu$m range indicate the likely presence of larger species, fully in line with the idea that smaller species on icy dust grains act as precursors for larger COMs. 

\subsubsection{Star formation}

The first step in the cosmochemical evolution of molecular species is the formation of H$_2$ via catalytic reactions on bare grain surfaces.\citep{watson1972, Hollenbach71, wakelam2017b} At later stages CO and other simple molecules are formed in the gas phase, in part through ion/radical-molecule reactions. In parallel, ice layers start growing on the cold dust grains, either through direct accretion or upon surface chemistry, offering a molecular reservoir that acts as a starting point for solid state reactions.
In collapsing dense cores, when the density of molecular hydrogen exceeds $\sim10^4$~cm$^{-3}$, the collision time between gaseous molecules and interstellar dust grains becomes shorter than the dynamical timescales (i.e. free-fall). As sublimation temperatures exceed the typical temperature of $\sim$10~K of star forming gas, most molecules carrying heavier elements (C, O, N) will freeze onto grain surfaces. As the most abundant molecule, H$_2$, is unemissive at these cold temperatures, astronomers rely on other calibrated probes (e.g. CO, HCO$^+$, CH$_3$OH, etc.) as tracers of key properties (density, temperature, velocity field, mass) of this hidden H$_2$.  Freezeout complicates this issue as the loss of key gas-phase tracers of H$_2$ (i.e. CO) has a significant impact on our ability to probe and characterize the initial phases of stellar birth. Such gas-phase depletions are now known to be wide-spread \cite{caselli1999, bergin02, Tafalla02} - and knowledge of the gas-grain interaction lies at the heart of our understanding of how to interpret the observations of the emission from disparate molecules during star formation \cite{Bergin07}. 
We stress that gas-phase tracers of H$_2$ are not only important to probe the initial phase of star formation. These molecules can have a direct impact on the dynamics of star formation. A prime example is the role of spontaneous dipole alignment during low temperature deposition of CO and other molecules from the gas phase. The spontaneous dipole alignment leads to the formation of a surface potential that impacts on ionization degree, which, in turn, determines the timescale of diffusion of magnetic fields\cite{rosu-finsen2016b}. 
Further, ever present, cosmic rays generate a steady production of H$_3^+$, the central engine of ion-molecule reactions in cold clouds \cite{Herbst73}.  The abundance of this molecule is held in control by abundant gas phase CO. The freeze-out of CO has two important effects.  First, the depletion of CO leads to an increase in the abundance of the readily observable N$_2$H$^+$ ion which becomes a direct probe of dense gas on its way to making a star \citep{Bergin07,vantHoff2017, bovino2019, feng2020}.  Further, the depletion of CO is a central facet in powering strong deuterium fractionation in cold (10~K) gas \cite{Bergin07, Caselli12, giannetti2014, bovino2019} leading to high deuterium enrichments in species such as water ice and other species assisted by solid state processes \citep{vanDishoeck13}.  These enrichments are widely used to explore the origins of Earth's oceans \cite{Ceccarelli14, vanDishoeck14, furuya2016}.

At 10~K, the diffusion of these adsorbed molecules is strongly limited, except for atomic hydrogen. For simplicity, and because of the difficulty in determing the mobility of species from laboratory measurements \cite{mispelaer2013, he2014, minissale2016b, kouchi2020, mate2020}, the diffusion barriers are very often defined as a fraction of the binding energies. In astrochemical models, this fraction is usually assumed to be the same for all molecules, except for H for which laboratory data have been determined \citep{al-Halabi2007,matar2008,watanabe2010,hama2012, iqbal2018}. Depending on the model, this fraction can be between 0.3 and 0.8 \citep{watson1972,hasegawa1992a,biham2001,chang2005,ruaud2016,he2018}. Sometimes, the choice of the value is driven by the necessity to have the species move on the surfaces to produce larger molecules. The formation of CO$_2$  on surfaces, for instance, requires atomic oxygen or OH, which have larger binding energies, or carbon monoxide to move in order to encounter and react \cite{ioppolo2011,minissale2014}. For cold dense cores, binding energies are important for the efficiency of non-thermal desorption processes \cite{minissale2016, chuang2018}, and thus the composition of the ice surface layers, which in turn determines the efficiencies of diffusion and surface reactions. In other words, binding energies could be dependent on the temperature and pressure at which species from the gas phase adsorb on the dust. In some conditions, molecules arriving on the surface could bind only with 1 or 2 other species on the surface, meaning that the binding energy would be substantially lower than the surface binding energy derived from experiments.  This could be the case for CO in starless cores \cite{cazaux2017}, where the weakly bound molecules, easily released into the gas phase through evaporation, would change the balance between accretion and desorption, resulting in a larger abundance of CO at high extinction. This could explain the abundances of CO in starless cores \citep{caselli1999, bergin02}, which are higher than predicted by freeze out models with non thermal desorption processes \citep{keto2010}.  Alternately,  this could indicate that the rates of these processes are overestimated.

\begin{figure*}[ht]
\centering
\includegraphics[width=\textwidth]{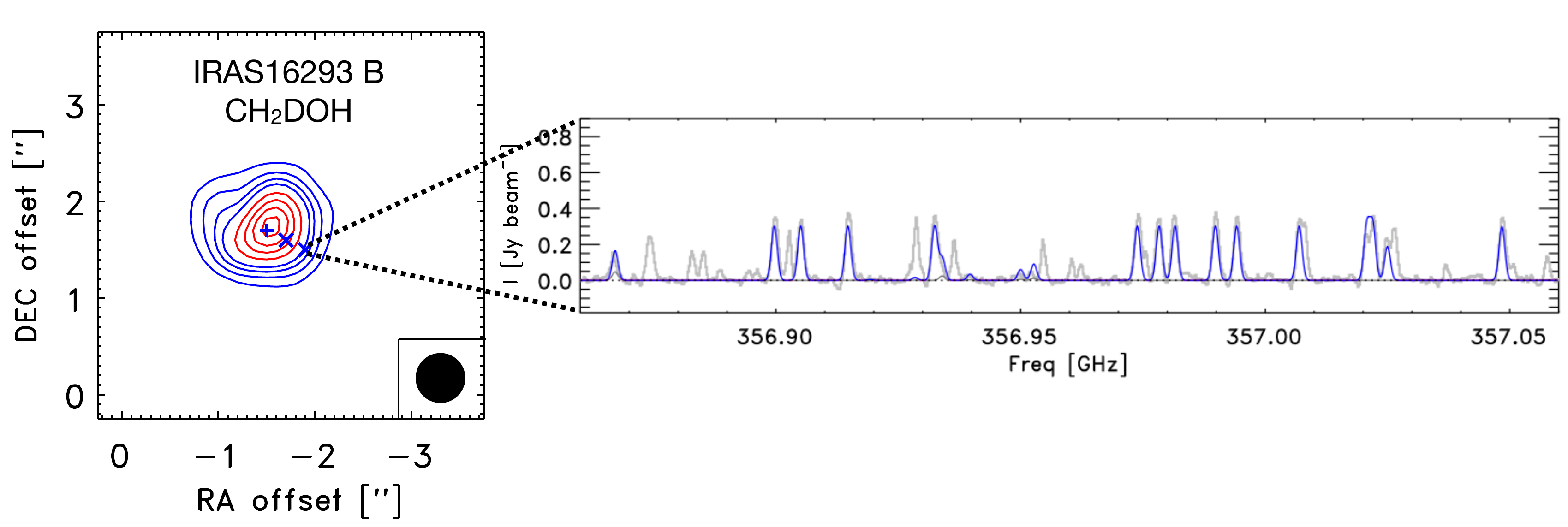}
\caption{Map and spectrum of deuterated methanol (CH$_2$DOH) observed towards the component B of the solar-type protostar IRAS~16293--2422 in the framework of the Protostellar Interferometric Line Survey (PILS, \cite{jorgensen2016}). The map shows the extent of the hot corino. Most of the complex organic molecules present similar spatial distributions. The observed spectrum (at a full beam offset position from the continuum peak) is indicated in grey, while a synthetic model of CH$_2$DOH is shown in blue. The other lines correspond to other complex organic molecules. Figure adapted from J$\o$rgensen et al.\cite{jorgensen2016}.}
\label{fig_i16293_pils}
\end{figure*}

Binding energies play an important role around protostars as well. In the early evolutionary stage, the protostar is surrounded by a thick envelope of dust and gas and the physical structure of the envelope is characterized by a gradient of density and temperature towards the central object. 
The molecules trapped in the icy grain mantles can consequently desorb in the gas phase when the temperature reaches a certain value dependent on the binding energy. In the warm inner regions where the temperature is higher than about 100 K, many molecules such as water and complex organic molecules are detected in the gas phase\cite{jorgensen2016, gelder2020, nazari2021}. 

Whereas the solid state formation of many smaller species, such as H$_2$O, NH$_3$ and CH$_4$ is well established, and therefore such species are expected to have directly desorbed from the grains, for COMs both solid state and gas phase formation pathways (involving desorbed species) have been proposed. \cite{fedoseev2017, skouteris2018, skouteris2019, cooke2019, vazart2020, ioppolo2021, shingledecker2021}.

These regions are called hot cores and hot corinos for the high-mass and low-mass star-forming regions, respectively \citep{ceccarelli2004, Caselli12,choudhury2015, Jorgensen20, sabatini2021}. There are well-known hot cores in the molecular clouds of Sgr B2 and Orion\citep{Blake87, Turner91, Belloche13, Neill14, Crockett14} while IRAS~16293--2422, NGC1333 IRAS2A, IRAS4A and IRAS4B were the first discovered hot corinos \cite{bottinelli2007,caux2011,jorgensen2016, Jorgensen20, coletta2020}.
Figure \ref{fig_i16293_pils} shows an example of the hot corino region observed towards the solar-type protostar IRAS16293--2422 B.

With low mass protostars, there is also evidence for sublimation fronts.
For instance, the sublimation radius for the more volatile CO is observed to be located at larger radii than that of water and other more complex species \citep{Anderl16, vantHoff18}.  At face value, this validates the basic understanding of volatility and its link to the ice-gas interaction governed by the bond strength of the volatile to the ice-coated grain surface.  
Ice sublimation is also used to trace the temporal variation of the luminosity of the central protostar. For some objects the sublimation front is located farther away from the central star than predicted by the radiation transfer model with the current luminosity; it suggests that the luminosity was higher but recently declined to the current value.\citep{Jorgensen20}.

\subsubsection{Planet formation}

The presence and location of the gas-ice transition at a sublimation front will directly influence the composition of planetary bodies as they form \citep{Oberg11b} (see Fig.~\ref{fig:01_01}). Such a transition region is known as a ``snowline'', within the snowline a specific species has been thermally desorbed and is found in the gas phase, beyond the snowline the species is still frozen.
The gas-ice transition at a snowline might also play an important role in planet formation. During the early stages of planetary assembly, ice coated silicates and carbonaceous refractory solids with typical sizes ranging from microns to up to pebble-sizes (mm/cm sizes) collide.  According to current theories of planet formation, planets can grow quite quickly via accreting pebbles as opposed to gravitationally fed collisions with many km-sized rocks \citep{Johansen17}.  Further, the size of the pebbles can influence the growth of planetary embryos \citep{Morbidelli15}.   Pebble formation and pebble growth is thus a key facet alongside the overall mass distribution of solids with the disk.

Oberg \& Bergin \citep{Oberg21} outline four ways that the snowline can affect planet formation.  These include  (1) the accumulation of solid mass density behind the snowline  due to the accumulation of abundant condensates such as water and CO \citep{Hayashi81}.   (2) The increase of solid particle size due to condensation of water ice \citep{Stevenson88, Ros13}.  (3) Changes in particle properties such as stickiness and/or fragmentation thresholds \citep{Guttler10, Gundlach15, Pinilla17} and (4) snowlines may induce changes in the pressure producing a localized dust trap \citep{Cuzzi04}.    In all of these cases the volatility of abundant ice carriers and the overall thermal structure is critical.

\subsubsection{Summary of Astrophysical Implications}

In summary, the actual physical value of the binding energy has a major impact on the interpretation and understanding of the physics and chemistry of star formation.  As discussed above, the strength of the gas grain interaction influences the ability of astronomers to probe the evolution of star-forming cores and the development of chemical complexity in the vast cold environs of interstellar space. The COMs formed on ice surfaces certainly have a direct impact on the composition of cometary bodies \cite{altwegg2019} and they may (or may not) be important for the development of life on habitable worlds.  Furthermore the binding energy directly influences the location of snowlines in planet-forming disks which plays a central role in the initial composition of forming planetary systems.

\section{2. Binding energy in experiments, chemical computations and astrochemical models}
\subsection{Experimental methods}

Experimentally, $E_\text{b}$ can be approached through equilibrium considerations where the rate of adsorption, $k_\text{ads}$, is equal to the rate of thermal desorption, $k_\text{des}$ \citep{attard1998}. Alternatively, it is approached through non-equilibrium (kinetic or dynamical) measurements of $k_\text{des}$ alone \citep{attard1998}. Generally, $k_\text{ads}$ is 
\begin{equation}
{k}_{\text{ads}}=S({n}_{i,\text{ads}})k_\text{coll}
\end{equation}
with $S(n_{i,\text{ads}})$ the sticking coefficient, which varies from 0 to 1 and is inherently a function of the surface concentrations of adsorbed species ($n_{i,\text{ads}}$) and $k_\text{coll}$ the collision rate which is given by
\begin{equation}
k_\text{coll}=n_{i,\text{gas}}{\left(\frac{{k}_{B}T}{2\pi
{m}_{i}}\right)}^{1/2}=x_{i,\text{gas}}{n}_{H}{\left(\frac{{k}_{B}T}{2\pi
{m}_{i}}\right)}^{1/2}.
\end{equation}
where $n_{i,\text{gas}}$ and $m_\text{i}$ are the number density in the gas phase and molecular mass of species $i$ respectively. Of course, the former can immediately be related to the fraction of species $i$, $x_{i,\text{gas}}$, and the number density of atomic hydrogen, $n_\text{H}$.
The rate of desorption, $k_\text{des}$, is given by the Polanyi-Wigner equation 
\begin{equation}
 {k}_{\text{des}}={\left({n}_{i,\text{ads}}/N_s\right)}^{o}\times \nu \left({n}_{i,\text{ads}}\right)\times{e}^{-{\lvert E }_{\text{des}}\rvert({n}_{i,\text{ads}})/\text{RT}}
 \label{eq:Polanyi}
\end{equation}
where $o$ is the order of desorption, N$_s$ is the number of sites per surface area, ${\nu}$ is the surface concentration dependent pre-exponential factor and $E_\text{des}$ is also generally found to be surface concentration dependent. A dependence on surface concentration reflects the presence of adsorbate-adsorbate interactions. 
One can note that for o$^{th}$ order of desorption ($o$ = 0), the desorption is independent of coverage, as in the case of multilayer desorption, while  first order desorption ($o$ = 1) corresponds to the thermal desorption of species already formed and interacting with the surface.

Combining $k_\text{ads}$ and $k_\text{des}$ within the limit of monolayer growth, where the order of desorption is unity as explained below, Langmuir was able to demonstrate how measurements of surface concentration \textit{versus} equilibrium gaseous number density (\textit{i.e.} adsorption isotherms) as a function of equilibrium temperature could be interpreted within a Clausius-Clayperon framework to yield isosteric (\textit{i.e.} coverage specific) enthalpies of adsorption, ${\Delta}_\text{ads}H$ \cite{attard1998}. The latter can be linked to $E_\text{b}$, which is equivalent to thermodynamic internal energies, through the classical thermodynamic relationship
\begin{equation}
{\Delta }_{\text{ads}}H={E}_\text{b}+{\Delta}(PV)
\end{equation}
where \textit{P} is the pressure and \textit{V} the volume, assuming ideal gas behaviour.
\begin{figure*}[ht!]
\centering
 \includegraphics[width=0.7\textwidth]{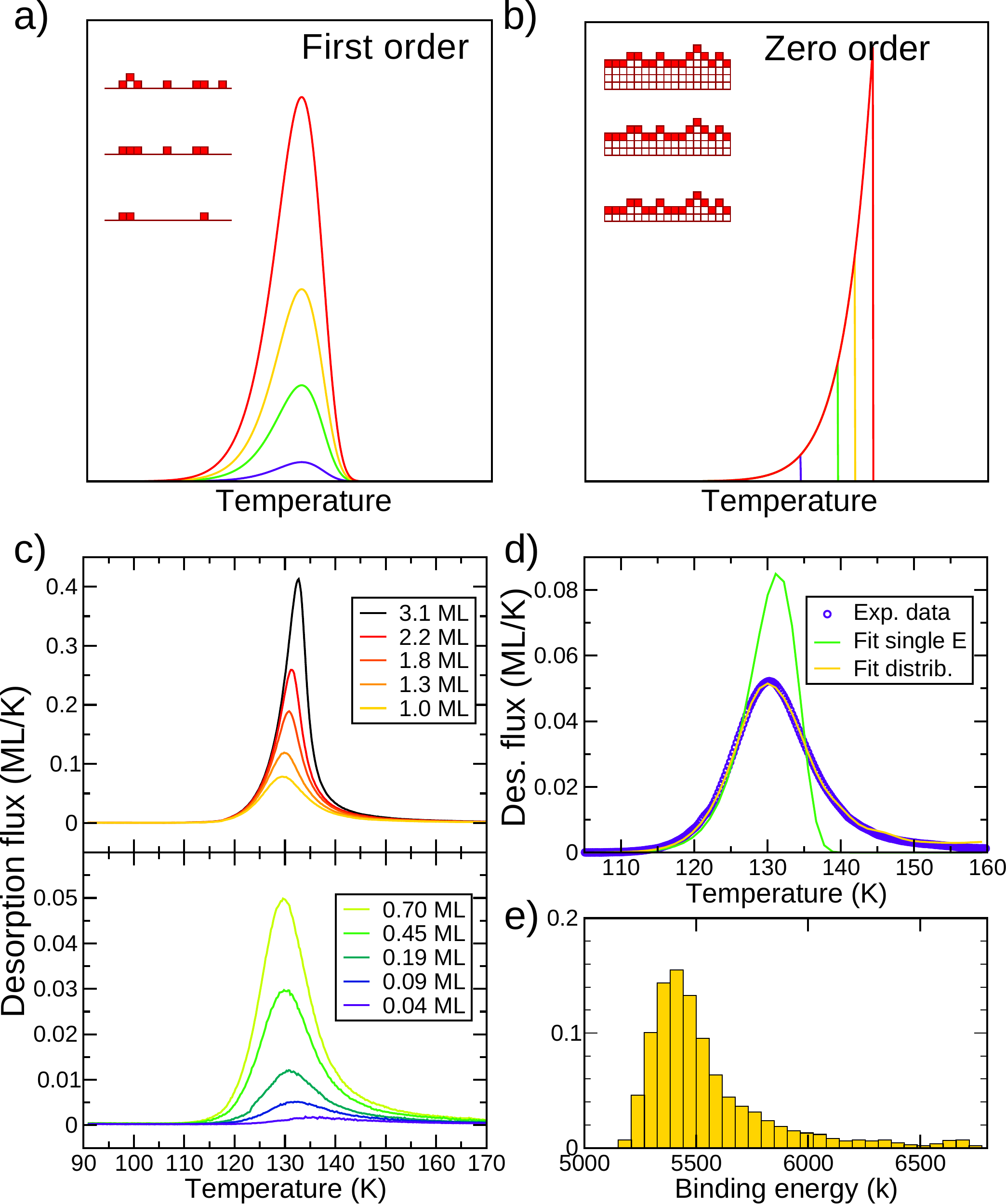}
 \caption{a) + b) Synthetic TPD spectra of first and zero order desorption, respectively, using Eq.~\ref{eq:Polanyi} for a single desorption energy. The different traces indicate different initial amounts of adsorbate. The insets schematically depict the corresponding physical pictures. The full blocks represent species available for desorption, since they are facing the vacuum; empty blocks are currently unavailable for desorption. c) Examples of multilayer (zero) and monolayer (first order) desorption for \ce{CH3OH} adsorbed on HOPG. Data is taken from Ref.~\citep{doronin2015}. d) Two fits to the 0.7 ML data are shown. Using a distribution of binding sites (e) results in a much better fit than applying only a single binding site.  }
 \label{fig2.2-TPD}
\end{figure*}
Many experiments have been performed at equilibrium, including pure gas isotherms (see the review of Fray \& Schmitt\cite{fray2009}). However, it is well known that many astrophysical media are not at thermodynamic equilibrium, and that the species are mixed and not pure. We will therefore consider here the more realistic case of non-equilibrium systems, especially those where the gas pressure is very low.
Turning to consider non-equilibrium studies, the goal is to measure $k_\text{des}$ and hence determine the
parameters $o$, $\nu$ and $E_\text{des}$; and hopefully the surface concentration dependence of the latter two. By far the most common method of doing so is \textit{temperature programmed desorption} (TPD). TPD is a long established technique in surface science \citep{woodruff1994, ioppolo2014} which has been revolutionised in recent years by the adoption of line-of-sight methods developed by  Hessey and Jones \citep{hessey2015}. Simply, under conditions of high pumping speed, changes in partial pressure measured by a quadrupole mass spectrometer can be shown to be proportional to $k_\text{des}$. Analysis of TPD data is well-described in the literature \citep{king1975}. However, the first step in any analysis is to determine $o$, the order of desorption. Strictly, as an empirical parameter, $o$ can take any value. However, there are two common behaviours that can be recognised when surface concentration dependant TPD data are available; zero and first order desorption. Examples of synthetic TPD spectra for both cases are given in Fig.~\ref{fig2.2-TPD}a) and b). These are obtained using Eq.~\ref{eq:Polanyi} with a single ${E}_\text{des}$ and using different initial adsorbate concentrations. There is a clear visual difference between both orders: first order has common peak positions, whereas zero order has common leading edges. First order desorption is associated with desorption from films of up to a complete monolayer while zero order desorption is observed from multilayer films. The cartoons in Fig.~\ref{fig2.2-TPD}a) and b)  depict the microscopic origin of the order. Note how in zero order desorption the number of adsorbate species in the top surface of the solid available for desorption remains constant. The impact of monolayer \textit{versus} multilayer behaviour in a real system is nicely illustrated in the case of methanol (\ce{CH3OH}) on highly-oriented pyrolytic graphite (HOPG) in Fig.~\ref{fig2.2-TPD}c)\citep{doronin2015}. Where simple visual recognition fails, an application of the initial rate method in a
leading edge analysis \cite{green2009} will yield $o$. 

Analysis of TPD data to yield $E_\text{des}$ and $\nu$ for a given $o$ can be approached through a hierarchy of simplifications \cite{king1975}. Monolayer $E_\text{des}$ distributions are obtained from the simplest assumption of surface concentration independent $E_\text{des}$ using Arrhenius analysis of TPD data assuming a fixed pre-exponential factor based on Redhead \cite{redhead1962} and Hasegawa, Herbst and Leung \cite{hasegawa1992a} of $10^{12}$ s$^{-1}$ \cite{collings2015}. More recently an optimisation method was proposed by Kay and co-workers \cite{tait2005a,tait2005b,tait2006,smith2016} to estimate both pre-exponential factor and monolayer $E_\text{des}$ distribution. 
Fig.~\ref{fig2.2-TPD}d) nicely illustrates the impact of the data inversion method on the simulation of the TPD of a sub-monolayer quantity of \ce{CH3OH} from HOPG with a single $E_\text{des}$ and a distribution of $E_\text{des}$ as given in Fig.~\ref{fig2.2-TPD}e) \cite{doronin2015}.

For the sake of completeness, we present below a brief overview of other experimental methods used to study adsorption-desorption mechanisms. In the 1970's, a technique relating the adsorbate structure determined by LEED (Low Energy Electron Diffraction) patterns with Clausius-Clapeyron analysis of equilibrium adsorption isosteres was commonly used to determine the binding energy\cite{tracy1969,madden1973}. More recently, desorption behavior of molecules was studied by the TD-XPS (Temperature Desorption X-ray Photoelectron Spectroscopy) method, a combination of desorption theory and surface spectroscopic techniques \cite{olivas1999}.
The main disadvantage of these techniques is that they can be mainly used for studying reversible adsorption systems. Surface calorimetry can be employed in the case of dissociative and reactive adsorption to measure the liberation of adsorption heat \cite{brown1998} (and references therein) and thus to derive the binding properties of the systems. Recently, a TPD-derived technique, i.e. the TP-DED (Temperature Programmed-During Exposure Technique) was employed to deal with adsorption-desorption behavior of reactive systems \cite{minissale2016b}. 
Collision-induced desorption is also used to look at desorption energetics using hyperthermal atom collisions with the surface bound species to cause desorption and measurement of the desorption cross-section using a surface sensitive spectroscopic method (RAIRS, XPS, TPD itself) to measure surface coverage as a function of time. Variation of the incident collider energy gives a threshold and using two colliders of different mass allows the binding energy to be determined. This is potentially a tool for measuring binding energies of radicals\cite{beckerle1990}.

Finally, nanoparticle mass spectrometry has been proposed as a method to study surface reactions, adsorption and desorption of gases on the surface of nanoparticles\cite{gerlich2000,schlemmer2001,esser2019}. This nondestructive method, based on optical detection, allows the determination of the absolute mass of charged particles in a three-dimensional quadrupole trap.

\subsection{Computational methods}

In general, one could state that the desorption data of stable species are relatively straightforward to determine experimentally, whereas for unstable species, like radicals, this is much harder, since these are more difficult to prepare and deposit and once deposited they are likely to react before desorbing. Also the chemical energy released upon reaction may trigger a desorption event.  Moreover, other characteristics of the surface (e.g., morphology and porosity) govern the thermal desorption process, inducing a coverage dependence of the binding energy or the distribution of binding sites. To this end, complementary computations are very useful, although  open-shell systems are computationally also harder to treat. 


The binding energy is typically defined as the desorption enthalpy at 0 K, $\Delta_\text{des}H(0)$, or the adsorption enthalpy at 0 K with opposite sign, $\--\Delta_\text{ads}H(0)$. These are related to the empirical binding energy, $E_\text{b}$, through  
\begin{align}
 \Delta_\text{ads}H(0) = H(0)_\text{S/M} - (H(0)_\text{S} + H(0)_\text{M}) &=  \nonumber\\ E_\text{S/M} + E^\text{ZPE}_\text{S/M} - \{(E_\text{S} + E^\text{ZPE}_\text{S})
 + (E_\text{M} + E^\text{ZPE}_\text{M})\} &=\nonumber\\
 \Delta_\text{ads}E + \Delta E^\text{ZPE}&= -E_b 
 \label{eq:Ebind_comp}
\end{align}

where $\Delta_\text{ads}E$ is the adsorption energy based on the absolute electronic energies, $\Delta E^{ZPE}$ the variation of the ZPE corrections, and where the subscripts refer to the surface/molecule complex (S/M), the isolated surface (S) and the isolated molecule (M) supposedly at infinite distance from each other (for more details on the nature of these expressions, the reader may refer to these papers \citep{boese2013,zamirri2017}). The ZPE corrections are typically carried out using frequency calculations in the harmonic approximation. Within these approximations, ZPE's are often overestimated, especially when dealing with relatively weakly bound systems which typically show a large anharmonicity in their interactions \cite{karssemeijer2014I}.

There is a whole arsenal of different computational techniques available, with different levels of accuracy and sophistication, to calculate the quantities in Eq.~\ref{eq:Ebind_comp}. In general, these techniques are classified by \emph{(i)} how they determine the energy: quantum mechanically or classically, \emph{(ii)} the description of the system: isolated molecule, cluster model, or periodic surface, \emph{(iii)} the size of the system and \emph{(iv)} dynamics of the system: only 0~K data or temperature dependence. Notice that Eq.~\ref{eq:Ebind_comp} does not include any temperature effects. Here, we will briefly discuss these different aspects.

A variety of different quantum chemical methods, also often referred to as electronic structure calculations, exists, which can be classified into two major groups: methods based on the wave function (i.e., Hartree-Fock (HF) and post-HF methods) and those based on the electron density (i.e., the so-called density functional theory (DFT) methods). Post-HF methods yield, in principle, the most accurate computational binding energies. The golden standard in this respect are coupled cluster calculations (CCSD(T) and higher)\cite{sherrill2010}, but these methods are prohibitively expensive to apply to adsorbate-substrate interactions; they are hence mainly used for dimers and small clusters. Although dimer calculations between, for instance, water and the adsorbate of interest, can be a starting point for a binding energy estimate \cite{wakelam2017}, one will miss important features like many-body electrostatic and dispersion effects and binding site distributions over a large substrate. 
DFT methods have become computationally cheaper alternatives to the wave function-based ones, in which well-designed DFT methods provide acceptable accuracy \cite{sousa2007}. The advantage of DFT over wave-function based methods is their applicability to both molecules and extended systems. However, the quality of the result heavily relies on the chosen functional. Despite many attempts to identify the ``best functional for everything'', the results remain ambiguous, as the trends are highly dependent on eventual peculiarities of the studied systems \citep{hao2013}. 
For closed-shell species, it has been shown that the B3LYP-D3 method performs very well for a wide range of molecules, while for open-shell methods the M06-2X functional provides accurate values \cite{ferrero2020}. In the same line, the functionals  wB97X-D and M06-2X have been shown to give very similar results for the interactions of OH, HCO and CH$_3$ on water ice surfaces \citep{sameera2017}.
Calculated binding energies are sensitive to the number of dangling-H and dangling-O atoms at the binding site that interact with the radical, where the electrostatic attractions, orbital interactions, and Pauli repulsion control the strength of the binding energy\cite{saamera2021}. Thus, a range of binding energies is possible for each radical species.

Standard DFT methods, however, do not account for long-range non-covalent (i.e., dispersion-based) interactions \citep{hobza1995,allen2002}. A pragmatic solution is to add these contributions to the DFT energies \textit{via} an \textit{a posteriori} correction term (the so-called DFT-D scheme) \citep{grimme2004,grimme2006,grimme2010,caldeweyher2017}. While this works reasonably well for larger molecules, the adsorption of small molecules is one of the biggest challenges in DFT, even with modern methods to account for electron dispersion~\cite{klimes2012}. Indeed, calculations of the adsorption energies of small molecules to graphene and benzene show serious overbinding~\cite{hamada2012,silvestrelli2014}.

Classical simulations that use analytical descriptions for the interaction potential are computationally much less demanding and can treat large amorphous surfaces with many different binding sites. With Molecular Dynamics (MD) simulations, it is further possible to include temperature, to perform statistical averaging and to study dynamical effects. Average binding energies, the related error (standard deviation) and binding energy distributions can be thus derived from the statistics, characterized by a most probable value and a distribution width that reveals information on the diversity of adsorption sites \cite{al-Halabi2004}. For some systems, information about surface diffusion can be obtained as well \citep{michoulier2018,al-Halabi2007}. The accuracy of these simulations depends heavily on the accuracy of the analytical interaction potential. \textit{Ab initio} molecular dynamics simulations (AIMD), which use DFT to calculate the forces on the atoms, offer a plausible solution. However, the computational cost of AIMD limits the size and time scales of the systems to be studied and MD using classical potentials remains a method of choice for systems containing thousands of atoms. If accurate coupled cluster calculations of dimer structures are used to construct these potentials, the accuracy can be better than DFT \cite{karssemeijer2014b}. 
Recent studies have also successfully combined statistical exploration and electronic structure calculations using hybrid QM/MM methods. Here a QM part, which is treated at the quantum mechanical level, is embedded in a larger region which is described using  computationally cheaper molecular mechanics (MM), or force fields. In addition, within the ONIOM scheme \cite{xu2008,xu2009,karssemeijer2012}, to avoid problems inherent to QM/MM methods, due in particular to the defaults of the classical force fields parametrization, a combination of various QM levels may be chosen with the important part of the system being treated at the high level and the rest at a lower, cheaper, level (this reads QMHigh:QMLow). When it is coupled with a MD sampling, several configurations can be explored resulting in distributions of the adsorption energies. This method has produced very satisfactory results for small molecules, atoms or radicals on both crystalline and amorphous ices. \cite{sameera2017,Duflot2021} 

Adaptive Kinetic Monte Carlo (KMC) is an alternative technique that combines the atomistic description of molecular dynamics simulations with the ability to probe the long timescales of KMC \cite{xu2008,xu2009,karssemeijer2012}. These timescales are in the order of milliseconds, as compared to nanoseconds for classical MD and picoseconds for AIMD. Time-averaged binding energies, binding energy distributions and even diffusion constants can be obtained for a larger number of systems and larger temperature range than for MD \citep{cuppen2013,karssemeijer2012,karssemeijer2014I,karssemeijer2014b}. 



\begin{figure*}[ht!]
\centering
\includegraphics[height=0.6\textheight]{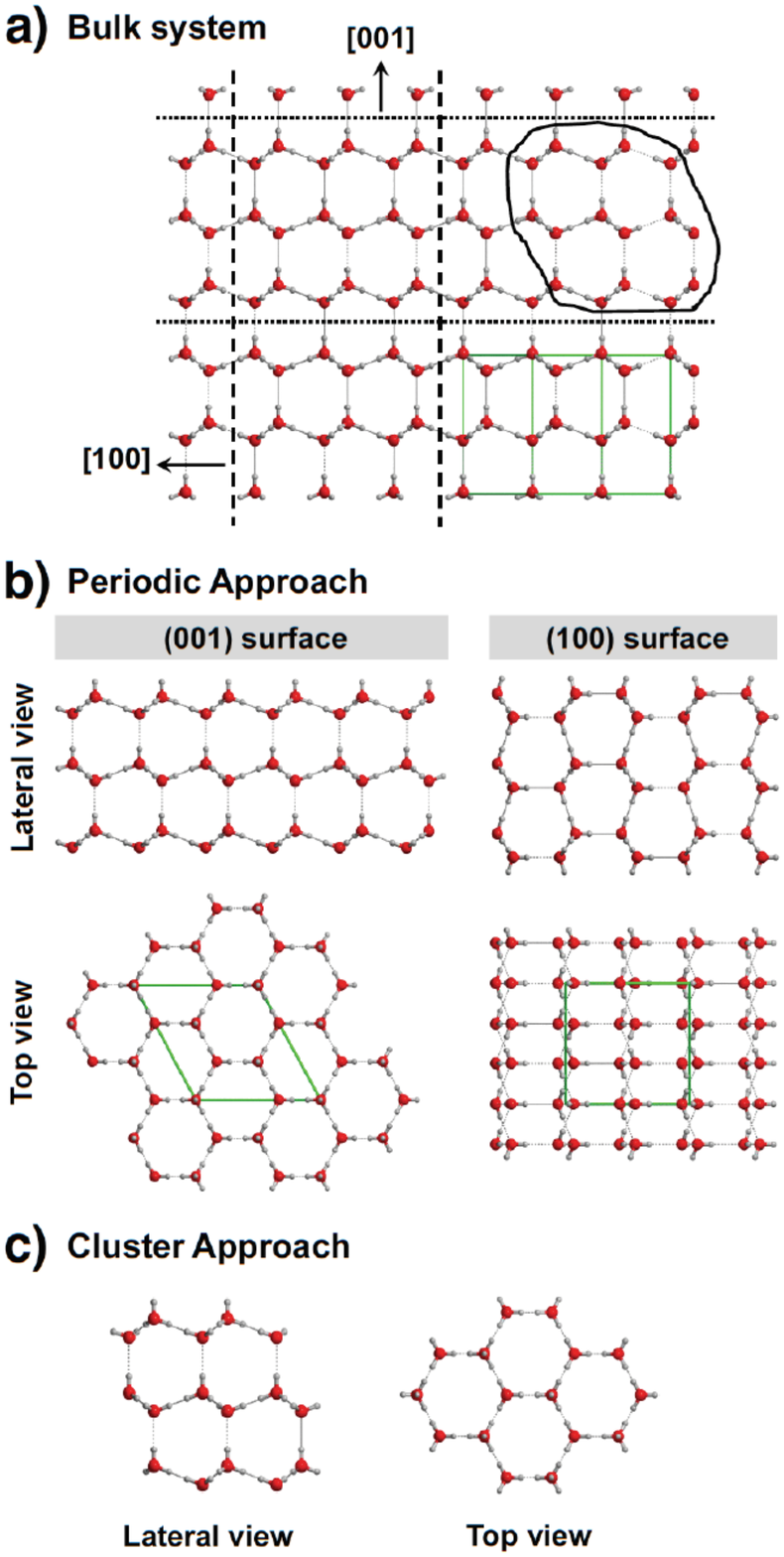}
\caption{Modelling strategies to generate periodic and cluster water ice surfaces from the crystalline bulk of hexagonal ice. a) Periodic bulk structure of hexagonal ice. Unit cell is shown in green, b) Top and lateral views of the periodic (001) and (100) water ice surfaces. They are generated by cutting out along the [001] and [100] directions (shown in a). Unit cells are shown in green. c) Top and lateral views of a cluster model extracted from the bulk structure. The fragment to extract is the structure inset shown in a).} \label{fig2.2_AR1}
\end{figure*}

\begin{figure*}[ht!]
\centering
\includegraphics[height=0.5\textheight]{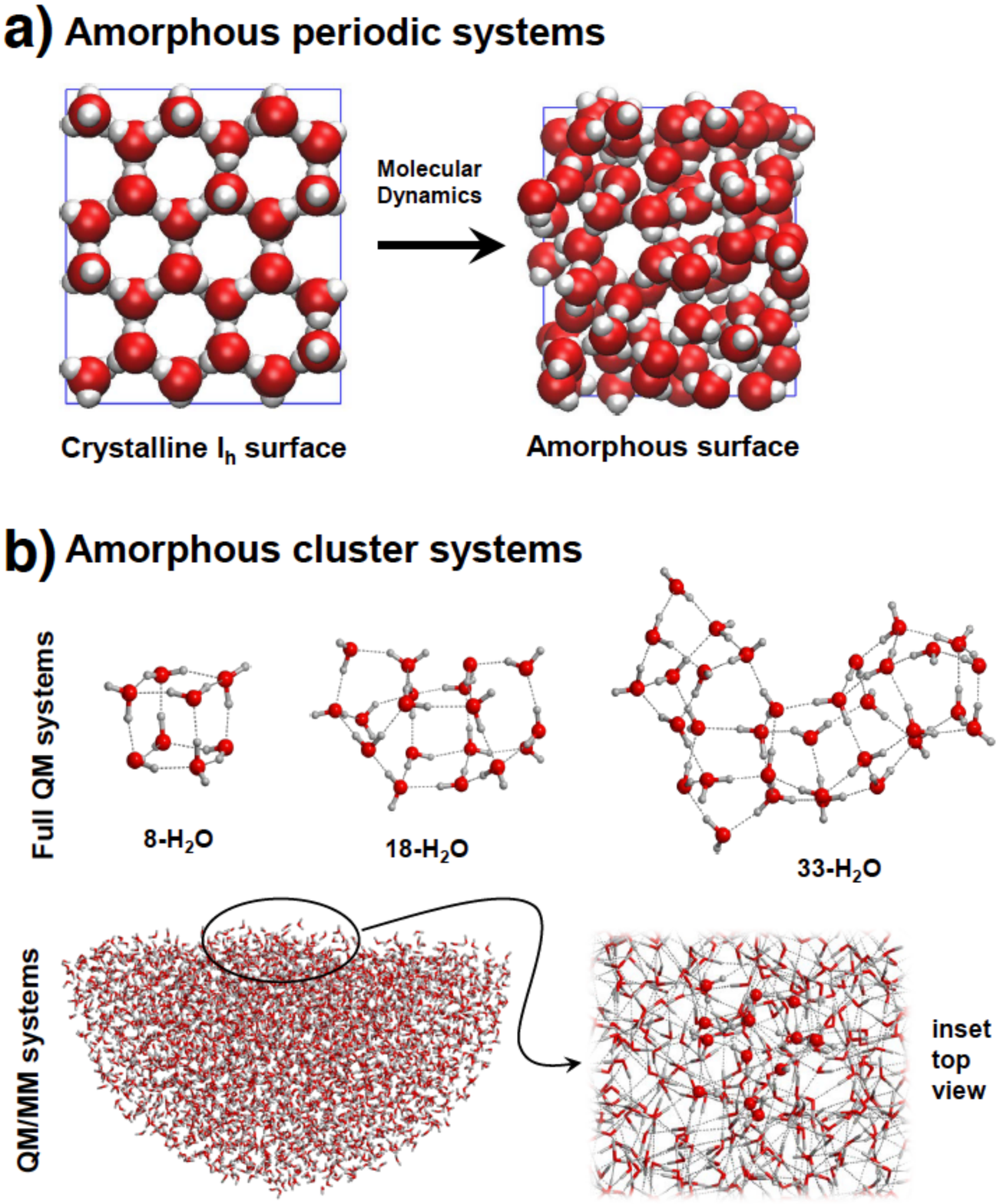}
\caption{Modelling strategies to generate amorphous structures for water ice surfaces adopting periodic (a) and cluster (b) approaches. a): Amorphisation of the crystalline I$_{h}$ surface by means of Molecular Dynamics simulations (adapted from \'Asgeirsson et al. 2017). b): Water clusters that are calculated at a full quantum mechanical (QM) level (top) or at a QM/MM level (bottom).} \label{fig2.2_AR2}
\end{figure*}

Since only a limited number of atoms can be treated, a surface is typically artificially enlarged through  periodic boundary conditions (PBC), where atoms on one boundary of the simulation cell interact with atoms on the other end. Surfaces are simulated using a ``slab model'', consisting of a finite number of atomic layers in one direction and PBC in the other two directions.The 3D bulk structure is shown in Fig.~\ref{fig2.2_AR1}a. Fig.~\ref{fig2.2_AR1}b  shows slab models for the (001) and (100) surfaces of hexagonal water ice.  Amorphous systems can also be modelled using a slab model, provided that the unit cell is large enough to ensure local disorder. As structural atomistic details are not available directly from experiments, the usual way to arrive at an amorphous material is to start from a crystalline bulk and simulate a melt/quench process using MD runs at relatively high temperature to ensure loss of order \citep{suter2006,buch2008}. An example of this procedure is shown in Fig.~\ref{fig2.2_AR2}a taken from \'Asgeirsson et al.\citep{asgeirsson2017}. Slab models can be used in combination with classical force fields and DFT using either Gaussian type functions or plane-waves to describe the electron density.

An alternative to PBC is using clusters, some paradigmatic examples for water ice being presented in \citep{maheshwary2001,ugalde2000,lamberts2016,kayi2011, buck2014,shimonishi2018}. This approach is particularly appealing in combination with post-HF  methods as implemented in molecular codes  which typically use atom-centered basis sets without periodicity. A cluster can either be generated by extracting a part from the bulk structure (see Fig.~\ref{fig2.2_AR1}c) or by combining individual molecules (see Fig.~\ref{fig2.2_AR2}b \citep{rimola2012,rimola2010,rimola2018}).  The main disadvantages of the cluster approach are the limited size of the clusters, a high surface-to-volume ratio and non-desired edge effects. A smart solution is the QM/MM approach as discussed previously.  An example of the application for an amorphous water ice system is shown in  Fig.~\ref{fig2.2_AR2}b \citep{song2016,lamberts2017}. Here, ice was initially generated through MD simulations with PBC. Then, a large cluster was created by cutting out a hemisphere from the bulk. Only the molecules visualized in ball-stick mode are at the QM level.
For more details on surface modelling of interstellar materials, we refer the reader to this review\cite{rimola2021}.

\subsection{Astrochemical models}
Astrochemical gas-grain models are typically based on integration of rate equations\cite{cuppen2017}. The thermal desorption rate is one of many terms included in these equations. Generally, this rate is included following a first order desorption process according to  Eq.~\ref{eq:Polanyi}, irrespective of the thickness of the ice. The surface concentration $n_{i, \rm abs}$ is typically taken as the concentration over the full mantle per square area. Some models reduce this concentration by a factor for the thickness of the ice. For a pure ice, desorption changes from zeroth order in the multilayer regime, to first order in the monolayer regime using the reduction factor. For mixed systems, this does not fully hold since it assumes homogeneous mixing at every time. Multiphase models,  like proposed by Hasegawa and Herbst \cite{hasegawa1993}, solve this to some extent. Here, the ice mantle is divided into two parts; a surface part from which desorption can occur and an inert inner part which can replenish the surface. Fayolle et al. \cite{fayolle2011} have shown that such a three-phase model can indeed capture some transitions from zero to first order.

Astrochemical databases for binding energies exclusively focus on tabulations of first order data as in Section 4. As explained above, experimentally the order is in many cases not limited to first order, depending on the system and the regime (monolayer vs. multilayer) that is probed. The presence of $o$ in Eq.~\ref{eq:Polanyi} complicates tabulating TPD data for direct use in astrophysical models. The unit of the pre-exponential factor, for instance, depends on the order of the desorption process. These pre-exponential factors can however be easily interconverted by placing the zero and fractional order data on a pseudo-first order basis using a suitable monolayer surface density, $n_\text{i,ads,0}$, to scale the pre-exponential factor. 

\citet{Penteado2017} studied the effect of uncertainties in the binding energies on an astrochemical two-phase model of a dark molecular cloud, using the rate
equations approach. The binding energy of \ce{H2} was found to be of critical importance for the chemistry of several simple ice species. The freeze-out of \ce{H2} that occurs for high \ce{H2} binding energies gives access to different parts of the reaction network, leading to a more methane-rich ice, instead of a methanol-rich ice that is found for lower binding energies. Grassi et al. 
\cite{grassi2020} found that the surface chemistry drastically changes when multiple binding sites are considered, instead of a single value binding site. They covered a range of different conditions, typical for the surroundings of a proto-star or in the disk mid-plane. A kinetic Monte Carlo study \cite{cuppen2011} showed for \ce{H2} formation that even the spatial distribution of the different binding sites over the surface can have a large effect on the chemistry. Implementation of different spatial distribution is not possible in rate equations and the implementation of a distribution of binding sites is computationally very costly. Moreover, the exact shape of the binding energy distribution  function  will  play  a  key  role  in  the  evolution  of  these chemical models.
\section{3. Case studies in models and experiments}
To further this pedagogical discussion, it is appropriate to consider some relevant case studies to illustrate both the issues that arise in experimentally and computationally exploring representative systems and the impact of such studies on astrophysical models.

\subsection{Thermal desorption of solid water}
Water (\ce{H2O}) is the most abundant ice component in astrophysical environments \citep{ehrenfreund2000,Boogert15}. \ce{H2O} is readily detected in the solid state through its 3.07 $\mu $m O--H stretching absorption band and has been studied extensively \citep[and references therein]{gillet1973,bar-nun1985,o'neill1999,collings2003,collings2004,tielens2013}. These data allow us to explore some of the complexity of this simple system.

The first issue that we have to address is the structural complexity of solid water that gives rise to a multitude of different forms of solid water produced at low temperature and under
high and ultrahigh vacuum (Table~\ref{tab3names} ). In the laboratory, \ce{H2O} is typically deposited onto a solid surface at low temperatures, about 10~K, where it is believed that film growth occurs by ballistic deposition. The incoming \ce{H2O} molecules ``hit and stick'' to the surface leading to a low-density, porous amorphous solid \ce{H2O} with a relatively short average O--O distance in the ice matrix \cite{stevens1999}. Warming of this \ce{H2O} ice to temperatures between 30 and 70~K leads to the coalescence of the pores\cite{raut2007, bossa2015, cazaux2015} or to their collapse to form high-density,
non-porous (np) or compact amorphous solid \ce{H2O}; this phase change is not apparent in infrared spectroscopy, but can be detected through various other analytical techniques \citep{narten1976,jenniskens1994,lu2001} and is accompanied by an increase in the average O--O distance in the solid matrix. Further annealing leads to the formation of the cubic crystalline phase above 140~K \citep{jenniskens1995,smith1997}. At higher pressures, approaching those of planetary atmospheres, crystallisation is into the well-known hexagonal phase of crystalline solid \ce{H2O}.

\begin{table*}
\centering
\caption{\ Overview of names for the different solid water phases upon deposition under low
pressure (i.e. high and ultrahigh vacuum) conditions and the temperatures at which they occur.
}
\begin{tabular}{cccc}
\hline
\hline
0--80~K & 80--120~K & 120--160~K & $>$ 160~K\\
\hline
Porous amorphous & Compact amorphous & Crystalline & Gas \\
\hline
HDA & LDA & I$_\text{c}$ (cubic) & \\
p-ASW & np-ASW    & I$_\text{h}$ (hexagonal) &\\
    &  c-ASW     & CSW & \\
\hline
\end{tabular}
\label{tab3names}
\end{table*}

In addition to thermal processing, radiation of various types can result in compaction of porous amorphous solid \ce{H2O} \citep{palumbo2006}. Radiation can also induce amorphisation of crystalline \ce{H2O} \citep{strazzulla1992,fama2010}. However, it should be recognised that in cold dense astrophysical environments, \ce{H2O} is reactively accreted from H and O atoms rather than growing by simple molecular deposition \citep{ioppolo2008,ioppolo2010,cuppen2010,dulieu2010}. In principle, the release of reaction enthalpy in forming \ce{H2O} should provide sufficient local thermal input to ensure that the resulting amorphous film is of a compact nature. Moreover, reaction enthalpy release from other molecular formation processes can also drive compaction \citep{accolla2011,accolla2013}. Thus, it is highly likely that \ce{H2O} ice in most astrophysical environments is of a compact, amorphous solid nature \citep{oba2009,accolla2011,accolla2013}.

\begin{figure}[b!]
\centering
 \includegraphics[width=0.7\textwidth]{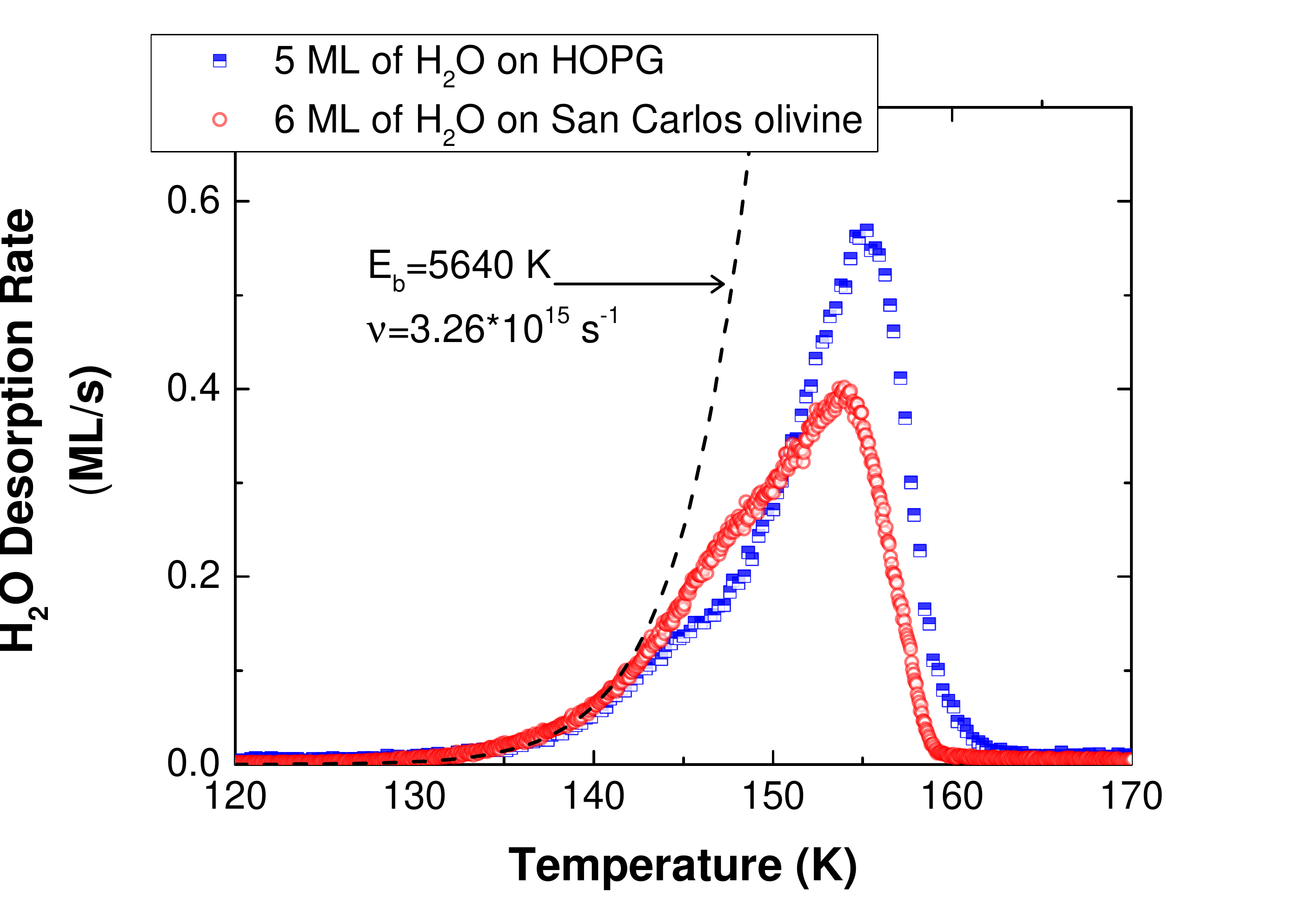}
 \caption{Comparison of TPDs of solid water from HOPG or San Carlos olivine surfaces. The dashed curve is an  exponential trend obtained by using an E$_b$=5640 K and a $\nu$=3.26 $\times$ 10$^{15}$s$^{-1}$.}
 \label{fig:3-1}
\end{figure}

High quality TPD data for solid \ce{H2O} exist that on some substrates demonstrates zero order desorption from the lowest of coverages \cite{collings2015,fraser2001,smith2014}, which is
consistent with de-wetting of the \ce{H2O} from the surface and formation of agglomerated water clusters as the film sublimes; \textit{i.e.} the \ce{H2O}-\ce{H2O} interactions are stronger than the \ce{H2O}-surface interactions. While on other surfaces, there is clear evidence of wetting behaviour, \textit{i.e.} the \ce{H2O}-\ce{H2O} interactions are weaker than the \ce{H2O}-surface interactions, and the formation of a monolayer occurs at the substrate surface on which multilayer growth occurs \cite{bolina2005}. Fig.~\ref{fig:3-1} illustrates these observations in the case of \ce{H2O} adsorbed on San Carlos Olivine and on HOPG. The leading-edge analysis (dashed curve) shows an example of how desorption parameters can be determined from TPD curves in the case of zero order desorption. One can note that such parameters are in good agreement with that presented in table 3 of section 4. 
Finally, we stress that both TPD curves in Fig.~\ref{fig:3-1} show evidence for the crystallisation of \ce{H2O} as a low temperature shoulder on the TPD profile (around 145 K).

Of course, analysis of the zero order behaviour of the multilayer films yields consistent activation energies for desorption and pre-exponential factors for both the amorphous and crystalline phase. This is consistent with the general view that desorption of multilayer films is unaffected by the nature of the underlying substrate.
On those substrates from which only zero order desorption is observed a logical question springs to mind. When does \ce{H2O} become mobile enough to agglomerate? Temperature programmed IR spectroscopy suggests that de-wetting might occur on a laboratory timescale at around 40~K \citep{collings2015}. However, more detailed time- and temperature-resolved IR studies reveal that agglomeration occurs on laboratory timescales at temperatures as low as 18~K \citep{rosu-finsen2016} and is accelerated by reaction enthalpy release \citep{rosu-finsen2018}. However, the latter is likely to have less impact in astrophysical environments where the rate of enthalpy release is much lower. This points to a picture of the initial phases of icy mantle development on a grain where three dimensional \ce{H2O} islands pepper the grain surface with bare grain surface between. This contrasts markedly with the widespread belief that icy mantles grow in a layer-by-layer fashion. 

\subsection{Thermal desorption of carbon monoxide on and from solid water}
After \ce{H2O}, carbon monoxide (CO) is the next most abundant species found in astrophysical ices. CO ice is readily detected in the infrared \textit{via} the C-O stretching
vibration at around 4.67~${\mu}$m. Three peaks are frequently evident in this spectral window - two relatively narrow features around 4.674 and 4.665~$\mu$m  and a broader component around 4.681~$\mu$m. The first is usually attributed to CO in an `apolar' environment, \emph{i.e.} an environment where weak van der Waals interactions dominate, as in a pure CO ice. The small feature at 4.665~$\mu$m is either ascribed to mixtures of  solid CO and CO$_2$ \citep{boogert2002a,broekhuizen2006} or to crystalline CO ice \citep{pontoppidan2003}. The broad feature \citep{tielens1991,chiar1995,pontoppidan2003} is generally attributed to CO in a hydrogen bonding environment \citep{sandford1988,allamandola1999}. This motivated many experimental studies on CO mixed with \ce{H2O}. However, recent observational studies show that the hydrogen-bonding environment is likely due to \ce{CH3OH} instead of water\citep{cuppen2011,Penteado2015}, where \ce{CH3OH} is formed through the hydrogenation of CO \cite{fuchs2009,watanabe2004, hiraoka2002,minissale2016}.

Equilibrium measurements of CO adsorption on ice have been reported by Allouche \textit{et al.} \citep{allouche1998}. Single values for the enthalpy of adsorption are reported for a number of samples of ice derived from isosteric measurement at temperatures between 43 and 48~K in the range 1060-1240~K. This is perhaps a hint towards the heterogeneity of the solid \ce{H2O} surface and a distribution of the CO binding energies. 
\begin{figure}[th]
\centering
 \includegraphics[width=0.5\textwidth]{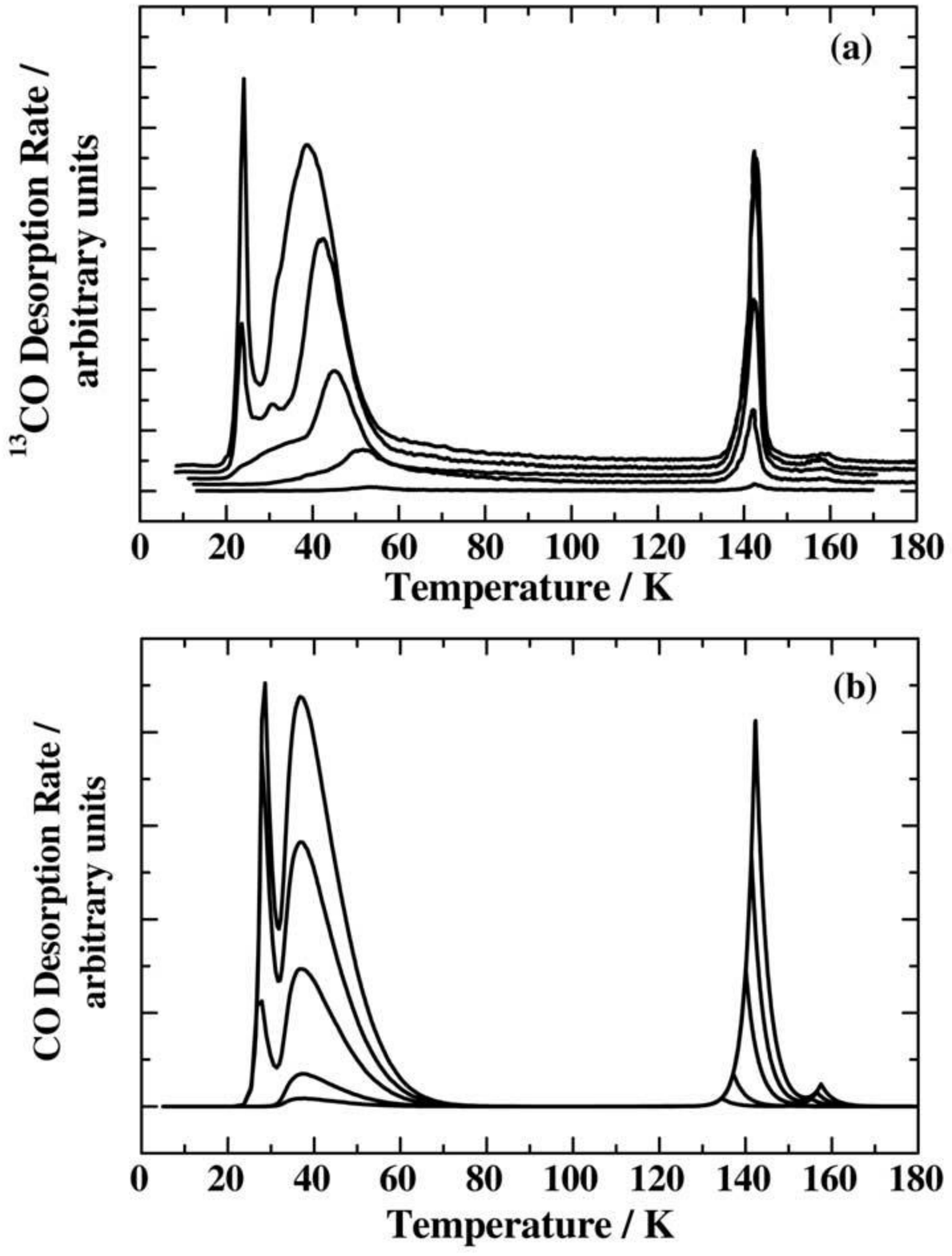}
 \caption{TPD of \textsuperscript{13}CO (1.28, 5.13, 25.6, 51.3, and 76.9 {\texttimes} 10\textsuperscript{14} molecules cm\textsuperscript{{}-}\textsuperscript{2}) adsorbed at 8~K on \ce{H2O} ($5.7\times 10^{17}$ molecules cm\textsuperscript{{}-}\textsuperscript{2}) adsorbed at 8~K, at a heating rate of 0.08~K s$^{-1}$, offset for clarity; (b) Simulated TPD of CO (1.28, 5.13, 25.6, 51.3, and 76.9 {\texttimes} 10\textsuperscript{14} molecules cm$^{-2}$) from \ce{H2O} (5.7$\times 10^{17}$ molecules cm$^{-2}$) using the model with rate equations listed in the paper using a heating rate of 0.08~K s$^{-1}$, initial populations of CO bound to water in pores are 1.28, 5.13, 20, 20, and 20 {\texttimes} 10\textsuperscript{14} molecules cm$^{-2}$ respectively, with the balance as solid CO. Reproduced from Collings et al.\cite{collings2003}}
 \label{fig:3-2}
\end{figure}
Detailed surface science studies of the CO-\ce{H2O} system followed \citep{collings2003,acharyya2007}. While the focus of these works was clearly on the role of trapping of CO in the ice, a kinetic analysis of the TPD data yielded desorption parameters for CO solid ($E_\text{b} = 830{\pm}25$~K and ${\nu} = 7\pm 2 \times 10^{26}$~cm$^{-2}$ s$^{-1}$) and for the monolayer of CO adsorbed on solid \ce{H2O} ($E_\text{b} = 1180{\pm}25$~K and ${\nu} = 5\pm 1 \times 10^{14}$~s$^{-1}$)\cite{collings2003}. A comparison of the detailed kinetic analysis of these experiments is shown in Fig.~\ref{fig:3-2}. 
Clearly, the single activation energy for desorption of CO from the solid \ce{H2O} surface ensures adequate reproduction of the experimental data. However, the tailing to high temperatures suggests that this might be a woefully inadequate description of the interaction of CO with a porous amorphous solid \ce{H2O} surface. Indeed, in Collings et al.\cite{collings2003}, this tailing is discussed in terms of CO diffusion into the porous structure of the substrate over the solid \ce{H2O} surface and out of the porous structure via gaseous diffusion\cite{he2014b, mate2020}. However, the commonality of the trailing edges and the lowering of the peak desorption temperature for CO leaving the solid \ce{H2O} surface is entirely consistent with first order desorption from a heterogeneous surface exposing a range of binding sites and hence a distribution of binding energies. Kay and co-workers have recently measured this distribution \cite{smith2016} and report binding energies ranging from 11 to nearly 2040 K with an optimised pre-exponential factor of $3.5 \times 10^{16}$ s$^{-1}$. 
Even in the simpler case of compact ASW (Amorphous Solid Water) where trapping and diffusion are limited, three independent studies \citep{noble2012,fayolle2016,he2016} have used fixed pre-exponential factors derived from the harmonic approximation by Hasegawa et al.\cite{hasegawa1992a} of $7.1\times 10^{11}$ s$^{-1}$,  or from the work of Redhead\cite{redhead1962} as $1.0\times 10^{12}$ s$^{-1}$; 
(see table \ref{Table:Pre-factor} for a more detailed list of pre-exponential factors), and obtained binding energy distributions ranging from 1000 to 1400 K. These values however differ from the one determined by more recent and detailed work of Smith and co-workers \cite{smith2016} (1503 K, 1.9$\times$10$^{13}$ s$^{-1}$).

Furthermore, the question as to which set of parameters to use is moot as it is possible to compare the data by recognising that, as proposed in reference \cite{chaabouni2018},
\begin{equation}\label{EqNUEq}
k(T)={ \nu_1e^{E_{1, des}/k_{B}T}}={ \nu_2e^{E_{2, des}/k_{B}T}},
\end{equation}
where ${\nu_1}$ and ${ \nu_2}$ are the pre-exponential factors, associated with the desorption energies ${\rm E_{1, des}}$ and ${\rm E_{2, des}}$, respectively, of the adsorbed molecules on the surface, and T is the experimental  temperature of the surface, at which the desorption is observed. To illustrate, we compare the apparently contradictory results from above. Re-scaling the pre-exponential factor of $3.5\times 10^{16}$ s$^{-1}$  to $7.1\times 10^{11}$ s$^{-1}$ corresponds to reducing the lower bound of the binding energy from 1320 to 1000 K  (assuming T = 30 K), and the upper bound from 2040 to 1560 K (assuming T = 45 K). In others words, the lower bound exactly matches, and the upper bound is slightly higher in the case of porous water ice than that of compact ice, which is exactly as reported in the literature \cite{noble2012,he2016}.

We underline here that experimental studies agree on the desorption rate,  but the choice of the coupled parameters, $E_{des}$ and $\nu$, may lead to apparent discrepancies. The use of equation \ref{EqNUEq} makes it possible to compare the derived coupled parameters. 

The final important point to make is that the surface morphology, as well as composition, directly influences the binding energy distribution of adsorbates on the water ice surface. We can see the effect of this surface heterogeneity on the water ice surface in relation to its significant impact on the position and width of the spectral bands of adsorbed CO \citep{taj2017}. Interactions of CO with other model substrates have been reported including amorphous silica \citep{collings2015,taj2017}. Typically on amorphous silica, the interaction with CO is considerably weaker and ranges from 720 to just over 1200 K assuming a pre-exponential factor of
$1.0\times 10^{12}$ s$^{-1}$, as is normally, if incorrectly, assumed for a physisorption interaction. This would clearly suggest a preference for adsorption of CO on solid water surfaces over the bare substrate if thermal desorption was the only desorption pathway available in astrophysical environments.


\section{4. Tables of recommended values}

Astrophysical models are all based on a parameterization of the physical properties of matter. 
In particular, in astrochemical models, physical parameters, such as the binding energy of chemical species on icy/bare surfaces are assigned to each species used in the model. The focus of this section is to review and discuss the values of binding energies commonly used in the astrochemical community.
Taking into account the diversity of astrophysical environments in terms of chemical composition and the characteristics of the dust grains, it is not possible to provide a complete overview here, and we have
restricted our selection to the most abundant molecules clearly identified as solid state species and to those species that hopefully will be identified on interstellar ices in the frame of the upcoming JWST programs. 

It has long been assumed that the binding energies of pure chemical species could represent the volatility of a species in different astrophysical environments. However, it is established that molecular ices are composite, and moreover that only molecules at the surface (or able to access the surface through cracks in ice cover) may desorb. Thus we have seen before that we can distinguish desorption of thick layers of pure bodies from the surface desorption of these same species. We have therefore chosen for this article not to recall the properties of pure substances, which can for example be found in Fray \& Schmitt\cite{fray2009} (and references therein), but to compile data on binding energy from model surfaces.
We have considered three types of astrophysically relevant surfaces:  {\it water ice}, and in particular compact ASW, since it is the major constituent of astrophysical ices; {\it silicate and carbonaceous surfaces} that represent bare grain surfaces (we propose a compilation of existing values of both silicates and/or carbonaceous surfaces). 

The objective of the tables is to list the existing experimental and theoretical values on the different surfaces and in the sub-monolayer regime, showing the disparity of the studies, and then to propose a single value, a simplified version recommended to those who would like to use a single value in the framework of a more complicated astrophysical model, which would not be able to deal with energy distributions for example.

\subsection{Description of the parameters}
\begin{figure}[h]
\centering
\includegraphics[width=10cm]{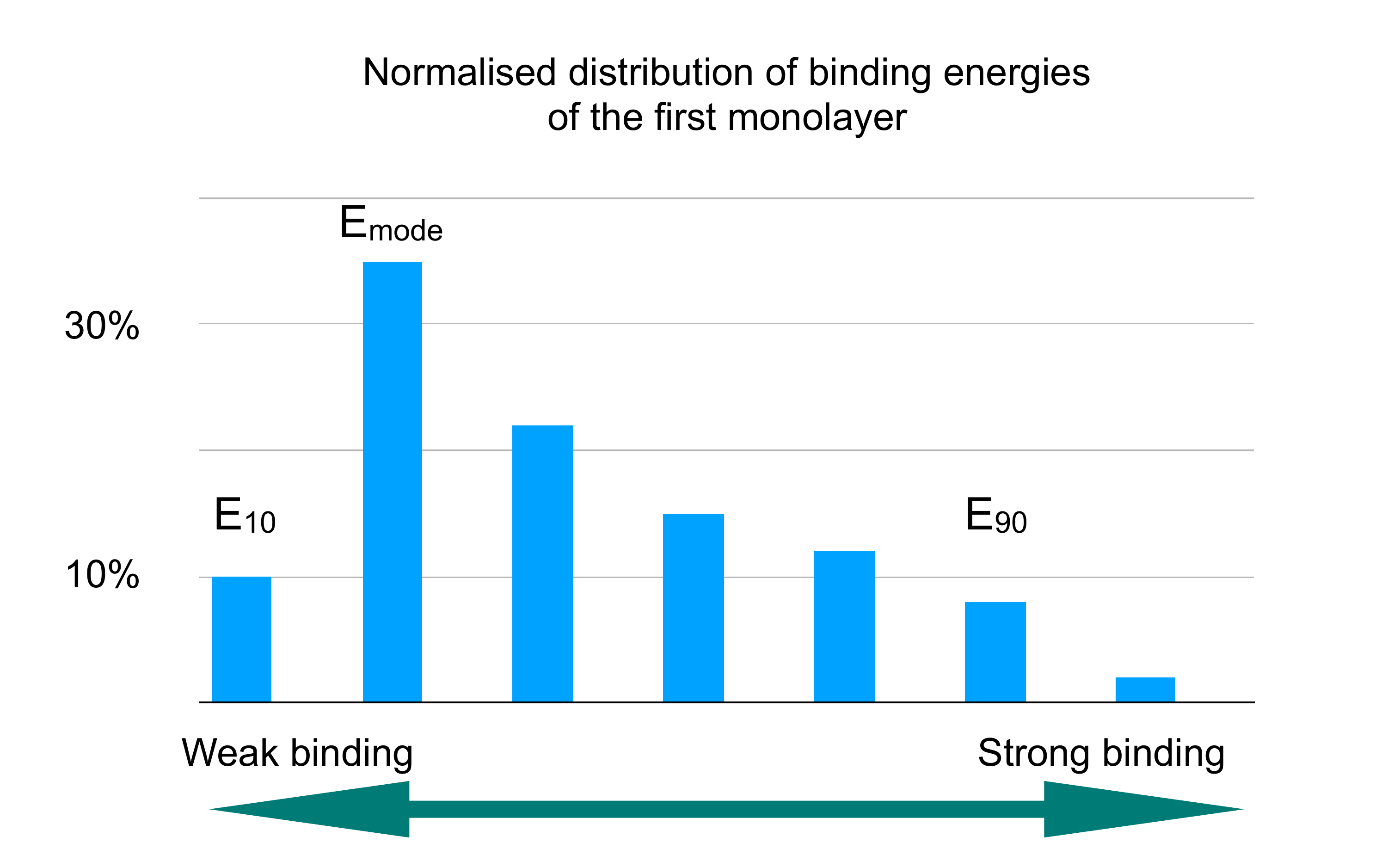}
\caption{Representation of the binding energy distribution of the first monolayer of adsorbate. E$_{10}$ (resp. E$_{90}$) represents the binding energy after 10\% (90\%) of desorption. E$_{mode}$ corresponds to the E$_b$ of the largest number of adsorption sites.} \label{fig:04fig01}
\end{figure}

We have seen in section 2 that there are many ways to represent a binding energy, but that one can concentrate on a couple of values (E$_b$, $\nu$) $i.e.$ binding energy and pre-exponential factor, if one approximates a desorption of order 1. However, E$_b$ is dependent on the fractional coverage of the surface and binding energies are in principle represented by a distribution. 
When this distribution is measured or calculated, we provide the values E$_{10}$ and E$_{90}$ which correspond respectively to 10\% and 90\% of the monolayer desorption (and therefore inversely for the coverage fraction), as well as E$_{mode}$ which corresponds to the maximum of the distribution. Since the E$_b$ increases with decreasing coverage, we will therefore need to have E$_{10}$ $<$ E$_{mode}$ $<$ E$_{90}$. Fig. \ref{fig:04fig01} illustrates our convention for the parameters.

In this review, we present a table containing E$_b$ and $\nu$ for different chemical species and surfaces. When the distribution is not measured, the average value and its uncertainty is given. The same is done for the pre-exponential factor $\nu$. We choose a value of 10$^{12}$ s$^{-1}$, if the value of $\nu$ is not clearly specified in the considered reference.

For the Tables in this section, we distinguish reliable and well-established values from simple educated guesses by using two safeguards: \textit{i)} each parameter is associated with a reference in the text that will quote as exhaustively as possible the related articles. When several sources exist, then we show the value {\bf in bold} and motivate our choice. When there is at least one study, then we use normal font, and when it is a default choice, or established with empirical rules we use {\it italics}. 
\textit{ii)} we divide explicitly the table into two parts. One is devoted to the collection of literature data, and it can be left empty if not existing. The second one is for recommended values and will be completed. We provide all relevant information in order for readers to make well motivated choices, as eventual issues are clearly mentioned as well. Generally speaking, when we have to pin point a value from a distribution to a "unique recommended value" (knowing that it is an oversimplification), we consider whether the species is abundant or not. If it is an abundant species, its surface coverage can be important, and therefore we chose the mean values (or E$_{mode}$). If the species is less abundant we consider the values of the low coverage, which is defined by the upper bound of the distribution. 

We stress that the recommended $\nu$ has been estimated through the Transition State Theory (TST) (see the extensive discussion in section 6) while by equalizing the desorption fluxes of the given species at T$_{peak}$:
\begin{equation}
    \nu_{TST}*e^{-\frac{E_b}{k_b T_{peak}}}=\nu_{LV}*e^{-\frac{E_{LV}}{k_b T_{peak}}}
\end{equation}
we can calculate the recommended E$_b$ using:
\begin{equation}
    E_{b}=E_{LV}-T_{peak}*ln(\nu_{LV}/\nu)
\end{equation}
where E$_{LV}$, $\nu_{LV}$, and T$_{peak}$ are the desorption parameters chosen from literature; T$_{peak}$ is the temperature at which the maximum desorption rate for a 1 ML TPD is found for a given species. 
We stress that the selected species are listed, both in the tables and in the guiding text, in ascending order of the recommended E$_b$ value. 

\subsection{Parameters for species thermally desorbing from compact ASW}
The first column of Table \ref{Table_bindings_ASW} lists our choice of the 21 most important molecules. The second set of columns entitled "Literature data" corresponds to values gathered from the literature and presented as a couple of pre-exponential factors and binding energies, in terms of a distribution if possible, or mean and spread or unique value if not. These can be experimental or calculated values. Then the last column, one of the legacies of this article, consists of recommended couples of pre-exponential factors and binding energies. The way we determine the pre-exponential factor will be explained in the next section.
Below we discuss separately the properties of the 21 selected species.
\\

\begin{description}

\item[H$_2$] There have been many studies reporting directly or indirectly the binding energy of H$_2$ on porous ASW, both experimentally and theoretically\citep{hixson1992,amiaud2006,amiaud2007,ferrero2020}. Sophisticated quantum statistical treatments have even established the difference in binding energy of ortho- and para- H$_2$, that was observed and calculated in previous work\citep{amiaud2008,hixson1992,buch1993}. We adopt the mean value $E_{mode}$ since H$_2$ is over-abundant at low temperatures ($<$ 15 K). We draw attention to the fact that reading the overview table could suggest that the binding energy of H is less than that of H$_2$. This is not clearly established. The lower limit for E$_b$ of H$_2$ from experimental work is shown in italics. However, given the temperatures of the measurements (usually at least 8 to 10 K) it is not possible to measure E$_b$ values much smaller than 300 K (because it would desorb rapidly). The multilayer E$_b$ of H$_2$ is estimated to be 111 K ($\nu= 10^{12} s^{-1}$) \cite{schlichting1993}.

\item[H] Despite the fact that H is certainly the most important reactive species to consider in astrophysics, experimentally measuring the binding energy of such a radical to a surface is a real challenge. Most experimental studies have focused more on diffusion than on desorption. However, these two processes are almost impossible to decouple. There is a reported experimental value E$_b$= 50 meV (580 K) for {\it porous} amorphous ice, reported in the article of Wakelam et al. \cite{wakelam2017b}. In this article, several experimental references are compiled\cite{manico2001,hornekaer2003,amiaud2007,matar2008,watanabe2010,hama2012}. Since binding energies of atomic H cannot be directly determined, most values originate from computational work. However the difficulty here is to define properly compact ASW from a theoretical point of view. Porous amorphous water ice can be simulated with cluster calculations \cite{Buch1992} because the binding energy is determined by the topology of the water ice surface at the molecular scale \cite{fillion2009}. Theoretical results obtained with crystalline ice match better with experimental findings on compact amorphous surfaces (and of course crystalline ones)  whereas calculations on amorphous surfaces compare very well with experiments made on porous amorphous surfaces \cite{senevirathne2017}. Therefore it is not a priori clear to decide which set of values should be recommended for use. The binding energy ranges from 20 to 24 meV for the case of crystalline ice, whereas it is 22-63 meV in the case of ASW. Despite the fact that experimental estimates are pointing to values lower than 29 meV for D atoms \cite{amiaud2007}, we adopt the conservative values of the calculations made on the ASW substrate by Senevirathne et al. \cite{senevirathne2017}, and which correspond to values determined experimentally for {\it porous} ASW, and to the conclusion in Wakelam et al. \cite{wakelam2017b}. The values recommended for future use are the conversion of the couple E = 580 K = 50 meV, to the new proposed value of $\nu$ (T = 12 K) (see next section for the determination of pre-exponential factors). We mentioned earlier that we would adopt the high value of the binding energy distribution in the case of an adsorbate of low surface concentration. So we need to justify why H is considered to have a low coverage on grain surface: it is due to its high mobility and high propensity to react even through tunneling, that makes this atom, while very abundant, quickly transformed at the surface of grains.

\item[N] Due to similar polarisabilities, C, O, and N binding energies have been established at 800 K in the pioneering paper of Tielens \& Hagen \cite{tielens1987}. The E$_b$ of N has been found to be relatively close to the initial value. Calculations have confirmed the lower value range of this E$_b$ \cite{shimonishi2018}.

\item[N$_2$] Different works measured values between  970 and 1155 K for the mono- or sub-monolayer regime \cite{fuchs2006,smith2016, fayolle2016, nguyen2018} with big discrepancies for the $\nu$ value: Smith et al.\cite{smith2016} measured 4.1$\times$10$^{15}$ s$^{-1}$, Fayolle et al. \cite{fayolle2016} 6-10$\times$10$^{11}$ s$^{-1}$, while Nguyen et al.\cite{nguyen2018} reported 10$^{13}$ s$^{-1}$. To calculate the recommended value, we consider the probable presence of co-absorbed CO that lowers the binding energy (see later paragraph on surface segregation effects). E$_b$ for this species on both crystalline and ASW has also been calculated theoretically \cite{ferrero2020}.

\item[O$_2$] Binding energies of O$_2$ span from 1500 to 900 K for coverages that go from 0 to 1 ML, respectively \cite{fuchs2006,noble2012, noble2015, he2016}. To calculate the recommended value, we consider the couple of values (1104 K, 5.4$\times$10$^{14}$ s$^{-1}$) proposed by Smith et al.\cite{smith2016}. E$_b$ for this species on both crystalline water and ASW has also been calculated theoretically \cite{ferrero2020}.

\item[CH$_4$] The methane parameters in the sub-monolayer regime have been measured both on ASW and on graphene \cite{smith2016} as well as computed theoretically \cite{ferrero2020}.

\item[CO] CO is an emblematic molecule for which desorption from   water has been widely studied, theoretically and experimentally (e.g. \cite{fuchs2006,collings2003,karssemeijer2014I, ferrero2020}). Distributions of E$_b$ have been independently established experimentally on compact ASW, leading to close conclusions \cite{noble2012,fayolle2016}. We note here that values on porous water ice can be much higher (up to 1500 K). 

\item[O] Tielens \& Hagen \cite{tielens1987} have proposed a value of around 800 K. In recent years, the binding energy of O has been challenged both experimentally and theoretically, and a value about twice higher (1300-1700 K) was estimated \cite{he2015, minissale2016b}.

\item[C$_2$H$_2$] A recent study \cite{behmard2019} has measured the binding energies of many small hydrocarbons, and we kept C$_2$H$_2$ as an example. In this paper, distributions, pre-exponential factors and variations from one molecule to another can be found.

\item[CO$_2$]  Different studies \cite{noble2012,andersson2004,edridge2013, sandford1990} have led to close results, also from a theoretical side\cite{ferrero2020}. As an example direct comparisons from the table are not easy since order of desorption and pre-exponential factor vary. 
The binding energy of the monolayer is actually very close to that of the multilayer. 

\item[H$_2$S] We use 2700 K, the empirical value derived by Wakelam et al. \cite{wakelam2017} that is in good agreement with the theoretical value derived by Oba et al.\cite{oba2018} and those by Ferrero et al.\cite{ferrero2020}. For the sake of completeness, we stress that two E$_b$ values (1800 and 2800 K) have been calculated depending on the bonding arrangement between H$_2$S and H$_2$O.

\item[H$_2$CO] This molecule has been studied in sub-monolayer regime\cite{noble2012b} and E$_b$ = 3260 K was measured. H$_2$CO is known to readily polymerise and therefore as the concentration rises during desorption when mixed with water will self-react prior to its desorption. E$_b$ for this species on both crystalline water and ASW has also been calculated theoretically \cite{ferrero2020}.

\item[CS] Wakelam et al \cite{wakelam2017} provided an estimation of E$_b$ of 3200 K ($\nu$= 10$^{12}$ s$^{-1}$) 

\item[HCN] We use the empirical value derived by Wakelam et al. \cite{wakelam2017}, which is in agreement with TPD curves published in Theule et al.\cite{theule2011}. E$_b$ for this species on both crystalline water and ASW has also been calculated theoretically \cite{ferrero2020}.

\item[NH$_3$] The multilayer E$_b$ of NH$_3$ is often reported (e.g. 3070 K, $\nu$= 2$\times$10$^{12}$ s$^{-1}$ \cite{martin-domenech2014} ), but there is no specific experimental study of NH$_3$ on ASW. Due to its H bonding, it may be partly integrated to the water network and its E$_b$ on ASW is actually higher. TPD data are shown in Collings et al\cite{collings2003}. A peak is seen at around 90 K, compatible with the multilayer desorption, and followed by a desorption tail. Although the initial substrate is not compact ASW but porous ASW, it is more or less what was measured from compact ASW (Dulieu, private com). So there is a broad distribution of binding energies starting from approximately 3000 K ($\nu$=10$^{12}$ s$^{-1}$) to 4800 K (ASW itself). These values are in agreement with the lower limit range of E$_b$ provided by Ferrero et al. \cite{ferrero2020} by means of quantum chemical simulations. A higher value has been reported (empirical method) by \citeauthor{wakelam2017}, but taking into account the desorption of water itself, we keep the ceiling value of 4600 K because of the low abundance of NH$_3$ ($<$ 0.1 of H$_2$O).

\item[OH] OH is a radical, therefore, during the heating ramp of conventional TPD, it has a large chance to diffuse and react, making the experimental evaluation of the desorption parameters almost impossible. There is however an experimental value reported but not detailed \cite{dulieu2013} of 4600 K, slightly less than the E$_b$ of water, but well above the reference value of OSU gas-phase database (2850 K). It has been confirmed by a systemic empirical approach\cite{wakelam2017b},and by quantum calculations where broad binding energy distributions have been found, also in an independent work\cite{ferrero2020}. Average values have been estimated to lie between 4300 and 4990 K \cite{sameera2017,miyazaki2020}.

\item[H$_2$O] It is intrinsically not possible to provide a value for the binding energy of water on ASW in the sub-monolayer regime, since it constitutes the surface. Moreover, we have seen that the water can undergo crystallisation before it desorbs, thus delaying the desorption (shoulder in Fig \ref{fig:3-1}). Therefore we can provide the binding energy of ASW itself, which is different and slightly lower than that of crystalline water. This value has been measured (E$_b$ = 5640 K, $\nu$= 3.26$\times$10$^{15}$ s$^{-1}$ ) \cite{speedy1996} and is in agreement with other measurements \cite{sandford1990,fraser2001} and theoretical calculations \cite{ferrero2020}. We remind the reader that this value is lower than that of crystalline (cubic) ice. See ref \cite{speedy1996} for comparison as well as for values of deuterated water. 

\item[CH$_3$CN] Together with its isomer CH$_3$NC, acetonitrile has been studied both experimentally and theoretically, and distributions of E$_b$ as well as pre-exponential factor have been provided.\cite{abdulgalil2013,bertin2017a,ferrero2020}

\item[CH$_3$OH] The desorption parameters of methanol have been studied both experimentally and theoretically\cite{bahr2008,ferrero2020} on ASW, even if it was a D$_2$O thin film. The authors report a value of 4800 K for $\nu$= 10$^{13}$ s$^{-1}$. It is close to the empirical value reported in Wakelam et al. of 5000 K, but this last one is discarded because it is slightly above the value of water, which is not what is observed. Methanol can partly co-desorb with the water, or in greater amounts prior to the ASW. 

\item[NH$_2$CHO] It has been shown that formamide desorbs after the water substrate, therefore there is no direct measurement possible. However, the E$_b$ for this species on both crystalline and ASW has been calculated theoretically \cite{ferrero2020}. We also provide in the table an estimate of the desorption energy of NH$_2$CHO, which is an average from the silicate and carbonaceous substrates\cite{dawley2014,chaabouni2018}. This value is slightly higher than the predicted empirical one.

\item[C] Wakelam et al.\cite{wakelam2017} suggested that C atoms should react with water, and so the binding energy of C has nothing to compare with physisorption and is probably an order of magnitude higher. This has been confirmed by a quantum calculation study \cite{shimonishi2018}, establishing the binding energy of C as chemisorption at 14300 K. This has not been confirmed experimentally and we keep this value.

\end{description}

\begin{table*}
	\centering
	\caption{Binding energies of sub-monolayer regimes on compact ASW. $\nu$ in s$^{-1}$ and E in K. The selected species are listed in ascending order of E value.}
	\label{Table_bindings_ASW}
\resizebox*{\textwidth}{!}{\begin{tabular}{l|c|ccc|c|c ||cc||c} %
		\hline
		\hline
	& \multicolumn{6}{c||}{Literature values} & \multicolumn{2}{c||}{Recommended values} & T$_{peak}$\\
	Species& $\nu$ & E$_{10}$ & E$_{mode}$ & E$_{90}$ & E$_{mean}$ $\pm$ $\Delta E$ & Reference & $\nu$ & E & K \\
    \hline
H$_2$	&10$^{13}$	& {\it300}&	450	&\bf{600}	&-	& \citep{hixson1992,amiaud2006,amiaud2007, wakelam2017b} & 1.98$\times10^{11}$&371 & 20	\\   
H   	&  10$^{12}$	&255 &	 -	&700	&- & \cite{manico2001,hornekaer2003,amiaud2007,matar2008,watanabe2010,hama2012,senevirathne2017}	& 1.54$\times10^{11}$&450	& 15\\
N   	&	10$^{12}$	& 640 &	 720 & 880	&- & \cite{minissale2016b}	& 1.17$\times10^{13}$ & 806 & 35	\\
N$_2$    &	10$^{13}$/4.1$\times10^{15}$	&970 &	-	&1350	& 1152 $\pm$50 &   \citep{smith2016}	& 4.51$\times10^{14}$ & 1074 &35	\\
O$_2$    &	5.4$\times10^{14}$	& -&	-	&	-&  1104$\pm$55	& \cite{noble2012, smith2016} & 5.98$\times10^{14}$ & 1107 &35	\\
CH$_4$	& 	9.8$\times10^{14}$	& - &	-	& -	& 1368$\pm$68& \citep{smith2016}	& 5.43$\times10^{13}$ & 1232 &47	\\
CO       &	7.1$\times10^{11}$	& 960& 1140&1310&-	& \cite{noble2012} & 9.14$\times10^{14}$& 1390 & 35	\\
O     	&	10$^{12}$	& 1300 &	-	&	1700 & 1586$\pm$ 480	&  \citep{wakelam2017, minissale2016b, he2016} & 2.73$\times10^{13}$ & 1751 & 50	\\
C$_2$H$_2$     &	3$\times10^{16}$	& - &	-	&	-& 3000$\pm$220  &  \cite{behmard2019}	& 4.99$\times10^{15}$ & 2877 &70	\\
CO$_2$	 &   9.3$\times10^{11}$/3.0$\times10^{7b}$	& 2236 &	2300	&2346	& 2105$\pm$902 & \cite{wakelam2017,edridge2013,noble2012}	& 6.81$\times10^{16}$ &  3196 & 80	\\
H$_2$S   & 10$^{12}$		& &		&	&2700 & \cite{wakelam2017,oba2018,collings2004}	& 4.95$\times10^{15}$ & 3426&85	\\
H$_2$CO  &	10$^{13}$	&- &	-	&-	&  3260 &  \cite{noble2012b}	& 8.29$\times10^{16}$ & 4117 &95	\\
CS  	&	10$^{12}$	&- &-		&-	& 3200 & \cite{wakelam2017} 	& 6.65$\times10^{16}$ & 4199 &90	\\
HCN      &	10$^{12}$	& - &	-	&	-& 3700 &  \cite{theule2011, wakelam2017}	& 1.63$\times10^{17}$ & 5344 &137	\\
NH$_3$ 	&	10$^{12}$	&- &-		&-	& 4600 & \cite{collings2003, martin-domenech2014, wakelam2017}	& 1.94$\times10^{15}$ & 5362 &105	\\
OH	     &	10$^{12}$	&- &-		&-	& 4600 &  \cite{dulieu2013, sameera2017, miyazaki2020} 	& 3.76$\times10^{15}$ & 5698 &140	\\
H$_2$O   &	3.26$\times10^{15}$	&- &-		&-	  & 5640 & \cite{sandford1990, speedy1996, fraser2001}	& 4.96$\times10^{15}$ & 5705 &155	\\
CH$_3$CN & 10$^{17}$	& 5802 &	6150&	6383			&- & \cite{bertin2017a,abdulgalil2013}	& 2.37$\times10^{17}$ & 6253 &120  	\\
CH$_3$OH &	10$^{12}$	& - &-		&	-& 5000 & \cite{wakelam2017}	& 3.18$\times10^{17}$ & 6621 &128	\\
NH$_2$CHO &	10$^{12}$	& -&	-	&-	& 6900 & \cite{wakelam2017,dawley2014,chaabouni2018}	& 3.69$\times 10^{18}$ & 9561 &176\\
C   	&	10$^{12}$	& -&	-	& -	& 14300	& \citep{shimonishi2018} & 7.38$\times10^{14}$ & 15981 & 300	\\
\hline
\hline
\multicolumn{10}{l}{$^a$see text for the choice of the preferred value. $^b$order of desorption is 1.07}
	\end{tabular}}
\end{table*}

\subsection{Parameters for species thermally desorbing from other surfaces}

In this subsection we comment on Table \ref{Table_bindings_othersurfaces} where we report the desorption parameters of atoms/molecules interacting with bare surfaces (i.e. not covered with water ice) of astrophysical interest, e.g. graphite or silicate. The table structure is similar to Table \ref{Table_bindings_ASW}. N and CS were not included in the present list since no data have been found in the case of physisorption. Binding properties of atomic nitrogen on bare surfaces were mainly studied in the chemisorption regime \cite{marinov2014, wang2021}. 
Finally, we stress that recommended $\nu$ values can differ up to 30-40\% with respect to those listed in Table \ref{Table_bindings_ASW} since T$_{peak}$ depends on the considered surface. 

\begin{description}
\item[H] We consider here only physisorption of H atoms. There are values derived from experiments and computations for many different types of surfaces. On silicates (i.e. olivine) it ranges from 290 K to 510 K depending on the morphology (crystalline vs amorphous)\cite{he2011,perets2007}. There is a rather converging estimate of H on graphite of 460 K \cite{ghio1980, bonfanti2007} slightly below the values of 660 K measured on amorphous carbon\cite{katz1999}, but discussed in the literature \cite{cazaux2004}. Moreover the main difference between amorphous or crystalline is found in the diffusion properties of the H atoms, that are much less diffusive in amorphous environments (see review and references within\cite{wakelam2017}). Having warned that the treatment of H adsorption/desorption shall include some aspects of chemisorption as well as some reactivity, we adopt a mean value of 500 K.

\item[H$_2$] Contrarily to the case of the water ice surface, the desorption of H$_2$ has not been studied specifically on bare surfaces, but through the study of H$_2$ formation\cite{pirronello1997,pirronello1999}. As with the water ice, the morphology of the surface is critical for establishing the binding energy. For the case of silicates the binding energy of D$_2$ peaks at 25 meV (290 K) ($\nu$=10$^{12}$ s$^{-1}$) whereas on amorphous surfaces it ranges from 290 to 900 K with a broad maximum at around 560 K, like shown in Fig. 4 of He et al. \cite{he2011}. We adopt this mean value, insisting on the importance of a broad distribution of E$_b$.

\item[O$_2$] Desorption of O$_2$ from HOPG was studied by  Ulbricht et al.\cite{ulbricht2006} and Smith et al.\cite{smith2016} who obtained similar couple of values [1440 K, 8$\times$10$^{13}$ s$^{-1}$] and [1419 K, 1.1$\times$10$^{13}$ s$^{-1}$] respectively. Noble et al. \cite{noble2012} reported a value of 1255 K of O$_2$ from SiO$_x$ surface. A similar value (between 1100 and 1347 K) was measured by Collings et al. \cite{collings2015} for 0.1 ML of O$_2$ on amorphous SiO$_x$. 

\item[N$_2$] Two works studied N$_2$ desorption from HOPG surfaces by measuring a couple of values of [1560 K,  5$\times$10$^{10}$s$^{-1}$]\cite{ulbricht2006} and [1395 K,  1.7$\times$10$^{13}$s$^{-1}$] K\cite{smith2016}. Ulbricht et al.\cite{ulbricht2006} find lower values for the pre-exponential factor with respect to TST theory or to similar molecules, such as O$_2$ or CO. For such reason, it seems more reasonable to use the Smith et al.\cite{smith2016} couple to calculate the recommended value. A value of 1058 -1347 K was measured by Collings et al. \cite{collings2015} for 0.3 ML of N$_2$ on amorphous SiO$_x$.

\item[CH$_4$] Methane binding energies on HOPG span from 2040 K (with $\nu=$4$\times$10$^{15}$ s$^{-1}$)\cite{ulbricht2006} to 1760 K \cite{ricca2006}. Smith et al.\cite{smith2016} reported a couple of values of [1792 K, 2.1$\times$10$^{13}$ s$^{-1}$]. 

\item[CO] As in the case of O$_2$, Ulbricht et al.\cite{ulbricht2006} and Smith et al.\cite{smith2016} reported similar values for CO desorbing from HOPG [1560 K, 2$\times$10$^{14}$ s$^{-1}$]  and [1503 K, 1.9$\times$10$^{13}$ s$^{-1}$]. A lower binding energy of 1418 K was measured for a 0.1 ML adsorbed on SiO$_x$ surface \cite{noble2012}. Recently Taj et al.\cite{taj2021} reported values between 750-1100 K for a monolayer of CO adsorbed on amorphous SiO$_x$.

\item[O] We used the value estimated by Minissale et al. \cite{minissale2016b} on oxidised HOPG. Similar results were obtained experimentally by He et al. on amorphous silicate \cite{he2014,he2015} or theoretically by Bergeron et al. \cite{bergeron2008} on a graphitic surface.

\item[H$_2$S] In spite of different works dealing with desorption properties of H$_2$S on a bare surface, only one study reports desorption parameters of hydrogen sulfide on HOPG (Puletti's PhD thesis \cite{puletti2014}). They reported an average value of 2290$\pm$360 K and $\nu$=9.6$\times$10$^{12}$ s$^{-1 }$\footnote{We stress that authors report a value of 10$^{13}$ (molec$\times$m$^{-2})^{0.1}$ s$^{-1}$ where n=0.1 is the order of desorption. We recalculate the pre-exponential factor for n=1}.

\item[C$_2$H$_2$] Acetylene desorption from bare surfaces has been scarcely studied. Behmard et al. \cite{behmard2019} investigated C$_2$H$_2$ desorption from a CsI window in the multilayer regime reporting a couple of values of [2800 K, 3$\times$10$^{16}$ s$^{-1}$]. Acetylene desorption was also studied by Collings et al. \cite{collings2004} on bare gold surface but no desorption parameters are clearly evaluated and reported. Peters and Morrison\cite{peters1986} estimated theoretically a binding energy between 2300 and 3010 K on a graphite surface.

\item[CO$_2$] Adsorbed CO$_2$ molecules on HOPG present binding energies between 2430\cite{edridge2013} and 2890 K\cite{ulbricht2006} and $\nu$ of 6.0-9.9$\times$10$^{14}$ s$^{-1}$. The binding energy is slightly higher on SiO$_x$ (3008 K)\cite{noble2012}

\item[NH$_3$] Desorption of ammonia in the multilayer regime from HOPG was studied by Bolina et al. \cite{bolina2005b}. They found E$_b$ = 2790 K and $\nu$ = 8$\times$10$^{25}$ molecules m$^{-2}$ s$^{-1}$. In the monolayer regime on HOPG, Ulbricht et al\cite{ulbricht2006} measured 3010$\pm$240 K and 5$\times$10$^{13}$ s$^{-1}$. More recently Suhasaria et al.\cite{suhasaria2015} obtained a similar value on a crystalline forsterite (Mg$_2$SiO$_4$) surface: 3103 K for the multilayer desorption and from 4100 to 3300 K for coverages of 0.2 and 1 ML, respectively.

\item[H$_2$CO] Studies in which this molecule is adsorbed on San Carlos olivine measured a E$_b$ = 3730 K ($\nu$ = 10$^{13}$ s$^{-1}$) in the sub-monolayer regime\cite{noble2012b}, a larger value than on ASW. 

\item[HCN]  No desorption parameters on bare surfaces of astrophysical interest have been found. To calculate the recommended values, we used the value of 3600 K measured by Noble et al.\cite{noble2013} on a copper mirror.

\item[CH$_3$CN] Abdulgalil et al.\cite{abdulgalil2013} measured binding energies of  4210-6010 K in the sub-monolayer regime for acetonitrile desorption from amorphous silica while Bertin et al.\cite{bertin2017b} report both E$_b$ and $\nu$ [5106 K,  8$\times$10$^{17}$ s$^{-1}$] on graphite. Recently, Ingman et al. \cite{ingman21} have reported values of 3464 - 4715 K for monolayer desorption from HOPG, with a pre-exponential value of 3.4$\times$10$^{15}$ s$^{-1}$].   

\item[OH] The study of the binding properties of OH is as puzzling as for all radical species. From an experimental point of view, the problem is even more complicated in the case of OH interaction with bare surfaces since the presence of hydroxyl radicals on the surface rapidly leads to the formation of ASW, thus preventing the study of interaction with bare surfaces. Dulieu et al. \cite{dulieu2013} report a binding energy of 4600 K on silicates. 

\item[CH$_3$OH] Methanol desorption from graphite was firstly studied by Bolina et al.\cite{bolina2005c} and Green et al. \cite{green2009} who report values of E$_b$ between 3970 to 5770 K going from 0 to 1 ML. More recently Doronin et al.\cite{doronin2015} measured a couple of values of [5454 K, 8$\times$10$^{16}$ s$^{-1}$] in agreement with values found by Ulbricht et al.\cite{ulbricht2006} on HOPG [5770 K, 3$\times$10$^{16}$ s$^{-1}$].

\item[H$_2$O] Water desorption from bare surfaces was extensively studied in the past. We cite here only a few cases: 4800 K on graphite\cite{bolina2005} and 4800 K on San Carlos olivine\cite{dulieu2013}. Ulbricht et al.\cite{ulbricht2006} 
 measured the following desorption parameters on HOPG: E$_b$ = 5530 K, $\nu$ = 9$\times$10$^{14}$ s$^{-1}$ 

\item[NH$_2$CHO] Formamide desorption from SiO$_2$ nanoparticles presents a value around E$_b$ = 7400 K.\cite{dawley2014}

\item[C] As in the case of C on ASW, we consider the chemisorption energy of 14300 K evaluated by quantum calculation study \cite{shimonishi2018}.

\end{description}

\begin{table*}
	\centering
	\caption{Binding energies of sub-monolayer regimes on other surfaces. $\nu$ in s$^{-1}$ and E in K. The selected species are listed in ascending order of E value.}
	\label{Table_bindings_othersurfaces}
\resizebox*{\textwidth}{!}{\begin{tabular}{l|c|ccc|c|c ||cc||c} %
		\hline
		\hline
	& \multicolumn{6}{c||}{Literature values} & \multicolumn{2}{c||}{Recommended values} & T$_{peak}$\\
	Species& $\nu$ & E$_{10}$ & E$_{mode}$ & E$_{90}$ & E$_{mean}$ $\pm$ $\Delta E$ & Surface/Reference  & $\nu$ & E & K \\
    \hline

\multirow{2}{*}{H}   	&10$^{12}$	&290 &	380	& 510	&-	& Olivine\cite{he2011, perets2007} & \multirow{2}{*}{1.15$\times10^{11}$}& 351 & 13	\\  
&10$^{12}$	&- &-		& -	&460	& Graphite\cite{ghio1980, bonfanti2007, katz1999} & & 431 & 13	\\  
\cdashline{1-10}
H$_2$	&10$^{13}$	& 290&	560	&	900&-	& Silicate\cite{he2011} &1.98$\times10^{11}$& 468 & 20		\\   
\cdashline{1-10}
\multirow{2}{*}{O$_2$}     &8$\times10^{13}$	& - &	-	& -	&1440	& HOPG \cite{ulbricht2006, smith2016}  & \multirow{2}{*}{ 5.98$\times10^{14}$} & 1522 & \multirow{2}{*}{41}	\\  
    &$7 \times 10^{11}$	&945 &1019		&1255	& &SiO$_x$\cite{noble2012}  & & 1385 & \\
\cdashline{1-10} 
N$_2$  &1.7$\times10^{13}$	& - &	-	& -	&1395	& HOPG \cite{ulbricht2006, smith2016}  & 4.11$\times10^{14}$ & 1538 & 41	\\  
CH$_4$	& 4$\times10^{15}$	& 1760 &	1800	& 2040	& -	& Graphite\cite{ricca2006,smith2016} &1.04$\times10^{14}$ & 1593 & 56	\\  
\cdashline{1-10}
\multirow{2}{*}{CO}     &2$\times10^{14}$	& - &	-	& -	&1560	& HOPG \cite{ulbricht2006, smith2016}  &\multirow{2}{*}{1.23$\times10^{15}$} & 1631& \multirow{2}{*}{47}	\\  
    &$10^{12}$	& 896&	1045	&1418	& &SiO$_x$\cite{noble2012}  &  & 1365& \\
\cdashline{1-10}
O     	&	10$^{12}$	&1400	&-	&1900	 &1650	& oxidised HOPG\citep{minissale2016b} & 3.10$\times10^{13}$ &  1821 & 55	\\
H$_2$S     	&	9.6$\times10^{12}$	&-	&-	&-	 &	2290$\pm$360 & HOPG\cite{puletti2014} & 4.91$\times10^{15}$ & 2616 & 75	\\ 
C$_2$H$_2$     &10$^{13}$	& 2300 & 2500		& 3010	& -& Graphite\cite{peters1986}  &  5.21$\times10^{15}$ & 2922 &70	\\
\cdashline{1-10}
\multirow{2}{*}{CO$_2$}	 & 9.9$\times10^{14}$	& 2430 & 2620	& 3000	&2890	& HOPG\cite{ulbricht2006, edridge2013} & \multirow{2}{*}{7.43$\times10^{16}$}& 3243& \multirow{2}{*}{82}	\\
& 10$^{13}$	& 2317 &2487	& 3008	&	& SiO$_x$\cite{noble2012} &   & 3738& 	\\ 
\cdashline{1-10}
\multirow{2}{*}{NH$_3$}  	& 5$\times10^{13}$	&  -&	-	&-	&3010$\pm$240	&HOPG\cite{ulbricht2006, bolina2005b} &\multirow{2}{*}{1.41$\times10^{15}$} & 3330& \multirow{2}{*}{96}	\\
& 10$^{13}$	&3300 &	-	&4100	&-	&Mg$_2$SiO$_4$\cite{suhasaria2015}  & & 4175 &	\\ 
\cdashline{1-10}
H$_2$CO  & 10$^{13}$	& -  &	- 	& -	& 3730	& Olivine\cite{noble2012b} & 2.98$\times10^{16}$& 4570& 105	\\ 
HCN      &10$^{13}$	& -&-		&-	&3600	& Cu\cite{noble2013} & 1.63$\times10^{17}$& 4909 & 135	\\ 
CH$_3$CN &8$\times10^{17}$	& -&	-	&-	&5106	& Graphite\cite{bertin2017b, abdulgalil2013} &2.73$\times10^{17}$& 4954 & 125	\\  
OH	    &	$10^{12}$ & - &	-	&	- &4600	& Olivine\cite{dulieu2013} & 3.76$\times10^{15}$ & 5698 &140	\\
CH$_3$OH & 8$\times10^{16}$ &- &	-	&-	&5454	& HOPG\cite{doronin2015, ulbricht2006, bolina2005c} &5.17$\times10^{17}$ &5728 & 147	\\  
\cdashline{1-10}
\multirow{2}{*}{H$_2$O}  &9$\times10^{14}$	& -&	-	&	-&5530$\pm$360	& HOPG\cite{ulbricht2006, bolina2005} &\multirow{2}{*}{4.96$\times10^{15}$} & 5792& \multirow{2}{*}{154}	\\ 
 &$10^{13}$	& -&	-	&	-&4800	& Olivine\cite{dulieu2013},  &  & 5755 & 	\\
\cdashline{1-10}
NH$_2$CHO & 10$^{13}$	& -&	-	&-	&7400	& SiO$_2$\cite{dawley2014} &9.53$\times10^{18}$ & 10539& 228	\\  
C   	&	10$^{12}$	& -&	-	& -	& 14300	& \citep{shimonishi2018} & 7.38$\times10^{14}$ & 15981 & 300	\\
\hline
\hline
\multicolumn{10}{l}{$^a$see text for the choice of the preferred value. $^b$order of desorption is 1.07}
	\end{tabular}}
\end{table*}

\subsection{Comments on recommended values}

In this paragraph, we comment on the uncertainty of the recommended values provided in Tables~\ref{Table_bindings_ASW} and \ref{Table_bindings_othersurfaces}.
The first source of uncertainty is directly coming from errors in the experimental measurements or quantum chemical estimations. The order of magnitude of absolute error in quantum chemical calculations is usually 1~kcal/mol$\sim$500 K, but the accuracy can be improved by comparison between different systems. In experimental measurements, where the desorption fluxes are measured as a function of temperature, the accuracy can go up to few percent of the estimated value, as shown by the reduced dispersion of results of the different studies on the same system (e.g. CO, CH$_3$OH…).
The second source of uncertainty comes from the difficulty of converting a desorption flux into E$_b$ and $\nu$ parameters, or, more precisely, by the fact that these two parameters are coupled and that there is a large latitude in the choice of the couple. We will have a full discussion of this subject in the next section. Here, we simply stress that the uncertainty in the estimation of one directly affects the calculation of the other. Nevertheless, since E$_b$ depends exponentially on the temperature, its estimation accuracy has a major impact on the results in the models. 
We restate that the recommended E$_b$ was found by equalizing the desorption fluxes at T$_{peak}$. This last parameter introduces the greatest source of uncertainty since it appears both when equalizing the desorption fluxes and in the determination of $\nu$ by TST. Clearly, the uncertainty in the T$_{peak}$ parameter affects the estimate of E$_b$, but the effect can be corrected by the benchmark on experimental values. We consider that the method increases the uncertainty, but that the relative uncertainty both for $\nu$ and E$_b$ remains below 10\% of the proposed values.
Finally, the last source of uncertainty is of physical origin and lies in the intrinsic distribution of E$_b$. For example, for H there is more than a factor of 2 between the lowest and highest estimates. For CO, which has been extensively studied, there is a variation of the E$_b$ with the coverage of almost 40\%. It is thus clear that the E$_b$ distribution remains the main uncertainty source in the choice of the E$_b$ and $\nu$ parameters for astrochemical models. In summary, we consider that the experiments can have intrinsic precision of a few \% and the methodologies for obtaining the couples limit the precision to 10\%. On the other hand, the main uncertainty is introduced by the variety of the compositions of the substrates, including the molecular neighbourhood, particularly for the most volatile species.

It should be noted that the choice of $\nu$ has an effect on the ranking of E$_b$. We can see in particular that on ASW for both H$_2$/H and O$_2$/O couples the atom is more bound to the surface than the molecule while it is the opposite for N and N$_2$. As noted earlier, this result is highly questionable for H and H$_2$ due to the large energy distributions. On the other hand, it is firmly established for oxygen, and only partially for nitrogen. More surprisingly, we see that CH$_3$OH and especially CH$_3$CN have higher E$_b$ than water when absorbed on ASW, even though these species are known to be more volatile. However,  when recalculating the desorption fluxes, we find that CH$_3$CN and CH$_3$OH desorb at lower temperatures than H$_2$O due to the different $\nu$ value. For much lower desorption flux conditions such as those of astrophysical environments, we will see in the final section that the snowline of methanol is now almost identical to that of water.
\section{5. Mixtures}

\begin{figure}[ht]
\centering
\includegraphics[width=8cm]{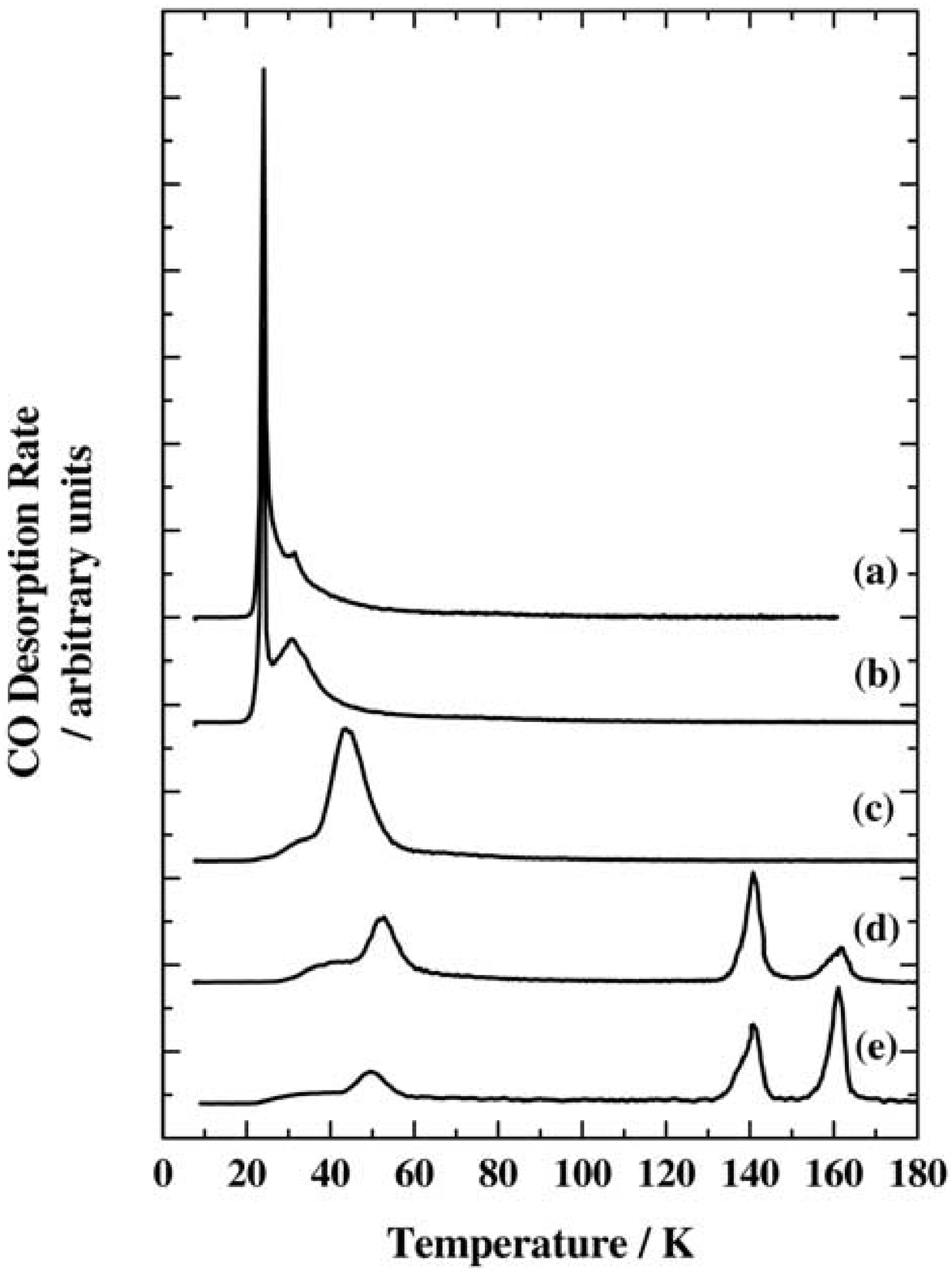}
\caption{The effect of water on the desorption of CO ice. 
TPD of $^{13}$CO deposited at 8 K (a) on a gold substrate; (b) on low density amorphous water ice deposited at 120 K; (c) on low density amorphous water ice deposited at 70 K; (d) on high density amorphous water ice deposited at 8 K; and (e) deposited simultaneously with \ce{H2O} at 8 K. Heating rate = 0.08 K s$^{-1}$. Traces are offset for clarity. Reproduced from Collings et al. \cite{collings2003}}
\label{fig:COwater}
\end{figure}

Whilst much of the discussion in this article has focused on pure ices on a range of surfaces (water ice, carbonaceous and silicaceous surfaces), real ices in astrophysical environments consist of a mixture of components. The physical chemistry of mixing is, in general, well understood\cite{atkins2014} at equilibrium. However, under the non-equilibrium conditions that exist in TPD experiments, the thermodynamics of mixing is less well understood. Depending on the region, the main component of astrophysical ices is usually water\cite{noble2013b, Boogert15}. Hence, understanding the desorption of ice mixtures containing water and other components is extremely important in an astrophysical context. The presence of mixed ices in astrophysical environments adds considerable complexity, with processes such as diffusion \cite{watanabe2010, mispelaer2013, minissale2014}, segregation \cite{acharyya2007,noble2015, nguyen2018}, and crystallisation \cite{smith1997, smith2011} all playing a part in the thermal processing and desorption of mixed ices. One of the results of this complexity is that it is generally not possible to derive specific binding energies for molecules in mixed ices in the same way as for pure ices (see section 4) as these different physical processes cannot usually be deconvoluted \cite{martin-domenech2014}. 
Since it is not possible to determine accurate binding energies for mixed ices, it is instead necessary to determine ways of describing the desorption (and the co-desorption) of the components of mixed ices so that these can be incorporated into astrophysical models in a realistic manner. We stress that the desorption behaviour of a mixed ice is not the same as the sum of the desorption behaviour of the individual components. Fig.~\ref{fig:COwater} (from Collings et al. \cite{collings2003}) shows an example of how the presence of water ice in a mixture affects the desorption of a common ice component, CO \cite{collings2003}. 
Clearly the presence of water ice has a big effect on the desorption of CO (from Fig.\ref{fig:COwater}b to Fig.\ref{fig:COwater}e)). In particular, CO is trapped within the pores of the water ice structure  and desorbs from the surface at a much higher temperature than observed for pure CO ice (Fig.\ref{fig:COwater}(a)). Hence, it is clear that the presence of water in an ice mixture adds a considerable amount of complexity which has to be accounted for in order to account for desorption in astrophysical models appropriately. 
The desorption of ice mixtures and thermal co-desorption can be considered as an efficient grain-gas bridge mechanism, likewise UV-induced photodesorption and chemical desorption\cite{ligterink2018}.
The question then remains as to how to accurately model desorption from water ice mixtures, since binding energies cannot easily be determined. An important step towards solving this problem was made by Collings et al.\cite{collings2004} by investigating the desorption of a broad range of astrophysically relevant molecules. The molecules were placed into categories, according to their behaviour in the presence of water ice. This behaviour depends on a number of factors, including their ability to hydrogen bond to water (so-called water-like species) and their ability to diffuse through the water ice and become trapped in the pores of the water ice structure. This latter effect depends on several factors, including the size of the molecule. 

\begin{figure}[ht]
\centering
\includegraphics[width=9cm]{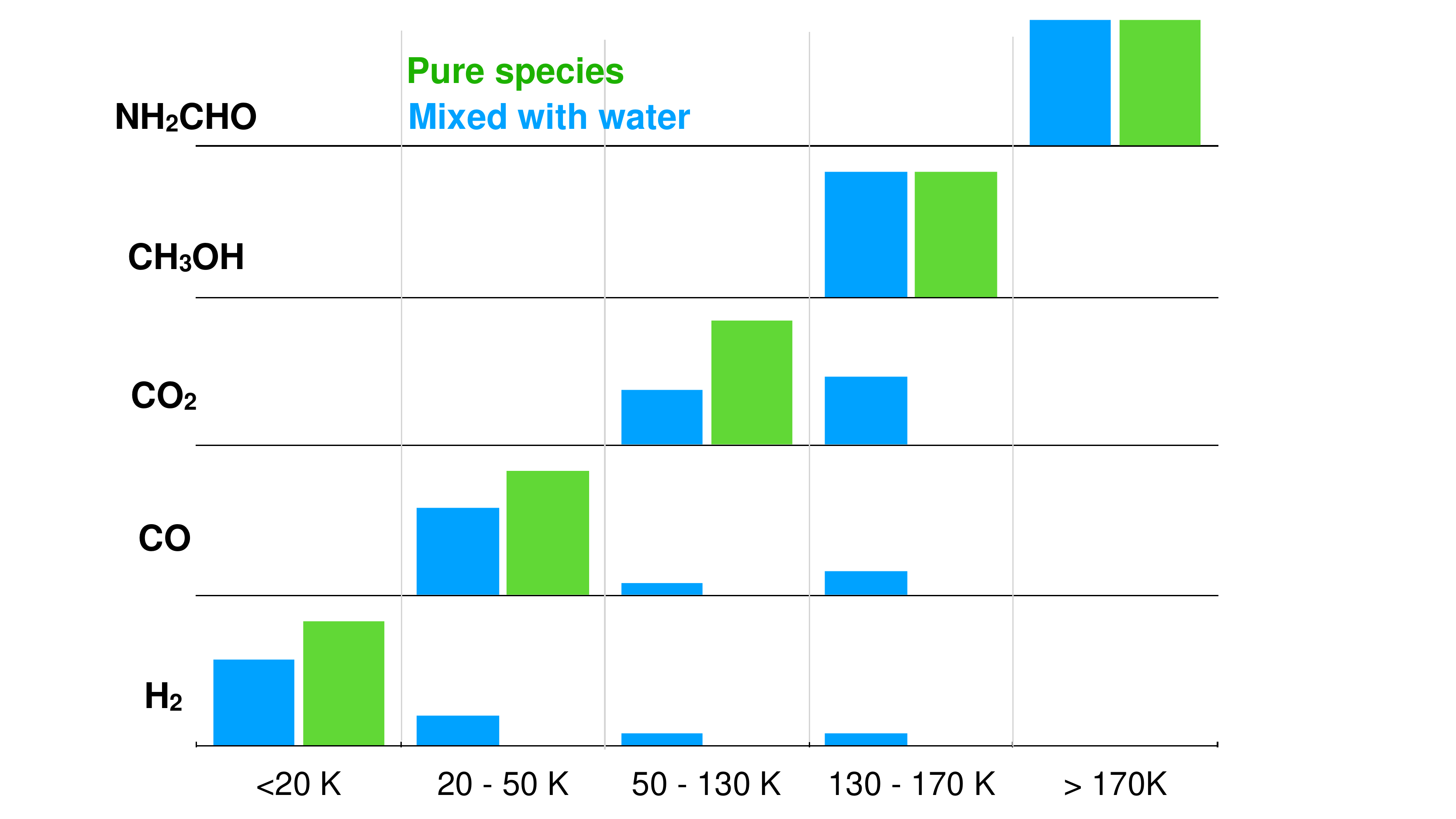}
\caption{Typical fraction of the desorbed species in the different temperature domains in usual TPD conditions (0.1 - 1 K/s) and for deposition temperatures lower than 20 K, leading to the formation of porous amorphous ice. The water desorption occurs in the 130-170 K window.}
\label{fig:histogram}
\end{figure}

Fig.\ref{fig:histogram} illustrates the general properties of molecules classified as volatile (so-called CO-like according to \cite{collings2004}, non-volatile (intermediate species according to Collings classification), refractory (species that have a higher desorption temperature for the pure ice and generally desorb after water ice \cite{congiu2012}) and hydrogen bonding species (classified as water-like by Collings et al). 
In general, as shown in Fig.\ref{fig:histogram}, volatile species diffuse into the water ice structure and become trapped in the pores of water ice, when found in a mixed ice, and only desorb when the water desorbs (Fig.\ref{fig:histogram}). These species also show a small amount of desorption that can be assigned to desorption of a monolayer of the molecules on the surface of the water ice. Common astrophysical species that fall into this category include CO, N$_2$, O$_2$ and CH$_4$. Water-like species, on the other hand, form strong hydrogen bonds to water and hence only desorb from a mixed ice when the water itself desorbs (co-desorption). This is illustrated in Fig.\ref{fig:histogram}. Molecules in this category include CH$_3$OH, NH$_3$ and HCOOH. Intermediate species (otherwise known as non-volatile species) fall in between these two categories and include molecules such as H$_2$S, CO$_2$, OCS, SO$_2$, CS$_2$, C$_2$H$_2$ and CH$_3$CN. These species can diffuse into the water structure and be trapped in the pores, but the extent of this is much less than for volatile species. These species also show evidence of the formation of a monolayer on the water ice surface, even when adsorbed in a mixed ice. The size of the molecule is particularly important for these molecules in dictating their diffusion through the pores of the water ice.

\subsection{Mixtures in the sub-monolayer regime and segregation effects}

\begin{figure}[ht]
\centering
\includegraphics[width=9cm]{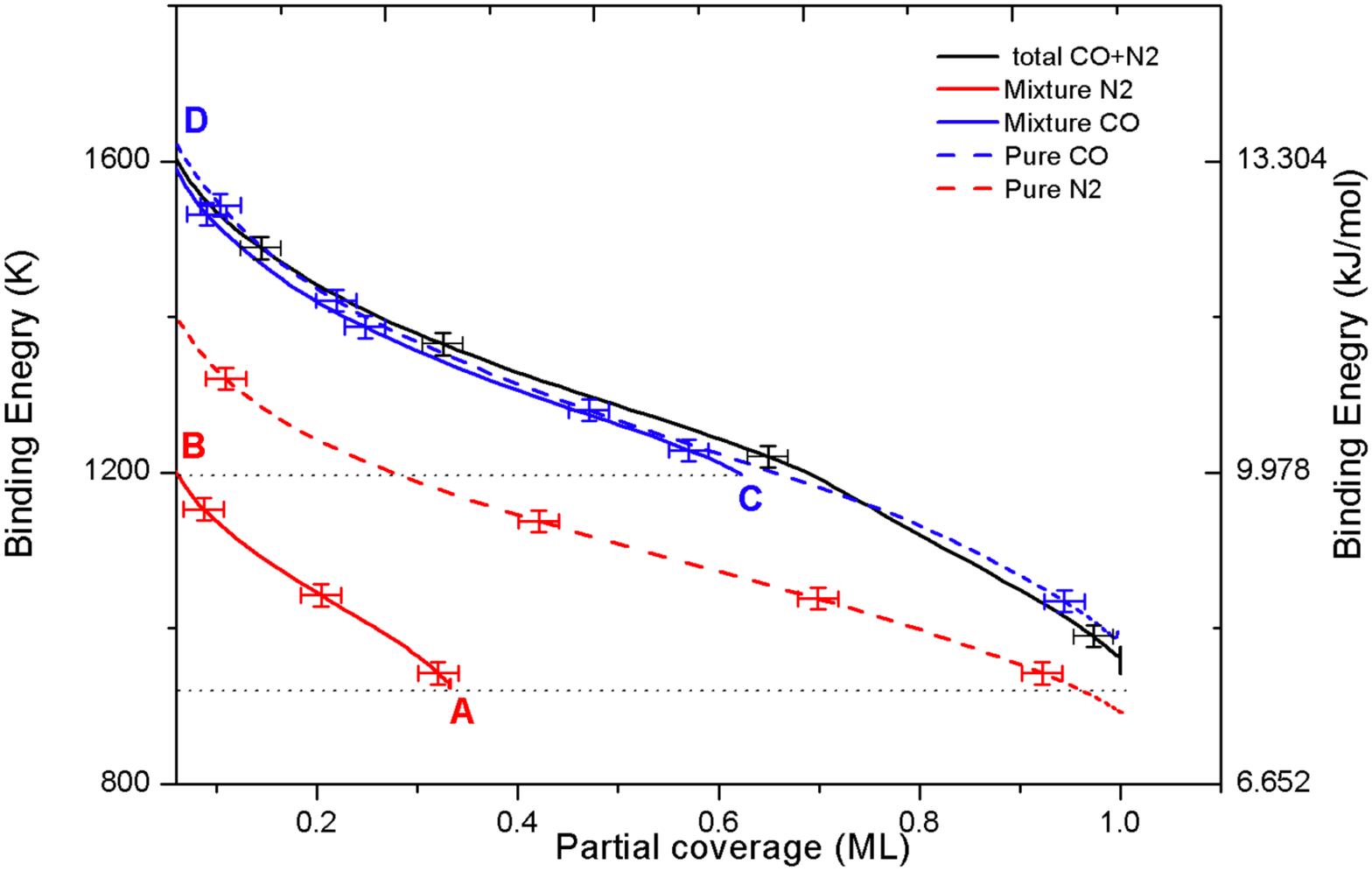}
\caption{Binding energy distributions of pure CO and N$_2$ (dashed lines) and of a 0.65 ML CO - 0.35 ML N$_2$ mixture (solid lines) on c-ASW obtained using the classical inversion method. Reproduced from Nguyen et al.\cite{nguyen2018}}
\label{fig:Segregation}
\end{figure}

Binding energies of pure ices strongly depend on the nature of the underlying substrate in the case of the submonolayer regime\cite{fayolle2016}. For example, on a non-porous ASW surface, the measured binding energy of a ML of CO is 870 K, while on porous ASW the binding energy is 980 K. Similarly for N$_2$ on np-ASW E$_b$ = 790 K while it is 900 K on p-ASW, an increase of 14\% \cite{he2016}. In the sub-monolayer regime, the coverage influences even more the binding energy with differences up to 50\% between a full monolayer and 0.1 ML \cite{noble2012, he2016}. Such strong dependence on the substrate and coverage has been also observed for mixtures in the sub-monolayer regime where segregation can further affect binding energies.
Segregation effects were observed for light species such as H$_2$; Dulieu et al.\cite{dulieu2005} show experimentally a strong isotopic segregation of molecular hydrogen on the surface of p-ASW when co-deposited with D$_2$. Similarly, segregation of O$_2$ and N$_2$ was observed when these molecules are mixed with CO. Contrary to the multilayer regime, Noble et al.\cite{noble2015} show that co-adsorbed O$_2$ and CO molecules present very different desorption behaviour when mixed and when adsorbed separately on an amorphous surface. Similar results have been observed in the case of N$_2$ and CO molecules\cite{fuchs2006,noble2015}. The two molecules present a similar shape of their binding energy distribution when deposited as pure species. On the contrary, when mixed, CO forces N$_2$ to lower energy adsorption sites, and nitrogen almost completely desorbs from the surface before CO desorption begins\cite{nguyen2018}. 
CO molecules are able to dislodge O$_2$ or N$_2$ molecules, thanks to a slightly higher affinity with the surface, by provoking a strong reduction of the binding energy of O$_2$ and N$_2$. Nguyen et al.\cite{nguyen2018} measured a decrease of around 150 K for 
N$_2$  molecules mixed with CO as shown in Fig.~\ref{fig:Segregation} (from Nguyen et al. \cite{nguyen2018}) where binding energies of CO and N$_2$ are compared in the case of CO:N$_2$ mixture (solid lines) and pure species (dashed lines) experiments.
The surface segregation mechanism should not affect the depletion for dust temperatures lower than 15 K. However, it appears that for temperatures between 15 and 30 K, where dust grains are partially covered with CO, O$_2$ and N$_2$ present on the grain, these species could be forced to thermally desorb.
Moreover, it seems that the segregation mechanism is likely to occur for molecules with similar properties.
As claimed by Noble et al.\cite{noble2012} in molecular clouds, where densities and temperatures continuously evolve, it is possible that the dislodging mechanism can strongly affect the gas and solid phase abundances of key species.
\section{6. Limitations and frontiers}


\subsection{Estimating the pre-exponential factor} 
\label{prefactor}
It is found experimentally that ln(\textit{k}) (where \textit{k} is the rate) of many processes involving atoms and molecules (reactions, desorption, diffusion) gives a straight line when plotted against 1/T. This behavior is usually described by the Arrhenius equation: 
\begin{equation}
     k = \nu \, exp\left(-\frac{E_a}{k_B T}\right)
\end{equation}
where $\nu$ is the pre-exponential factor, $E_a$ the activation energy, and $k_B$ the Boltzmann constant. $\nu$ is often called the frequency factor or attempt frequency since it is related to the frequency of vibration of a particle (atoms or molecules) in a potential well or, in other words, the number of attempts per second to overcome a given activation barrier. The Arrhenius equation is also used to describe the desorption process and the following law\cite{tielens1987,hasegawa1993} is usually used to evaluate the pre-exponential factor:
\begin{equation}\label{equation:HH}
   \nu = \sqrt{ \frac{2 N_s E_a }{\pi^2 m_X}}
\end{equation}
where $N_s$ is the number of sites per surface area and
$m_X$ is the mass of species X. For typical values of the binding energy in the case of physisorption, this formula leads to prefactor values of about $10^{12} -  10^{13} s^{-1}$. This approach works to evaluate the pre-exponential factor of atoms or small molecules such as Ar or N$_2$. However, several experimental works have pointed out that bigger molecules may exhibit prefactor values several orders of magnitude higher than what is predicted by the previous formula, sometimes reaching $10^{20} s^{-1}$ \cite{muller2003, tait2005b, doronin2015, slater2019}. This discrepancy originates from the fact that the previous formula neglects the rotational partition function of the desorbing molecules. Indeed, the pre-exponential factor reflects the entropic effect associated with the desorption kinetics, that is, in the Transition State Theory (TST), it can be written as:
\begin{equation}
\nu_{TST}=\frac{k_B T}{h}\frac{q^{\ddagger}}{q_{ads}}
\end{equation}
where  $q_{ads}$ and  $q^{\ddagger}$ are single-particle partition functions for the adsorbed (initial) state and the transition state, respectively, calculated at the temperature T.
The partition function is obtained only considering the rotational and the translational degrees of freedom, since both electronic and vibrational partition functions are equal for adsorbed and TS particles - mainly because typical desorption temperatures that we consider here ($<$ 200 K) are not sufficient to populate excited vibrational or electronic states. To evaluate the ratio, we consider the limit of a completely immobile particle in the adsorbed state: 
Tait et al.\cite{tait2005b} have proposed a method for estimating this partition function ratio, using the following equation
\begin{equation}\label{eq:nuTST}
\nu_{TST}=\frac{k_B T}{h}q^{\ddagger}_{tr, 2D} \; q^{\ddagger}_{rot, 3D}
\end{equation}
with $q_{ads}$ = 1. We use a 2D-translational partition function instead of a 3D equivalent since we subtract the contribution from translational motion perpendicular to the surface. The 2D translational partition function parallel to the surface plane is given by:
\begin{equation}
q^{\ddagger}_{tr, 2D}=\frac{A}{\Lambda^2}
\end{equation}
where A is the surface area per adsorbed molecule and $\Lambda$ is the thermal wavelength of the molecule, calculated by
\begin{equation}
\Lambda=\frac{h}{\sqrt[]{2 \,\pi\, m \,k_B \,T_{peak}}}
\end{equation}
where $m$ is the mass of the particle, $T_{peak}$ is the approximated desorption temperature that may be associated with the temperature of the TPD peak. A is fixed to 10$^{-19}$ m$^2$ (the inverse of the number of sites per unit area) for all molecules in Table~\ref{Table:Pre-factor}, except for coronene (C$_{24}$H$_{12}$) for which we used 1.1$\times$10$^{-18}$ m$^2$ \cite{thrower2013}. We stress that $T_{peak}$ can vary with the coverage and heating ramp as discussed in section~2. Such variations can lead to a variation of 20-30 \% in the determination of the pre-exponential factor, but basically will not alter the order of magnitude of the final result. The rotational partition function given in \equationame~\ref{eq:nuTST} is calculated using
\begin{equation}
q^{\ddagger}_{rot, 3D}=\frac{\sqrt[]{\pi}}{\sigma \,h^3} (8\, \pi^2 \,k_B\, T_{peak})^{3/2} \sqrt[]{I_x\, I_y \,I_z} 
\end{equation}
where I$_x$, I$_y$, and I$_z$ are the principal moments of inertia for rotation of the particle, obtained by diagonalizing the inertia tensor. The symmetry factor, $\sigma$, can be thought of classically as the number of different but indistinguishable rotational configurations of the particle. 
For diatomic molecules, we only consider the partition function in two dimensions
\begin{equation}
q^{\ddagger}_{rot, 2D}=\frac{\sqrt[]{\pi}}{\sigma \,h^2} (8\, \pi^2 \,k_B\, T_{peak}) \,\sqrt[]{I_y \,I_z} 
\end{equation}
For atoms, the rotational partition function is negligible and only the translational partition function is considered.
To evaluate the principal momentum of inertia, we used the chemical structure provided by the ChemSpider database (http://www.chemspider.com).
The main parameters used to calculate the rotational partition functions for a collection of molecules can be found in Table~\ref{tab:molpar}.

\begin{table*}
	\centering
	\caption{Symmetry numbers, $\sigma$; principle moments of inertia, I$_x$, I$_y$, and I$_z$; monolayer peak desorption temperatures, T$_{peak}$; thermal length of the molecule at T$_{peak}$, $\Lambda$; rotational partition functions, $q^{\ddagger}_{rot, 3D}$ and translational partition functions, $q^{\ddagger}_{tr, 2D}$, calculated for T$_{peak}$. The selected species are listed in ascending order of mass value.}
	\label{tab:molpar}
\resizebox*{\textwidth}{!}{	\begin{tabular}{l|cc|ccc|ccc|ccc} 
		\hline
		\hline
Species name	&	Mass	&	$\sigma$	&	I$_x$	&	I$_y$	&	I$_z$	&	T$_{peak}$	& Surface &	$\Lambda$	&	$q^{\ddagger}_{rot, 2D}$ & $q^{\ddagger}_{rot, 3D}$  	& $q^{\ddagger}_{tr, 2D}$  	\\

	&	(amu)	&		&	 \multicolumn{3}{c|}{(amu \AA$^2$)}		&	(K)	&		&	(pm)	& &	&	\\
    \hline
H	        &	1	& 1	&  	-- &--	&--	&	15$^a$	&	--	& 450	& --	& -- & 0.49 \\
H$_2$	    &	2	& 2 	&   --	 & 0.28		&	0.28	&	20$^b$	&	ASW	&	276	& 0.36	& -- & 1.31\\
C	        &	12	& 1	&  -- &  --	 &  --	&	$\ge$300$^c$	&	--	&	29.1 & -- & -- & 118.11\\
N	        &	14	& 1	&  --	 &	-- & --	& 35$^d$		&	ASW	&	78.8	& --& -- & 16.08\\
O	        &	16	& 1	&  	 --	 & --	& --	&	50$^d$		&	ASW	&	61.7	& --& -- & 26.25\\
CH$_4$  	&	16	&	12	&	3.17	&	3.17	&	3.17	&	47$^e$ & ASW	&	73.5	& --	& 2.25	&	24.67 	\\
OH	        &	17	&	1	&	--	&	0.91	&	0.91	&	140$^c$	&	ASW	&	35.1&  16.50	&-- & 78.08	\\
NH$_3$  	&	17	&	3	&	2.76	&	1.71	&	1.71	&	105$^c$	&	ASW 	&	41.3&--	& 15.11	& 58.56	\\
H$_2$O  	&	18	& 2 	&   1.83	 & 1.21		&	0.62	&	155$^c$	&	ASW	&	33.0& --	& 16.77	& 91.54	\\
D$_2$O  	&	20	& 2 	&  1.85 	 & 1.21		&	0.64	&	155$^c$	&	ASW	& 31.3&	 -- & 17.14	& 101.71	\\
C$_2$H$_2$	&	26	& 2	&  --	 &	 13.2 & 13.2	&	70$^p$	&	ASW	&	39.5 & 59.8	&  & 58.7 \\
HCN	        &	27	& 1	&  3.36$^1$	 &	 10.81 & 10.81	&	137$^f$	&	Gold	&	28.7&--	& 471.48 & 121.36 \\
CO      	&	28	&	1	&	--	&	8.60	&	8.60	&	35$^g$	& ASW &	55.8	&	38.98	 & -- &	32.15	\\
N$_2$   	&	28	&	2	&  --	&	8.43	&	8.43	&	35$^g$	& ASW &	55.8	&	19.26	& -- & 	32.15	\\
H$_2$CO	    &	30	&	2	& 14.36		&	12.88	&	1.48	&	95$^h$	&	HOPG	&	32.7&--	&113.67	  & 93.51	\\
C$_2$H$_6$	&	30	&	6	&	6.46	&	25.30	&	25.30	&	75$^e$ & ASW	&	36.8	& --& 	103.29	&	73.82	\\
O$_2$	    &	32	&	2	& --	&	9.85	&	9.85	&	35$^g$	& ASW &	52.2 	& 22.33	&-- &	36.75	\\
CH$_3$OH	&	32	& 1 &	21.06	&	20.24	&	4.01	& 128$^i$	& HOPG &	26.4 &--	&	983.73	&	143.38	\\
H$_2$O$_2$	&	34	&	2	& 16.40		&	16.14	&	1.67	&	170$^c$	&	ASW	&	22.9&--	& 345.75	& 189.64	\\
H$_2$S & 34	&  2	&	3.62	&	1.93	&	1.69	&	85$^j$	& ASW &	30.7& --	& 23.592	&	105.97	\\
Ar	&	40	&	1 & --	&--	&	--	&	26$^k$	&	ASW	&	54.1	& -- &--	&	34.12 	\\
CH$_3$CN	&	43	&	3	&	50.60	&	50.59	&	3.19	&	120$^l$	&	ASW	&	24.9	& -- &	587.60	& 161.42\\
CH$_3$NC	&	43	&	3	&	45.60	&	45.60	&	3.19	&	120$^l$	&	ASW	&	24.9	& -- &	529.54	& 161.42\\
C$_3$H$_8$	&	44	&	2	&	17.70	&	58.40	&	66.60	&	93$^e$	& ASW  & 	27.3	& -- &	1745.88	& 134.26	\\
CO$_2$  	&	44	&	2	&	42.45	&	42.45	&	2.47	&	80$^m$	&	HOPG	&	29.4	&--  & 354.18	& 115.49	\\
CS      & 44	&  1 	&	--	& 23.24		&	23.24		&	90$^f$	& ASW &	 27.7&  273.02 &--	&	129.93		\\
NH$_2$CHO	& 45	 & 1 	&	46.59	&	39.87	&	6.72	&	176$^n$	& ASW  &    19.6	& -- &	3870.91	&	259.85		\\
Xe	&	84	&1 &	--& --  & --	&	39$^k$	&	ASW	&	30.5	&	--& --	&	107.48	\\
Kr	&	131	&1 &	--	& --	& --	&55$^k$	&	ASW	&	2.07	&	--& --	& 232.78\\
C$_{24}$H$_{12}$	&	300 	&	12	&	3011.38	&	1505.70	&	1505.68	&	400$^o$	&	HOPG	& 5.03 & -- & 8.17$\times 10^{5}$ & 3937.14	\\
\hline\hline
\multicolumn{12}{l}{$^a$\cite{wakelam2017b}, $^b$\citep{wakelam2017}, $^c$ \cite{dulieu2013}, $^d$ \cite{minissale2016b}, $^e$\citep{tait2005a}, $^f$ \citep{theule2011},  $^g$\citep{smith2016}, $^h$\cite{minissale2016c}, $^i$\citep{doronin2015}, $^f$ \citep{collings2004}, $^j$\cite{puletti2014}, $^k$\cite{doronin2015b}, $^l$\citep{bertin2017a}, $^m$\citep{noble2012}, $^n$\citep{chaabouni2018}, $^o$\cite{thrower2013}, $^p$\cite{behmard2019}}
14	\end{tabular} } 
\end{table*}

\begin{figure*}[ht]
\centering
\includegraphics[width=\textwidth]{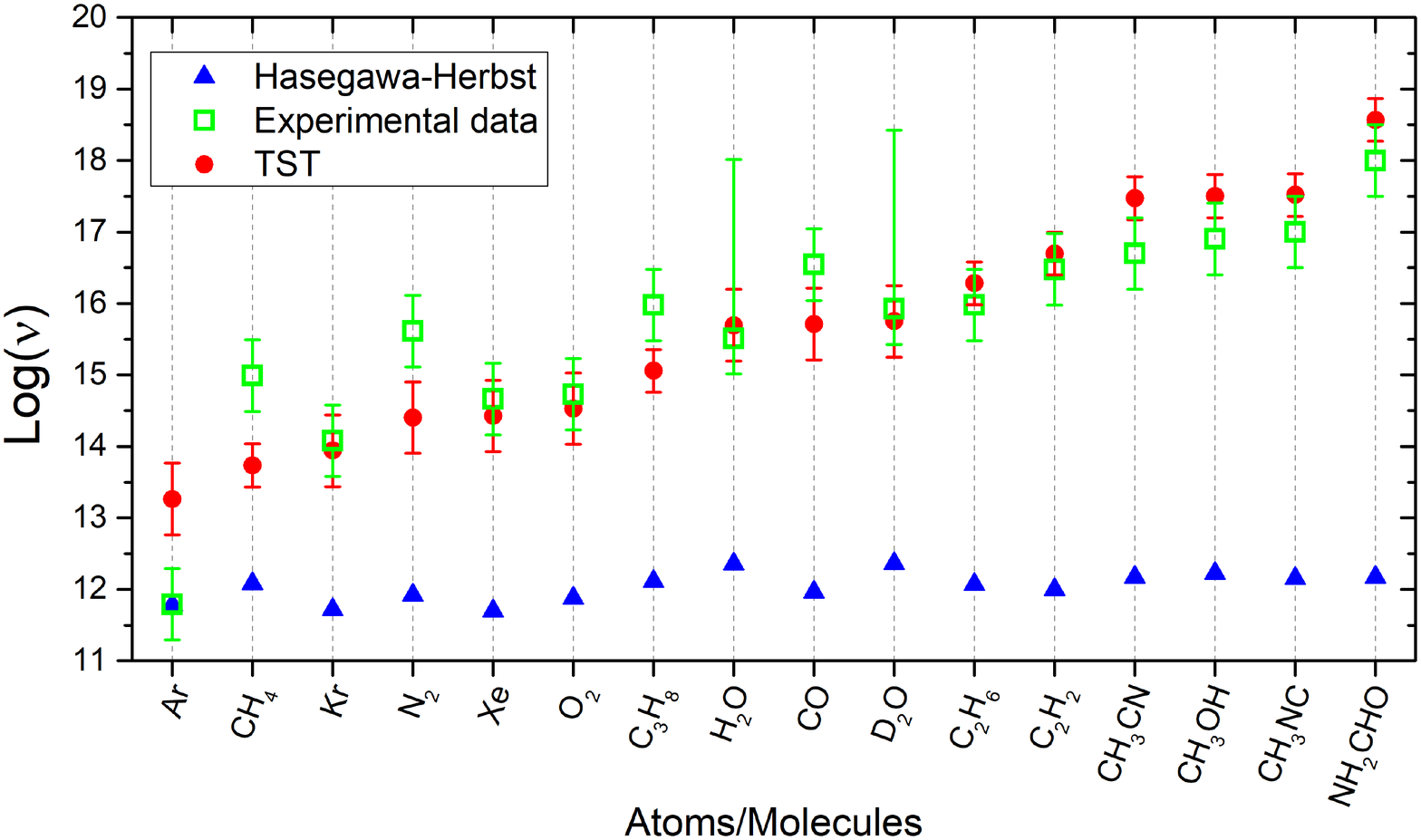}
\caption{Theoretical values for pre-exponential factors of a collection of species adsorbed onto water ices or graphite surface  (blue) using the Hasegawa \& Herbst formula \citep{hasegawa1993} and (red) using the proposed simplified Tait formula from the Transition State Theory - Eq. \ref{eq:nuTST}. A comparison is made with experimental values (green) derived from TPD experiments in ref. \cite{tait2005a, speedy1996, smith2011, smith2016, bertin2017a, doronin2015, chaabouni2018}.}
\label{fig_tait}
\end{figure*}

In Table~\ref{Table:Pre-factor}, we give a list of pre-exponential factors $\nu_{TST}$ calculated using Eq.~\ref{eq:nuTST} using the approximations depicted above, together with the experimental values of the prefactor $\nu_{exp}$ as derived from TPD studies. For these species, the prefactor using the Hasegawa \& Herbst (HH hereafter) formula (Eq.~\ref{equation:HH} - $\nu_{HH}$) commonly used in the astrochemical models, was also calculated. A comparison of these values can be found in Fig.~\ref{fig_tait}. As can be seen from the values in Table~\ref{Table:Pre-factor}, the rotational part of the partition functions accounts for an increasing weight in the total prefactor as the molecular complexity and size increases. Since this is neglected in the HH formula, we expect this formula to underestimate the pre-exponential factor. Indeed, Fig.~\ref{fig_tait} shows that the experimental values for the prefactors are always higher, sometimes by several orders of magnitude than the values given by the HH formula, except for the simplest system which is Ar. The proposed formula (Eq.~\ref{eq:nuTST}), derived from ref. \cite{tait2005a}, gives much better agreement with the measured values. This relatively simple formula for estimating pre-exponential factors based on a knowledge of the molecular parameters of the relevant species should therefore guarantee better modeling of thermal desorption by taking into account the entropic contribution to the process. 

\begin{table*}
	\centering
	\caption{List of pre-exponential factor measured experimentally, calculated through the Hasegawa \& Herbst\citep{hasegawa1993} formula using the listed binding energy and through the Transition State Theory formula adapted from Tait et al. \citep{tait2005b} in the case of rotation and rotation + translational motions. All values for pre-exponential factors are given in s$^{-1}$. The selected species are listed in ascending order of mass value.}\label{Table:Pre-factor}
\resizebox*{\textwidth}{!}{\begin{tabular}{l|cccc c} 
		\hline
		\hline
Species name	&	$\nu_{exp}$	 &	$\nu_{HH}$ & E$_{HH}$(K)&	$\nu_{rot}$ &	$\nu_{TST}$ \\
    \hline
H	        &--	 & 2.25$\times$10$^{12}$	& 300	& --	&	1.54$\times$10$^{11}$	\\
H$_2$		&--		&1.59$\times$10$^{12}$ & 300	&1.51$\times$10$^{11}$	&	1.98$\times$10$^{11}$\\
C	        &--		 & 3.75$\times10^{12}$	& $\ge$10000	&  --	&	 7.38$\times10^{14}$	\\
N	        &--		 & 9.31$\times10^{11}$ 	& 720	& --	&	 1.17$\times10^{13}$	\\
O	        & --	 	 & 1.22$\times10^{12}$& 1410 & --	&	 2.73$\times10^{13}$		\\
CH$_4$		& 9.80$\pm4.90\times10^{14a}$	& 1.20$\times10^{12}$	& 1368 &	2.20$\times10^{12}$	&	5.43$\times10^{13}$	\\
OH	     	&--		& 2.14$\times$10$^{12}$& 4600	&	4.81$\times$10$^{13}$&  3.76$\times$10$^{15}$\\
NH$_3$		&--		&1.65$\times$10$^{12}$ & 2760	&	3.31$\times$10$^{13}$&  1.94$\times$10$^{15}$\\
H$_2$O		& 1.1$\pm0.1\times$10$^{18b}$/3.26$\times$10$^{15c}$	&2.29$\times$10$^{12}$ & 5600	&	5.41$\times$10$^{13}$&  4.96$\times$10$^{15}$\\
D$_2$O  	&	5.3$\pm0.4\times$10$^{18b}$/8.45$\times$10$^{15c}$ 	 & 	2.17$\times$10$^{12}$ & 5600		& 5.53$\times$10$^{13}$	&	 5.63$\times$10$^{15}$	\\
C$_2$H$_2$ &	3$\pm$2.5$\times$10$^{16}$	 	 &	1.56$\times$10$^{12}$	&	3000	&	8.36$\times10^{13}$	&	5.21$\times10^{15}$ \\
HCN	        &	--	 	 &	1.86$\times$10$^{12}$	&	4200	&	1.34$\times10^{15}$	&	1.63$\times10^{17}$ \\
CO		& 3.50$\pm1.75\times10^{16a}$	&9.24$\times$10$^{11}$	& 1416 &	2.84$\times10^{13}$	& 9.14$\times10^{14}$	\\
N$_2$		& 4.10$\pm2.05\times10^{15a}$	&8.34$\times$10$^{11}$& 1152	&  1.4$\times10^{13}$	& 	4.51$\times10^{14}$	\\
C$_2$H$_6$	& 9.50$\pm4.75\times10^{15a}$	&1.18$\times$10$^{12}$& 2484&	1.61$\times10^{14}$	&	1.19$\times10^{16}$	\\
H$_2$CO		& --	&1.34$\times$10$^{12}$ & 3200	&	9.05$\times$10$^{14}$&  8.29$\times$10$^{16}$\\
O$_2$		& 5.40$\pm2.70\times10^{14a}$	&7.63$\times$10$^{11}$&	1104 &	1.63$\times10^{13}$	& 5.98$\times10^{14}$	\\
H$_2$O$_2$	& --	&1.72$\times$10$^{12}$ & 6000	&	1.22$\times$10$^{15}$&  2.32$\times$10$^{17}$\\
H$_2$S	        &	 --		 &	1.11$\times$10$^{12}$	& 2500		&	4.67$\times10^{13}$	& 4.95$\times10^{15}$	 \\
CH$_3$OH	& 8.0$\pm3.0\times10^{16d}$	& 1.70$\times$10$^{12}$& 5512	&	2.36$\times$10$^{15}$	&	3.18$\times$10$^{17}$		\\
Ar	    	& 6.2$\pm3.1\times10^{11a}$ & 6.04$\times$10$^{11}$& 864	&	--	&	1.84$\times$10$^{13}$ 	\\
CH$_3$CN	&  1.0$\pm0.5\times10^{17e}$	& 1.54$\times$10$^{12}$& 5802&	1.47$\times$10$^{15}$	&	2.37$\times$10$^{17}$\\
CH$_3$NC	& 5.0$\pm2.5\times10^{16e}$	&1.48$\times$10$^{12}$& 4874 	& 1.32$\times$10$^{15}$	&	2.13$\times$10$^{17}$		\\
C$_3$H$_8$	& 1.10$\pm0.55\times10^{16a}$	&1.29$\times$10$^{12}$& 3156 	&	3.38$\times$10$^{15}$	&	4.54$\times$10$^{17}$	\\
CO$_2$		& 3.26$\times$10$^{15}$	&9.31$\times$10$^{11}$ &	2260    &	5.90$\times$10$^{14}$&  6.81$\times$10$^{16}$	\\
CS	        &	-- 	 &	1.01$\times$10$^{12}$	&	2700	&	5.12$\times$10$^{14}$&  6.65$\times$10$^{16}$	 \\
NH$_2$CHO	& 1.0$\times10^{18^g}$ 	& 1.49$\times$10$^{12}$& 7460	&	1.42$\times$10$^{16}$	&	3.69$\times$10$^{18}$			\\
Kr		& 1.2$\pm0.6\times$10$^{14^h}$ &5.25$\times$10$^{11}$	&1368	&	-- & 8.73$\times$10$^{13}$		\\
Xe		& 4.6$\pm2.3\times$10$^{14^h}$	& 5.01$\times$10$^{11}$	&1956   &    --	 & 2.67$\times$10$^{14}$ \\
C$_{24}$H$_{12}$	&	3.5$\times10^{18\pm 1g}$	&9.55$\times$10$^{11}$ &	16.2$\times$10$^{3}$    &	9.14$\times$10$^{18}$&  4.45$\times$10$^{23}$	\\
\hline\hline
\multicolumn{6}{l}{$^a$ \citep{smith2016}, $^b$\citep{smith2011}, $^c$\citep{speedy1996}, $^d$\citep{doronin2015}, $^e$\citep{bertin2017a}, $^f$\citep{chaabouni2018}, $^g$\citep{thrower2013}, $^f$\citep{doronin2015b}  }
\end{tabular}} 
\end{table*}

\begin{figure}
    \centering
    (a)\includegraphics[width=0.5\textwidth]{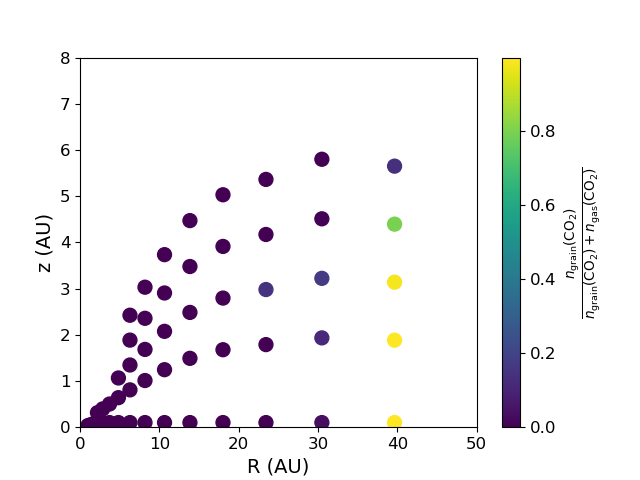}
    (b)\includegraphics[width=0.5\textwidth]{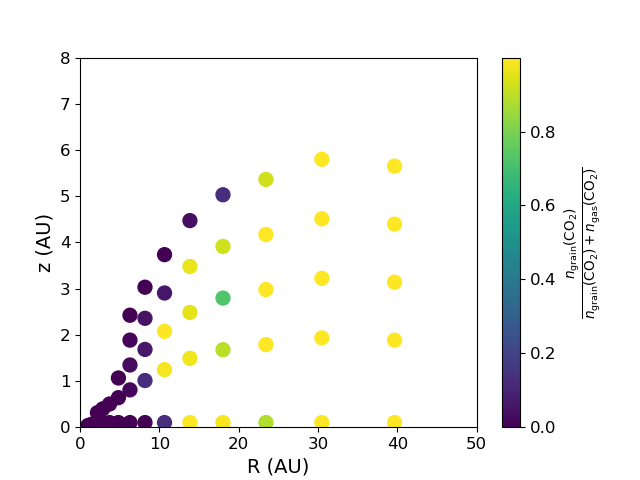}
    (c)\includegraphics[width=0.5\textwidth]{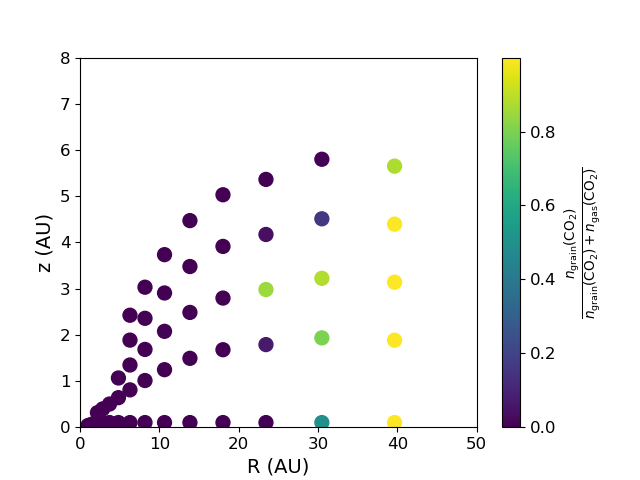}
    \caption{Fraction of \ce{CO2} in the ice at different points in a protoplanetary disk with a mass of 1.0~$M_{\astrosun}$, a sound speed of 0.26~km~s$^{-1}$ and a uniform rotation rate of $10^{-14}$~s$^{-1}$. The panel (a) shows the results with the $E_\text{mode}$ binding energies and $\nu_\text{HH}$, (b) with the recommended values reported from Table~\ref{Table_bindings_ASW} and $\nu_\text{HH}$, and (c) with the recommended binding energies and $\nu_\text{TST}$.  }
    \label{fig:CO2-disk}
\end{figure}

Fig.~\ref{fig:CO2-disk} shows the fraction of \ce{CO2} in the ice at different points in a protoplanetary disk with a mass of 1.0~$M_{\astrosun}$, a sound speed of 0.26~km~s$^{-1}$ and a uniform rotation rate of $10^{-14}$~s$^{-1}$. The same two-dimensional (2D) chemodynamical model set-up as in Visser et al.\cite{Visser2011} was used. The chemistry was described by the model by \citet{Drozdovskaya2014,Drozdovskaya2015} with latest updates in the chemistry from \citet{Fredon2021}. The first panel shows the results with the original binding energies -- typically $E_\text{mode}$-- and $\nu_\text{HH}$, the second panel shows the results with the recommended values reported from Table~\ref{Table_bindings_ASW} and $\nu_\text{HH}$, and finally, the third panel shows the effect of changing both the binding energy and the prefactor $\nu_\text{TST}$. Prefactors of all molecules in the model have been updated to $\nu_{TST}$, as well as all binding energies listed in Table~\ref{Table_bindings_ASW}.
We stress that the second case is shown only for the "pedagogical" purpose of showing the effect of changing the desorption parameters. In fact, the choice of the $E_\text{mode}$ and $\nu_\text{HH}$ value pair in the model is not based on any realistic physical assumption.
In the original model of Visser et al.\cite{Visser2011} a binding energy of 2300~K was used and a prefactor according to equation \ref{equation:HH}. The higher binding energy has clearly shifted the \ce{CO2} snow line much closer in, whereas the addition of the increased prefactor moves the snowline almost completely back. This shows that it is important to always use the combination of prefactor and binding energy. The results further show that the new recommended values shift the \ce{CO2} snowline further in. This is generally true for the species in Table~\ref{Table_bindings_ASW}.
For the sake of completeness, we stress that the prefactor can be calculated using an alternative to the approach presented above; Campbell \& Sellers \cite{campbell2012} used the thermodynamic formulation of TST, where the prefactor is linked to the standard-state entropies of molecularly adsorbed species that linearly track entropies of the gas-phase molecule. 

\subsection{\textit{f}: the desorption/diffusion ratio}

Diffusion on dust grains is a key process in understanding the chemical evolution of molecular clouds \cite{tielens2010, herbst2014}. Many molecular species are believed to be formed through the diffusive Langmuir-Hinshelwood mechanism on dust grain surfaces and icy mantles. Diffusion is often the rate-limiting step in the increasing of molecular complexity both at low temperatures, when the surface chemistry is dominated by hydrogenation reactions, and at higher temperature, when larger species become mobile. Furthermore, diffusion processes can strongly influence the conditions under which species are released back into the gas phase as the cloud collapses. 
Despite its critical importance, diffusion is a poorly understood process in the field of solid-state astrochemistry \cite{cuppen2017} and diffusion energy barriers are often not well defined. In astrochemical models, the diffusion of species is very often directly and simply assumed to be linked to the desorption energy through the \textit{f} ratio: \textit{f} = E$_{dif}$/ E$_{des}$. Since all solid-state chemistry is ruled by diffusion/reaction and diffusion/desorption competitions, this strong assumption has far going consequences.
The use of the \textit{f} ratio is a key limitation for the models, because there is no fundamental physical argument
for such a universal ratio to even exist.
Such ratio can depend on different physical-chemical parameters such as the temperature, the substrate, the surface coverage, and clearly the diffusing species. 
The \textit{f} ratio is very poorly constrained and values between 0.3 and 0.8 are used by the modeling community \citep{hasegawa1992a, cuppen2017,  Garrod2013}. Currently, the \textit{f} ratio has been determined only for few species and we give in Table~\ref{Table_ratio} a list of measured and calculated \textit{f} values.
\begin{table}[ht]
	\centering
	\caption{List of \textit{f} ratios, E$_{dif}$/E$_{des}$, for various atomic and molecular species. }
	\label{Table_ratio}
\resizebox*{0.48\textwidth}{!}{	\begin{tabular}{l|c|c} %
		\hline
		\hline
Species & \textit{f} ratio & Surface\\
    \hline
  
\multirow{2}{*}{O}      &  0.51$^a$ & ASW$^e$ \\
                        &  0.53$^a$ & Oxidised graphite\\
                        & & \\
N                       &  0.50-0.80$^a$ & ASW \\
CH$_4$	                &  0.34-0.50$^b$ & ASW \\ 
\multirow{5}{*}{CO}     & & \\
                        &  0.31-0.56$^b$ & ASW \\
                        &  0.30-0.40$^c$ & ASW \\
                        &  0.21-0.24$^d$ & CO$_2$ ice \\
                        & & \\
N$_2$                   &  0.34-0.57$^b$ & ASW \\
O$_2$                   &  0.29-0.48$^b$ & ASW \\
Ar                      &  0.31-0.48$^b$ & ASW \\
CO$_2$                  &  0.30-0.40$^c$ & ASW \\
\hline \hline
\multicolumn{3}{l}{$^a$\cite{minissale2016b} $^b$\cite{he2018} $^c$\cite{karssemeijer2014c} $^d$\cite{cooke2018}, $^e$ASW is compact for all the cases listed in this table. }
\end{tabular}}
\end{table}

In particular, the CO molecule has been extensively studied: Karssemeijer \& Cuppen \cite{karssemeijer2014c} have estimated theoretically a \textit{f} ratio lying within the 0.3-0.4 range for CO on np-ASW not far from the value measured by He et al. \cite{he2018} of 0.31-0.56. On the other hand, Cooke et al. \cite{cooke2018} reported \textit{f} values ranging from 0.21 to 0.24 in the case of CO diffusing on CO$_2$ ice. The \textit{f} ratio has been estimated even in the case of radicals, like O and N atoms. Minissale et al. \cite{minissale2016b} have determined a \textit{f}$\sim$0.57 for O atoms on np-ASW and 0.53 on oxidized graphite, while they reported a big range for N atoms on np-ASW (0.5-0.8). We stress that such a large range is a direct consequence of our poor knowledge of the diffusion mechanism of N atoms.
Such diffusion parameters are obtained by a macroscopic law (Fick’s law), but do not agree with microscopic approaches such as Monte Carlo simulations. For the diffusion of CH$_4$ within water ice, these experimental results can be reproduced by Monte Carlo simulations only using a diffusion rate at least 50 times higher.\cite{mate2020}

\subsection{Neighbors and clusters}
The binding energy of a chemical species depends on the number of other species it is attached to, e.g. number of neighbours. For water, the binding energy has been assumed to increase linearly with the number of water molecules in its neighborhood, estimated at 0.22 eV per hydrogen bond \cite{brill1967,isaacs1999,dartois2013}.
However, hydrogen bond strengths can vary with the number of water molecules involved as donors and acceptors in the neighborhood \cite{Hus2012} as well as with the presence of hydrogen bond cooperativity/anticooperativity\cite{guevara-vela2016}. In theoretical models, such as kinetic Monte Carlo simulations \cite{Cuppen2007,Garrod2013}, the binding energy of each species has been calculated as being the sum of the pair-wise interaction potentials with its neighbours. While van der Waals interactions between two molecules have been extensively studied, and the binding energy between one molecule and a surface can be measured experimentally, the step from the interaction between one species and an identical species (a dimer) to a surface represents a considerable challenge. Calculations have been made to simulate an icy surface as being a cluster of water ice. In such calculations \cite{das2018,Rimola2014}, the binding energies of many species converge toward the experimentally obtained value as the size of the water cluster increases. Such calculations illustrate how the binding energy of one chemical species changes with the number of other species it is attached to.

In astrochemical models, the chemistry occurring on dust and icy surfaces is determined by using binding energies of species derived from TPD experiments. In such experiments, binding energies are measured by increasing the temperature of the surface/ices until the species evaporates, which means that the ices are already re-organised when the measurement is performed. The binding energies measured from TPD experiments correspond to species bound with a well-organized structure of ices, where the adsorbates could have thermally diffused to the most bound sites before their desorption. Thus, TPD is more sensitive to the highest values of adsorption energies, and thus will provide higher limits to realistic binding energies values.

While such binding energies are a good approximation for an environment subject to thermal and photo processes (Hot Cores, PDRs), they do wash away the structure of the ices, and may not be correct for shielded and cold environments such as molecular clouds or starless cores \cite{cazaux2018}. In such environments, species sticking to the icy surface may have binding energies much lower than the one derived by TPD. This could have very important implications for the formation of ices and the depletion of species observed in cold environments. In starless cores, the temperature can be so low that weakly bound molecules do not re-organise and stay weakly bound to dust grains. This could be the case for CO depletion \cite{caselli1999,bergin02}, which has been observed as a function of $A_V$ \cite{keto2010}. The observed depletion is much less severe than estimated by models when using binding energies from TPD. This suggests either that processes release frozen CO back into the gas phase such as non-thermal desorption processes (CR impact, chemical desorption, electron impact, UV photolysis)\cite{hasegawa1993} or that CO has lower binding energy \cite{cazaux2017}. Considering weakly bound molecules in the CO ices changes the CO depletion in starless cores and allows a less severe depletion \cite{cazaux2017}. As a result, the morphology of the ice (i.e., the presence of pores) and the kinetics of reactions on surfaces would be very different. As a conclusion, the binding energies of species at very low temperature in shielded environments could depend on the conditions at which they have been deposited (temperature, pressure). In such conditions, the binding energies derived by TPD cannot be confidently used.

\subsection{Limitations of the TPD technique}
The TPD technique is by far the most common experimental method to determine the thermodynamic parameters involved in the thermal desorption process. From this, the binding energies can be derived, usually approximating it to the desorption energy. Although it is a powerful and well adapted technique for probing and measuring adsorption and thermal desorption dynamics and parameters, the TPD method still suffers from some limitations, that cannot be overcome, but should be considered when it comes to using the quantitative data that are extracted from the experiments. Probably the main drawback of the technique is that it is by essence destructive: the system is modified simultaneously as it is probed since it has to be warmed-up to the desorption point. This leads to several effects that may limit the pertinence of the data. 

First, it is important to stress that the adsorption energies and prefactor that are provided by the method are only strictly valid for "high" temperature systems, that is to say at temperatures where the species are desorbing. No information can be extracted for adsorption energies of systems whose temperatures are kept very low and far from the onset of thermal desorption, i.e. where no thermal rearrangement of the system could have taken place. As discussed above, the parameters which are provided by TPD are perfectly adapted for evaluating the desorption flux of an ice which is constantly warmed-up, as is the case in regions of the ISM close to a thermal source (hot cores, frontier regions of protoplanetary disks...). On the contrary, the provided energies are not characteristic of very cold systems where the molecules freeze on the surface without enough thermal activation for them to diffuse and probe the most bound adsorption sites, as can be the case for instance in water or CO snowlines in clouds or disks. As a consequence, one can consider that TPD will be more sensitive to the highest adsorption energies, i.e. the tightest bound sites, and therefore will supply a higher limit for the adsorption energies. Nevertheless, it is still possible to probe less bound adsorption sites by studying coverage close to unity, where it is impossible for the adsorbates to diffuse to more bound sites which are already occupied during the warming-up. However,  this raises the question of the relevance of such systems since the derived energies will take into consideration lateral interactions between adsorbates which have little chance to occur on interstellar ices due to the very low quantities compared to the water in the icy mantles. This point justifies by itself the importance of using several and complementary approaches, such as the experimental-theoretical comparisons depicted above, in order to supply realistic values for the thermal desorption parameters.

Secondly, the destructive character of the TPD technique is a very strong limitation in the case of adsorption of molecules on water ice. As discussed earlier, the water ice surface is a fundamental surface to consider when one wants to model the gas-grain interactions on interstellar icy grains. With TPD, it is possible to access the thermal desorption parameters for a molecule which desorbs before the water substrate, but, when the adsorbate-water interaction is higher than the cohesion of the water ice itself, then the sublimation of the supporting water surface will occur first and it is impossible to probe the molecule-water binding energy (e.g., for acetic acid, methanol, ethanol, and many other organic species that can create several hydrogen bonds with the H$_2$O surface \cite{collings2004, lattelais2011, burke2015}). This can also be seen as an advantage, since it may be a route explaining the smooth transition (i.e. non-destructive) from solid to gas phase. Obviously, this is only a possibility in higher temperature regions. We stress that such a mechanism cannot explain COMs in the gas-phase at really low temperatures. 
Today, theoretical simulation inputs are the only way to evaluate properly adsorption energies of such molecules on water ice. 

Moreover, many TPD studies focus on very simple cases where the molecules are deposited on top of a closed and pure water ice surface. In reality, the adsorbates can also be present within the water ice, or mixed with other abundant species (CO, CO$_2$...) on surfaces of interstellar dust grains. It becomes then very complicated to accurately describe the thermal desorption process. Desorption can indeed compete with other phenomena, such as diffusion-driven thermal segregation or ice reconstructions. These effects are dynamic effects dependent on the initial coverage, which are very difficult to quantify, and which may have a strong impact on the desorption dynamics. Some studies exist which aim at quantitatively understanding the thermal desorption from mixed systems\cite{martin-domenech2014}, but a lot of work still needs to be done to rationalize the conclusions (eg. \cite{noble2012, may2013}).
Finally, TPD studies are not suitable for studying the E$_b$ of radicals or reactive systems, even if alternatives have been proposed\cite{minissale2016b}. Although it is very difficult to do simulations with open shell systems, computation seems to be a faster progressing alternative. Moreover, we pinpoint that TPD analysis can be particularly complicated when studying complex systems. In fact, the quadrupole mass spectrometer, used to perform TPD experiments, needs electron impact with desorbed species that can cause fragmentation. The correct assignment of these fragments need specific calibrations and comparison with database (e.g., NIST database). Furthermore, we stress that, in some specific experimental configurations, the ionizing electrons of the quadrupole mass spectrometer may impact directly with the solid-state species inducing undesired processes (electron-induced chemistry or desorption).

\subsection{Frontiers and limitations of computational simulations/ chemistry}
In computational chemistry providing highly accurate data (and hence reliable results) is, obviously, fully desired. This also applies for gas-grain binding energies. Many computational chemical methods are available at our fingertips (e.g., only limited to the DFT domain, hundreds of functionals have been developed), each one having its own performance. Accordingly, to find the ``right" method requires benchmarking, that is, comparing the performance of a method with a reference benchmark, which can be another method (considered a valid one) or an experimental observable. In computational chemistry, benchmarking is strongly associated with chemical accuracy (i.e., to within 1 kcal mol$^{-1}$ of experimental values) in calculation of thermodynamical parameters and it can be done in two ways: i) by comparing a theoretical method (normally computationally cheaper) with a benchmark method (computationally expensive), giving rise to the so-called \textit{theory benchmarking theory (tbt)}, and ii) by comparing the theoretical method with appropriate experimental observables, giving rise to the \textit{experimental benchmarking theory (ebt)}. A lot of work has been done adopting both approaches but mainly limited to gas phase \cite{mata2017}. Several databases, protocols and systems have been developed to facilitate the task of identifying the best method for a particular chemical problem but for gas-phase molecular systems. In contrast, for solids and surfaces, benchmarking has not received as much attention because making tests for solids is much harder than for molecules. Indeed, it is usual in the \textit{tbt} approach, that the benchmark method is CCSD(T)/CBS since it is considered the ``gold-standard". However, this method is impractical for solid systems and accordingly the \textit{ebt} approach is uniquely applicable. For crystalline solid systems, structural and energetic observables are logical choices to carry out benchmarking. However, in the particular case of astrochemical gas/grain systems the majority of the surfaces are amorphous in nature, hence ruling out the application of structural techniques for benchmarking and limiting the procedure to energetic observables. The most obvious are adsorbate binding energies and their distributions but this is not a straightforward task since difficulties to establish relations between what has been measured and the results of the calculations can arise. This is particularly true when comparing calculated binding energies for a single adsorbate species in different surface binding sites with desorption energies derived from TPD experiments, in which the desorption of adsorbate layers is measured. Due to technical limitations, adsorption of single adsorbates cannot be reproduced by experiments. Thus, for reliable benchmarking, calculations have to simulate what actually happens in the experiment, namely, desorption of layers of adsorbates. This is key for the future in this discipline. Moreover, a quantitative comparison of theoretical and experimental binding energies is complicated by several problems: 
1) experimental binding energies are mostly obtained by simulating experimental curves with the Polanyi-Wigner equation. The value of the binding energy depends on the value (used or measured) of the pre-exponential factor. In the case of theoretical binding energies, these values are not related to the pre-exponential factor. 
2) Theoretical data often do not correspond to a distribution but to individual values at individual sites. Experimentally, an overall distribution is obtained, and it is not trivial (especially because of the amorphous nature of the samples) to measure individual values on individual sites. 
Despite such problems, we are aware that a comparison between experimental and theoretical values is necessary for effective improvement of astrochemical models. Hopefully, the fruitful discussions conducted in the framework of this work open up perspective scenarios for conducting research in this direction.
\\

An additional aspect for the reliability of computed gas/grain binding energies is that the surface models mimicking the external faces of the grains have to be as realistic/representative as possible of the actual grain surfaces. In Section 2, we have presented the two usual approaches for modeling surfaces and the corresponding pros and cons.  For crystalline surfaces, since they are based on the experimental crystal structure, they are de facto realistic and the main aspect to consider for a reliable binding energy distribution is by accounting for the most extended surfaces of the crystal system, which can have different  adsorptive features. A particular difficulty when dealing with amorphous systems (as interstellar grains) is that surfaces exhibit a very rich chemical complexity and structural diversity such as surface morphologies and defects, which are fundamental in their adsorptive features.  Therefore, surface models have to take this variability into account for an accurate binding energy distribution. Obviously, to generate a unique surface model accounting for all the chemical complexity is unpractical so the usual strategy is to generate different surface models , each one paying attention to one (or a limited number) of these complexities. In the case of water ice, models based on both dense and porous materials are employed. In the former model the chemical surface features to focus on are the H/O dangling bonds present in the external surfaces which dictate their adsorptive properties. In the latter, the focus is on the porous, whose adsorptive features are different to dense external faces. The same procedure is employed for surfaces of bare dust. In olivines (i.e., interstellar silicates with general formula of (Mg,Fe)$_2$SiO$_4$), different surfaces have to be used, each one addressing particular surface defects; i.e., metal under-coordination, vacancies, and substitutions (e.g., Fe$^{2+}$ by Mg$^{2+}$).

Finally, some of the computational simulations are dramatically expensive if they are performed at a quantum mechanical level so that they are only affordable by adopting classical molecular mechanics. This is the case, for instance, when investigating dynamic properties of the adsorbate on the grain surfaces (e.g., diffusion, desorption), as they require execution of molecular dynamics and/or kinetic Monte Carlo simulations, in which several trajectory simulations with long time scales are compulsory to have statistically representative results. As mentioned in Section 2, classical molecular mechanics are based on force fields. The most recurrent force fields are well parametrized to properly simulate biological systems and the properties of bulk materials but are very limited as far as gas/grain interactions are concerned. Accordingly, reparameterization of the force fields is mandatory. The usual way to do that is by taking quantum mechanical results as reference values. However, one has to be sure that the description of the gas/grain interactions provided by quantum mechanics is accurate enough (i.e., all the interaction contributions are well described) to have a reliable force field. This is critical since it is well known that traditional DFT methods tend to fail at describing non-covalent interactions dictated by dispersion, polarizability and quadrupole moments. A possible solution to overcome these limitations, if applicable, is to reduce the gas/grain system into a model gas-substrate small (even dimer) molecular system in such a way that the reference values for reparametrization are those provided by the “gold-standard” CCSD(T) for the model system. This has been done, for instance, for CO$_2$/H$_2$O interactions in which the CO$_2$-H$_2$O dimer was used as model system for subsequent reparametrization. However, reducing the gas/grain interactions to a model system brings inherently associated limitations such as omission of cooperative effects and boundary effects, which can greatly affect the actual binding energy values. 

However, developing force fields capable of treating bond breaking or formation remains a difficult task, despite recent developments \citep{senftle2016}. In view of these limitations, new embedding methods based on QMHigh:QMlow schemes are being developed with PBC-based simulations. This approach has been successfully tested in the adsorption of propane and its cracking reaction inside the porous of acidic zeolites \cite{Berger2021}. Results indicated that the periodic embedded MP2:(PBE-D2) method reaches a high chemical accuracy ($\pm$ 4 kJ mol$^{-1}$) in detriment of the standard full PBE-D2 (which give significantly larger errors, between 16 - 20 kJ mol$^{-1}$). In the same line, high accuracy can also be obtained by adopting an ONIOM-like approach with PBC simulations. That is, the entire periodic system is computed at full DFT level, while the binding site region is computed at a higher level, e.g., CCSD(T), considering that region as a cluster system. Derivation of the binding energy values is achieved by treating the calculated energies according to the ONIOM scheme. This has been applied to derive accurate binding energies of a set of astrochemically-relevant molecules on a crystalline water ice surface model \cite{ferrero2020}.

\subsection{Equilibrium (temperature and pressure) vs. dynamics (out-equilibrium, planetary science)} 

In this work we have focused on low pressure ($\sim$10$^{-15}$~bar) kinetic chemical effects in terms of adsorption (deposition) and desorption (sublimation) governed by the binding energy of a given molecule to an ice-coated dust particle and the local temperature. In the context of planet formation, with a focus on rocky materials and minerals, an additional perspective has long been provided by elemental condensation within thermochemical equilibrium \citep{Grossman72, Lewis72}, which is also influenced by the local temperature under higher pressure conditions ($\gtrsim$10$^{-3}$ bar).  

\begin{figure*}[ht]
\centering
\includegraphics[width=0.7\textwidth]{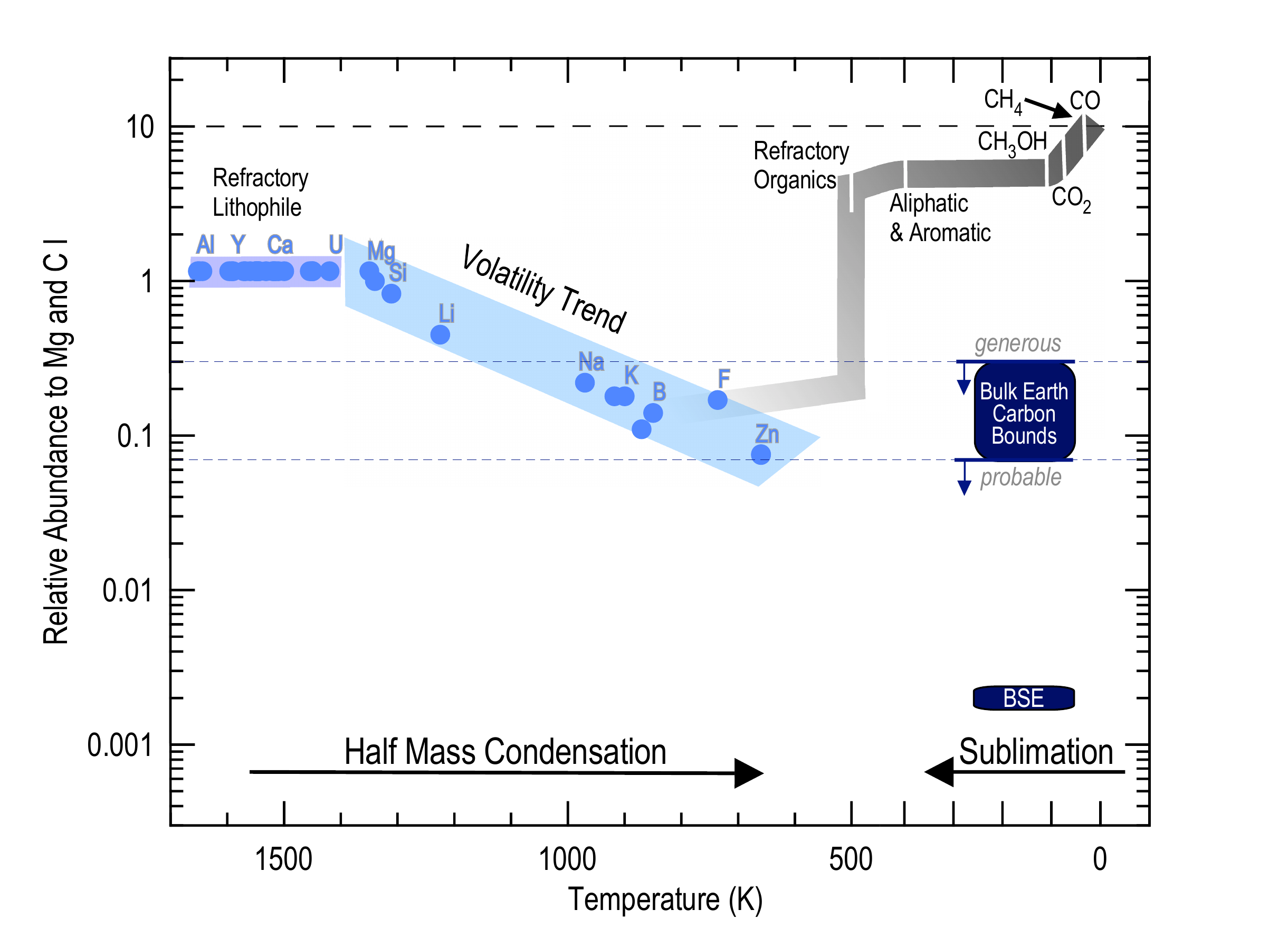}
\caption{Elemental relative abundances on Earth compared to measurements in CI chondrites (representing solar abundances) in both cases measured relative to Mg. The {\bf volatility trend associated with condensation (blue-shaded band)} describes the relative abundances of lithophile (``rock loving'') elements in the ``bulk silicate Earth'' (its mantle and crust; BSE) as a function of their half-mass condensation temperatures.  On the low-temperature side, {\bf the sublimation sequence of carbon (gray thick line)} traces the falling relative abundance of condensed carbon in the solar nebula  as the disk warms up with refractory and icy carbon carriers released into the gas.  Upper limits for the carbon content of the Bulk Earth and the estimated value for the BSE are also given.  Based on \citet{Li21}.}\label{fig:tcond}
\end{figure*}

In the condensation model, which is an important theory for the origin of Earth's composition, all elements are initially assumed to be in chemical equilibrium within hot ($>$2000~K) gas. As the gas cools various minerals form in a sequence dictated by their condensation temperature and precipitate  \citep{Grossman72}. At minimum, two pieces of evidence support this picture.  First, according to chemical equilibrium, the first solids to condense will be minerals containing calcium and aluminum and  calcium-aluminum rich inclusions (CAI) in meteorites represent the oldest solids in these bodies \citep{Amelin02}.
   Second, when normalized to Mg and solar composition there is a trend in relative abundances for lithophile elements (rock loving elements found in the mantle and not in the metal dominated core)  with their half-mass condensation temperature, the temperature at which 50\% of the mass in an element is condensed from the gas into solids (blue-shaded band in Fig.~\ref{fig:tcond}).

As a contrast, sublimation may be viewed from the perspective of ices that are provided to a young disk after forming in a cold (10--20~K) environment and then are heated to higher temperatures.  This leads to a sublimation sequence that is illustrated for the main elemental carbon carriers \citep[][]{Gail17}, both refractory and ices, in Fig.~\ref{fig:tcond} \citep{Li21}.  The primary difference in these perspectives, which might operate in tandem at different locations and/or times, is pressure.  Throughout the majority of the star formation process, and through the initial assembly of the building blocks of terrestrial worlds, the pressure is estimated below 10$^{-3}$ bar \citep{DAlessio95}, that is the (uncertain) threshold where thermochemical equilibrium is established. The exact pressure of this threshold is well established \citep[][]{Prinn81}. However, for the purposes of this review, we focus on the low pressure environments where deposition and sublimation are dominant factors in the ice to gas interaction.

\begin{figure*}[ht]
\centering
\includegraphics[width=1.0\textwidth]{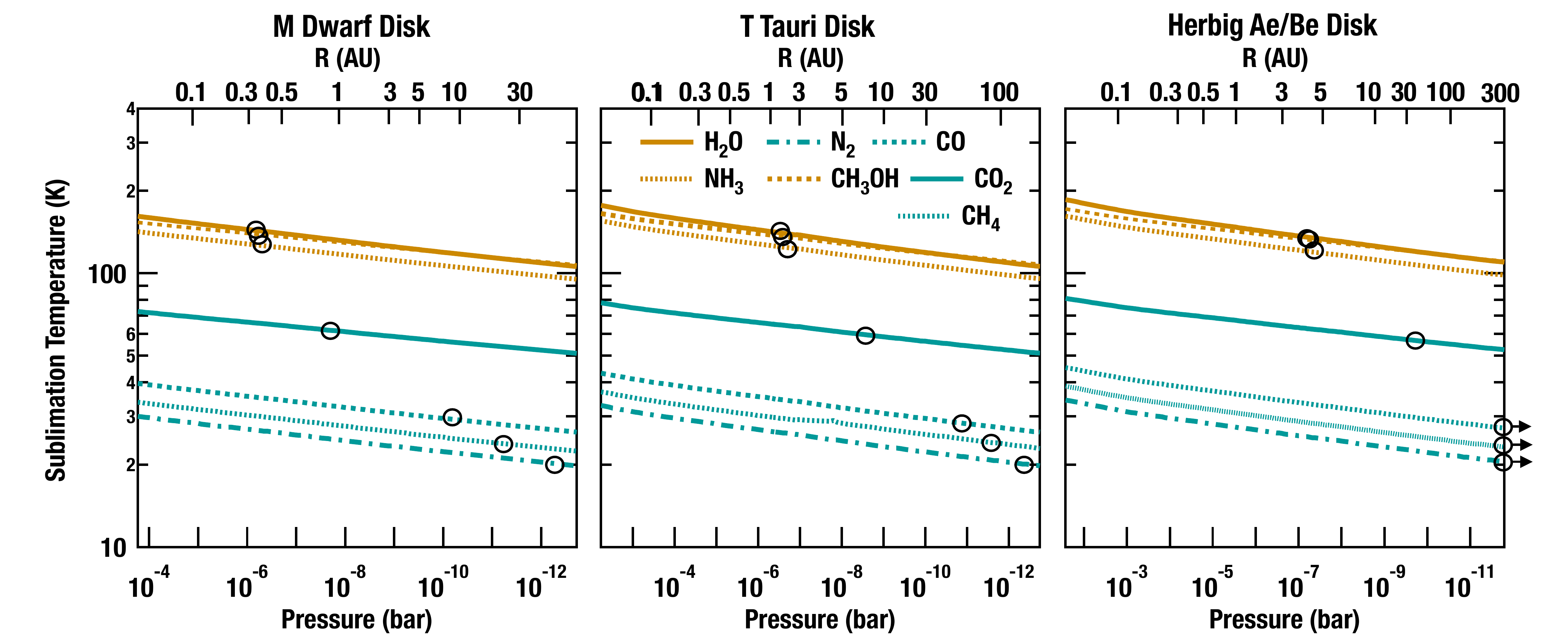}
\caption{Sublimation temperatures of key volatile carriers of C, O, and N as a function of pressure and radius using E$_b$ and $\nu$ for sub-monolayer regimes on ASW from Table~2. Three models of the midplane pressure distribution are given appropriate for disks surrounding an M dwarf ($T_{eff} = 3000$~K), a T Tauri star ($T_{eff} = 4000$~K), and an Herbig Ae/Be star ($T_{eff} = 10000$~K). Models,\textbf{ based on Eq 4 ( section 1)}, are kindly provided by  C. Walsh from published work\citep{Walsh15}.  Vapor-ice (snowline) transitions for a given molecule are denoted as open circles on the P-T line.  A central facet in comparison between these models is how the snowline moves inwards as the stellar luminosity (effective temperature) decreases.   We note that the M Dwarf disk model extended only to 50 au and we present extrapolated models.  }\label{fig:tsub}
\end{figure*}

In Fig.~\ref{fig:tsub} the import of sublimation and snow lines becomes clear by looking at how the fundamentals of the bond strength to an ASW surface varies between different molecular carriers within a planet-forming disk.   In this plot 3 models are shown as appropriate for a disk surrounding an M dwarf star, a T Tauri star, and an Herbig Ae/Be star.  The snowline locations for a given disk are labelled with circles.   Connecting this plot back to Fig.~\ref{fig:01_01} shows that planets born in different locations in the disk will (at birth) receive variable amounts of different molecular carriers of abundant elements (C, O, N).  These locations will change depending on the stellar luminosity and the strength of accretion heating in the inner disk\cite{Walsh15} all determined by the physics of the gas-grain interaction.  This illustrates the full import of the binding energy as a fundamental parameter in planet formation.

\section{7. Concluding remarks}

The evolution of star-forming regions and their thermal balance are strongly influenced by their chemical composition \cite{hocuk2014}, that, in turn, is determined by the physico-chemical processes that govern the transition between the gas phase and the solid state, specifically icy dust grains (e.g., particles adsorption and desorption). Gas-grain and grain-gas transitions as well as formation and sublimation of interstellar ices are thus essential elements of understanding astrophysical observations of cold environments (i.e., pre-stellar cores) where unexpected amounts of chemical species have been observed in the gas phase \cite{bacmann2012}. The parameterization of the physical properties of atoms and molecules interacting with dust grain particles is clearly a key aspect to interpret astronomical observations and to build realistic and predictive astrochemical models. 

In this consensus evaluation, we focus on parameters controlling the thermal desorption of ices (non-thermal processes are not discussed in detail) and how these determine pathways towards molecular complexity, define the location of snowlines and ultimately influence the planet formation process.\\
We review different crucial aspects of desorption parameters both from a theoretical and experimental point of view. We critically assess the desorption parameters (the  binding energies $E_b$ and the pre-exponential factor $\nu$) commonly used in the astrochemical community for astrophysically relevant species and provide tables with recommended values. Taking into account the range of different astrophysical environments in terms of chemical composition and surface type, we have deliberately restricted the list of molecules to those which are the most commonly encountered in the ISM, and only to two surface types (ASW and bare grain). The aim of these tables is to provide a coherent set of critically assessed desorption parameters for common use in future work. We show how a non-trivial determination of the pre-exponential factor $\nu$ using the Transition State Theory can affect the binding energy value. \\
The primary focus is on pure ices, but also the description of the desorption behavior of mixed, i.e. astronomically more realistic, ices is discussed. This allows us to discuss segregation effects. 
 Finally, we conclude this work by discussing the limitations of theoretical and experimental approaches currently used to determine the desorption properties, with suggestions for future improvements. We discuss the difficulty of benchmarking experimental results and numerical simulations, even if many points of convergence have been achieved in recent years. It appears that many of these parameters (i.e., E$_b$ and $\nu$) have been measured or calculated, but that there are still many studies to be conducted. Contrary to other aspects of astrochemical models, binding energies are starting to be globally well constrained, and we propose a compilation of values, which demonstrate the progress of this field.
However, we have shown that the E$_b$ of a pure molecular species is probably not a notion that can be directly applied to astrophysical media.
The detailed studies of pure ices now has opened the way to also study mixed ices that are more representative of 'real' inter/circumstellar ices. In the years to come, it is important to aim for experimental and theoretical studies that also allow the study of such ices at the level of accuracy that meanwhile has been reached for pure ices. 

\section{Acknowledgement}

\textbf{All authors thank the EPOC lab, and the Institute of Advanced Study of the CY Cergy Paris Universit\'{e} for their support during the meeting that catalyzed the writing of this manuscript. 
YA acknowledges support by Grant-in-Aid for Scientific Research (S) 18H05222, and Grant-in-Aid for Transformative Research Areas (A) 20H05847. AC acknowledges financial support from the Agence
Nationale de la Recherche (grant no. ANR-19-ERC7-0001-01). AR, FD and PU acknowledge funding from the European Union’s Horizon 2020 research and innovation programme under the Marie Skłodowska-Curie grant agreement No 811312 for the project "Astro-Chemical Origins” (ACO). PU acknowledges support from the Italian MUR (PRIN 2020, Astrochemistry beyond the second period elements, Prot. 2020AFB3FX).
FD, MM and VW acknowledge the french national programme "Physique et Chimie du Milieu Interstellaire" (PCMI) of CNRS/INSU with INC/INP co-funded by CEA and CNES.
FD thanks MM for accepting the difficult task of managing this project, and VW and EB to made it possible.}

\bibliography{biblio}

\
\end{document}